\documentclass[prd,aps,showpacs,twocolumn,superscriptaddress,floatfix,nofootinbib]{revtex4-1}
\usepackage[utf8]{inputenc}
\usepackage{bm}
\usepackage{bbold}
\usepackage{mathtools}
\usepackage[toc,page]{appendix}
\usepackage{dsfont}
\usepackage{combelow}
\usepackage[colorlinks=true,linkcolor=blue,citecolor=blue, urlcolor=blue]{hyperref} 
\usepackage[a4paper,top=2.4cm,bottom=2.4cm,right=1.0cm,bindingoffset=-1.2cm]{geometry}
\usepackage{tikz}
\usepackage{graphicx}
\usepackage{subfigure}
\usepackage{scalerel,amssymb}
\DeclareMathAlphabet{\pazocal}{OMS}{zplm}{m}{n}
\usepackage{ulem}

\renewcommand{\d}{\mathrm{d}}
\newcommand{\epn}{\frac{\d E_\perp^0}{\d \eta}}
\newcommand{\F}{\pazocal{F}}
\newcommand{\xT}{{\mathbf{x}_\perp}}
\newcommand{\pT}{\mathbf{p_\perp}}
\newcommand{\txT}{\tilde{\mathbf{x}}_\perp}

\newcommand{\vT}{\mathbf{v_\perp}}

\newcommand{\eavg}[1]{\left\langle #1 \right\rangle_\epsilon}

\begin{document}
\title{Opacity dependence of transverse flow, pre-equilibrium and applicability of hydrodynamics in heavy-ion collisions}
\author{Victor E. Ambru\cb{s}}
\affiliation{Institut f\"ur Theoretische Physik, Johann Wolfgang Goethe-Universit\"at, Max-von-Laue-Strasse 1, D-60438 Frankfurt am Main, Germany}
\affiliation{Department of Physics, West University of Timi\cb{s}oara, \\
Bd.~Vasile P\^arvan 4, Timi\cb{s}oara 300223, Romania}
\author{S.~Schlichting}
\affiliation{Fakultät für Physik, Universität Bielefeld, D-33615 Bielefeld, Germany}
\author{C.~Werthmann}
\email{cwerthmann@physik.uni-bielefeld.de}
\affiliation{Fakultät für Physik, Universität Bielefeld, D-33615 Bielefeld, Germany}
\affiliation{Incubator of Scientific Excellence-Centre for Simulations of Superdense Fluids, University of Wrocław, pl. Maxa Borna 9, 50-204 Wrocław, Poland}
\date{\today}

\begin{abstract}
We evaluate the full opacity dependence of collective flow in high-energy heavy-ion collisions within a microscopic kinetic description based on the Boltzmann equation in the conformal relaxation time approximation.  By comparing kinetic theory calculations to hydrodynamic and hybrid  
simulations for an average initial state, we point out shortcomings and inaccuracies of hydrodynamic models and present modified simulation setups to improve them.
\end{abstract}

\maketitle

\tableofcontents

\section{Introduction}\label{sec:intro}

Relativistic heavy ion collisions have proven to be an important tool for probing the dynamical properties of QCD matter in and out of equilibrium. Many current efforts are concerned with using experimental data to assess the conditions under which a quark gluon plasma (QGP) forms in the collision, as well as its properties \cite{Nijs:2020roc,Gardim:2019xjs,JETSCAPE:2020mzn}. Since the QGP itself cannot be directly observed, its properties have to be inferred by studying suitable aspects of the data and comparing with model descriptions. Hydrodynamics has proven to be a powerful tool for simulating the QGP dynamics~\cite{Teaney:2009qa,Song:2010mg,Gale:2013da,Heinz:2013th,Luzum:2013yya,Jeon:2015dfa} and can accurately describe data for transverse flow, which is an important indicator of collective behaviour. Modern Bayesian inference frameworks based on simulations using hydrodynamics are able to provide significant constraints on the transport properties of the medium created in the collision~\cite{Putschke:2019yrg,JETSCAPE:2020mzn,Nijs:2020roc}.

However, the conditions for applicability of hydrodynamics to describe hadronic collisions is still an open question. It is doubtful whether it can be applied to small systems with a dilute medium and large local gradients. Certainly it cannot describe the far-from-equilibrium stage right after the collision. The system will quickly approach equilibrium and start behaving hydrodynamically, 
but the time scales of its applicability in realistic systems are yet unclear.
The topics of applicability to small systems and the properties of the pre-equilibrium stage have been the focus of many recent endeavors, as described below.

In an effort to find clear distinctive features that indicate the presence or absence of a QGP, small systems have been extensively studied in experiment and have proven to feature non-vanishing transverse flow~\cite{ALICE:2014dwt,ATLAS:2017hap,CMS:2017kcs,Dusling:2015gta,Loizides:2016tew,Nagle:2018nvi} and therefore display an onset of collective behaviour. There have been many efforts in simulating these systems in hydrodynamics~\cite{Bozek:2011if,Bozek:2012gr,Bozek:2013df,Bozek:2013uha,Bozek:2013ska,Bzdak:2013zma,Qin:2013bha,Werner:2013ipa,Kozlov:2014fqa,Schenke:2014zha,Romatschke:2015gxa,Shen:2016zpp,Weller:2017tsr,Mantysaari:2017cni,Schenke:2019pmk}, which have produced reasonable results. However, in contrast to nucleus-nucleus collisions, such calculations are subject to much larger uncertainties, where in addition to the poorly constrained initial state geometry~\cite{Schenke:2014zha,Schenke:2021mxx,Demirci:2021kya}, one may question the theoretical justification for employing a hydrodynamic description for a system which features a very short lifetime and consists of very few degrees of freedom. Hence, alternative descriptions with a more sound motivation of their applicability have been put forward. For example, it has been studied whether initial state effects as described by the color glass condensate model could be the source of collective flow in small systems~\cite{Schenke:2015aqa,McLerran:2015sva,Schenke:2016lrs,Dusling:2017dqg,Dusling:2017aot,Greif:2017bnr,Mace:2018vwq,Mace:2018yvl,Kovner:2018fxj,Greif:2020rhi,Agostini:2021xca,Carrington:2021qvi}. However, it turns out that these dynamics fail to describe the important systematics~\cite{Schenke:2022mjv}.

On the other front, significant progress has been made in pushing the theoretical understanding of the dynamics in the pre-equilibrium stage and the approach to hydrodynamic behaviour and eventually equilibrium in large and small systems~\cite{Kurkela:2015qoa,Heller:2018qvh,Kurkela:2020wwb} (see also~\cite{Schlichting:2019abc,Berges:2020fwq} for recent reviews). Descriptions of Bjorken flow have been found to exhibit universal behaviour across different dynamical models and initial conditions~\cite{Berges:2013fga,Berges:2013lsa,Heller:2015dha,Spalinski:2017mel,Strickland:2017kux,Strickland:2018ayk,Spalinski:2018mqg,Giacalone:2019ldn,Kurkela:2019set,Denicol:2019lio,Almaalol:2020rnu,Heller:2020anv,Du:2020zqg,Blaizot:2021cdv,Chattopadhyay:2021ive,Du:2022bel}. The far-from-equilibrium behaviour 
depends on the setup,
but the approach to equilibrium proceeds in the same way, by 
means of 
an attractor solution that has been studied extensively. This concept has since also been applied to systems with trivial and nontrivial transverse expansion~\cite{Behtash:2019qtk,Dash:2020zqx,Ambrus:2021sjg,Ambrus:2021fej}. Phenomenologically, it has been shown that the pre-equilibrium stage has a non-negligible influence on final state observables~\cite{NunesdaSilva:2020bfs,Ambrus:2021fej,Gale:2021emg,Liyanage:2022nua,daSilva:2022xwu} and it is therefore crucial to employ realistic descriptions thereof.

An appropriate alternative dynamical model for small systems as well as pre-equilibrium is kinetic theory, which is a mesoscopic description of the phase space distribution of interacting particles and is therefore less constrained in its applicability to very dilute systems and far-from-equilibrium dynamics. Applications of kinetic theory to heavy ion collisions have been proposed already 30 years ago~\cite{Ko:1987gp,Blattel:1988zz,Li:1989zza,Botermans:1990qi} and have been used in different model scenarios to various levels in complexity. Among others, this has lead to the simulation code Boltzmann Approach to Multi-Parton Scattering (BAMPS)~\cite{Xu:2004mz,Xu:2007jv}. Several efforts have succeeded in describing transverse dynamics and the buildup of transverse flow within this description~\cite{Heiselberg:1998es,Borghini:2010hy,Romatschke:2018wgi,Kersting:2018qvi,Kurkela:2018ygx,Borghini:2018xum,Kurkela:2019kip,Kurkela:2021ctp, Borghini:2022qha,Borrell:2021cmh,Bachmann:2022cls}, in some cases even with event-by-event simulations~\cite{He:2015hfa,Greif:2017bnr,Kurkela:2020wwb,Roch:2020zdl}.

The success of relativistic hydrodynamics in describing experimental observables, demonstrated repeatedly during the past two decades \cite{Jacak:2012dx}, is heavily dependent on the various theoretical models that lead to such an effective description of strongly-interacting matter. Key ingredients include initial-state generators such as IP-Glasma \cite{Schenke:2012wb} or MC-Glauber \cite{Miller:2007ri},
QCD equation of state and realistic transport coefficients \cite{Auvinen:2020mpc}, hadronization models \cite{Cooper:1974mv}, as well as particle-based hadronic transport such as UrQMD \cite{Petersen:2008dd}. All of these stages introduce sometimes unquantifiable uncertainties, while statistical approaches such as the Bayesian analysis can be used to pinpoint the most probable parameter-set values of each of these models \cite{Bernhard:2019bmu}.

In our previous work~\cite{Ambrus:2021fej}, we found that for final state observables related to transverse flow, results from purely hydrodynamic simulations are in disagreement with results from kinetic theory even at very large opacities due to differences 
 between
the dynamics in these two theories during the pre-equilibrium phase. Even though equilibration proceeds on arbitrarily short timescales for sufficiently large opacities, conversely the rate of change of observables in this period increases, such that it still has a tangible effect on their final values. We also examined how at early times, even an inhomogeneous system obeying boost invariance can be described locally by 0+1D Bjorken flow and used the corresponding universal attractor solution to predict the time evolution before the onset of transverse expansion. This also allowed us to describe the discrepancies between hydrodynamics and kinetic theory due to pre-equilibrium in quantitative detail and verify that the size of this effect matches with the described discrepancy of final state observables.

Motivated by these results, the aim of this paper and its companion paper~\cite{Ambrus:2022qya} is to examine how in practice simulations of heavy ion collisions based on hydrodynamics can be brought into agreement with kinetic theory simulations. In the present paper, we perform an in-depth theoretical analysis of the non-equilibrium dynamics in different time evolution models based exclusively on mid-central collision events, while a broad phenomenological analysis inferring conclusions for the applicability of hydrodynamics in small systems is presented in Ref.~\cite{Ambrus:2022qya}.

The time evolution is modeled in a simplified description based on the relaxation time approximation (RTA) of conformal kinetic theory. 
 In such a simplistic model, the ultrarelativistic equation of state $\epsilon = 3P = a T^4$ can realistically describe the quark-gluon plasma only in the ultra-high temperature phase, when interactions become negligible \cite{Yagi:2005yb}. Furthermore, the bulk viscous pressure vanishes identically for a conformal fluid, while Bayesian studies indicate that bulk viscosity can play a significant role on final-state observables~\cite{Ryu:2015vwa}. Also, our conformal model gives a constant shear viscosity to entropy density ratio, $\eta / s = {\rm const}$, which is a crude approximation for the expected temperature variation of this ratio \cite{Bernhard:2016tnd,Bernhard:2019bmu}. Nevertheless, due to its simplistic evaluation of the collision kernel, the RTA has the clear advantage of being computationally cheaper over more realistic collision kernels (e.g. AMY \cite{Arnold:2002zm}). Such kernels are typically too expensive to be implemented in deterministic solvers, such as the lattice Boltzmann approach that we employ in this paper \cite{Ambrus:2018kug,Ambrus:2022adp,Ambrus:2021fej}. Previous implementations of higher-dimensional dynamics (e.g. BAMPS \cite{Xu:2004mz,Xu:2007jv}) therefore rely on a test particle algorithm and thus suffer from statistical noise. Furthermore, the first- and second-order transport coefficients computed for the RTA can be readily implemented in the relativistic hydrodynamics solver, allowing for a well-defined comparison between the two theories.
Within this model, we 
perform an analysis of the circumstances under which hydrodynamics becomes applicable as a function of opacity and time, as determined by comparing results for a set of observables related to cooling and transverse flow to kinetic theory. 
Due to the above simplifications our simulation results cannot be expected to 
realistically describe
experimental data, nevertheless we expect that our conclusions regarding the applicability of hydrodynamics also hold for more realistic models. One argument for this is that the low-momentum behaviour close to equilibrium - which is the relevant part for a comparison to hydrodynamics - should be qualitatively similar between all collision kernels.
The model setup, initial conditions and the set of observables are introduced in Sec.~\ref{sec:init}.

Apart from kinetic theory and hydrodynamics, in our work we also used other evolution models, which are discussed in Sec.~\ref{sec:evol}. We employed an expansion scheme of kinetic theory that linearizes in opacity and should agree with full kinetic theory in the limit of small interaction rates. We also employed K{\o}MP{\o}ST~\cite{Kurkela:2018wud,Kurkela:2018gitrep} as an alternative to using a full kinetic theory simulation of the pre-equilibrium phase. Switching from this description to hydrodynamics for the equilibrated system in a hybrid simulation framework is one way to properly include pre-equilibrium dynamics. K{\o}MP{\o}ST is an approximation of the dynamics of kinetic theory, which we were able to verify in quantitative detail 
by comparing to full kinetic theory, the results of which are presented in Sec.~\ref{sec:validation_of_kompost}.

Before presenting our results, we first 
discuss
in detail how pre-equilibrium is described 
in hydrodynamics and kinetic theory,
pointing out the differences between the two theories. 
To this end, in Sec.~\ref{sec:early} we introduce the 1D Bjorken flow attractor solution. This description is valid locally also for early times in 3D simulations assuming boost invariance. We use it to make predictions of the pre-equilibrium behaviour in both evolution models, including a prediction of pre-flow. Based on our results for the differences of kinetic theory and hydrodynamics in this phase, we then introduce a scaling scheme for the initial condition of hydro that can counteract these differences. This scheme relies on a timescale separation of equilibration and the onset of transverse expansion.

In Sec.~\ref{sec:space-time_evolution}, we discuss the 
time evolution of the system 
at three example opacities. On the basis of transverse profiles, we indicate how the picture changes from a close-to-free-streaming to an almost fully equilibrated system in kinetic theory. We compare the time evolution in kinetic theory and viscous hydrodynamics as well as in hybrid schemes. 
Within these hybrid schemes, the first part of the system's evolution is modelled using kinetic theory. Afterwards, we switch to hydrodynamics to model the remainder of the evolution.
For sufficiently large opacities, our proposed scaling scheme indeed brings hydrodynamics into agreement with kinetic theory after pre-equilibrium. Based on the system's equilibration, we present a useful criterion for the applicability of hydrodynamics, which can be used to define the switching times for hybrid schemes. This criterion is reached at later evolution times for smaller opacities and in some cases is never fulfilled. We find that when switching sufficiently late, hybrid schemes are also in good agreement with kinetic theory. K{\o}MP{\o}ST + viscous hydro simulations yield similar results as simulations with full kinetic theory + viscous hydrodynamics.

The range of applicability of the different schemes can best be assessed by studying the opacity dependence of final state observables. In Sec.~\ref{sec:opacity_dependence}, we compare first naive and scaled hydrodynamics to kinetic theory and establish $4\pi\eta/s\lesssim 3$ as the opacity range where the scaling scheme brings agreement. We then show results from the two hybrid simulation schemes, which can improve on scaled hydro results in the intermediate opacity range around $4\pi\eta/s\sim 3$.

In Sec.~\ref{sec:conclusion}, we present our conclusions and give a brief outlook. Appendix~\ref{app:RLB} summarizes  the details regarding the relativistic lattice Boltzmann solver that we employ for solving the kinetic equation. Appendices~\ref{app:setup_numerical_LO} and~\ref{app:details_numerical_LO} provide further details on how the linearized results in opacity expansion were obtained, while in Appendices~\ref{app:tevo_gdep} and~\ref{app:time_evolution_hybrid_kompost} we discuss some additional results for the time evolution of the system.

\section{Initial state and observables} \label{sec:init}

We will describe the time evolution of the plasma created in a collision
under the assumption of boost invariance in the longitudinal direction, 
when the phase-space distribution $f \equiv f(x, p)$ of single particles depends only
on the difference of the pseudorapidity $y = {\rm artanh}(p^z / p^t)$ and the spacetime rapidity $\eta= {\rm artanh}(z /t)$. We also assume that at initial time $\tau_0$, the particles comprising the fluid have an isotropic distribution in transverse momentum $\mathbf{p}_\perp$ and vanishing momentum along the longitudinal direction, or in other words, the longitudinal pressure $P_L$ 
measured in the local rest frame
vanishes~\cite{Mueller:1999pi}. For the latter assumption to be valid in kinetic theory simulations, we choose the initialization time $\tau_0$ to be small enough for the system to start from the early-time free-streaming attractor of kinetic theory~\cite{Kurkela:2019set}. 
Further assuming that the interparticle interactions can be modeled in the relaxation-time approximation (RTA), 
and describing only a reduced distribution function with no dependence on total momentum~\cite{Kurkela:2019kip,Ambrus:2021fej},
the initial state is fully determined by the initial transverse energy density per unit rapidity, $\d E^{0}_\perp / \d\eta \d^2\xT$. The detailed reduced distribution functions are given in Appendix~\ref{app:RLB}.

\subsection{Initial state}\label{sec:init:init}

We will use a realistic average initial condition for the $30-40\%$ most central Pb-Pb collisions (see also our companion paper~\cite{Ambrus:2022qya} for a comparison of hydrodynamization in different centrality classes). This initial condition was generated numerically on a transverse grid of size $512\times512$ in the following way. A saturation model based initial state generator was used to generate $8\times 10^6$ events with aligned directions of the impact parameter, which were then divided into centrality classes. Then the pointwise average of all events in each centrality class was taken. We made sure that in the resulting event averages statistical fluctuations are sufficiently suppressed by checking that they feature no local peaks above an energy density level of $10^{-6}$ times its maximum. More details on this event generation procedure can be found in~\cite{Borghini:2022iym}. 

Given this initial condition for $\d E^0_\perp / \d\eta \d^2\xT$, the full initial state can be constructed according to the model assumptions.  Enforcing at initial time $\tau_0$ a vanishing longitudinal pressure $P_L$ and ignoring possible initial-state transverse-plane dynamics,
the initial energy-momentum tensor 
is diagonal and has the following components:
\begin{align}
    T^{\mu\nu}(\tau_0,\xT)= {\rm diag}(\epsilon_0,\epsilon_0/2,\epsilon_0/2,0)\;,
    \label{eq:init_Tmunu}
\end{align}
where the initial energy density $\epsilon_0 \equiv \epsilon(\tau_0, \xT)$ is given by
\begin{equation}
 \epsilon(\tau_0, \xT) = \frac{1}{\tau_0} \frac{\d E^{0}_\perp}{\d\eta \d^2\xT}\;.
 \label{eq:init_eps0}
\end{equation}
In order to characterize the initial energy distribution, we define the total transverse energy per rapidity $\d E_\perp^0 / \d\eta$
\begin{equation}
 \frac{\d E^{0}_\perp}{\d\eta} = \int_{\xT} \tau_0 \epsilon_0 
  \label{eq:initial_dEdeta}
\end{equation}
and effective radius $R$
\begin{equation}
 R^2 \frac{\d E^{0}_\perp}{\d\eta} = \int_{\xT} \tau_0 \epsilon_0 
 \mathbf{x}_\perp^2\;,
 \label{eq:initial_R}
\end{equation}
where $\int_{\xT} \equiv \int \d^2 \xT$,
as well as the eccentricities $\epsilon_n$
\begin{equation}
\epsilon_n(\tau) = -\frac{\eavg{x_\perp^n\cos\left[n(\phi_x-\Psi_n)\right]}}{\eavg{x_\perp^n}}\;,
\label{eq:obs_eccn}
\end{equation}
where $\Psi_n$ are event plane angles
and the energy density weighted average over transverse space is defined as
\begin{align}
\eavg{\mathcal{O}}(\tau)=\frac{\int_\xT \mathcal{O}(\tau,\xT) \epsilon(\tau,\xT)}{\int_\xT \epsilon(\tau,\xT)}\;.
\end{align}

Based on the definitions in Eqs.~\eqref{eq:initial_dEdeta} and~\eqref{eq:initial_R}, we introduce the opacity of a system with shear viscosity to entropy density ratio $\eta /s$ via
\begin{equation}
 \hat{\gamma} = \frac{1}{5 \eta/ s} 
 \left(\frac{R}{\pi a} \frac{\d E_\perp^0}{\d\eta}\right)^{1/4},
 \label{eq:ghat}
\end{equation}
where $a$ is related to the equation of state via
\begin{equation}
 a = \frac{\epsilon}{T^4} = \frac{\pi^2\nu_{\rm eff}}{30},
\end{equation}
where $T$ is the local temperature and $\nu_{\rm eff} = 42.25$ represents the effective number of degrees of freedom of high temperature QCD~\cite{HotQCD:2014kol,Borsanyi:2016ksw}. The characteristic properties for the initial condition we use are summarized in Table~\ref{tab:profiles}. As we use a fixed profile, the parameters $R$ and $\d E_\perp^0 / \d\eta$ are also fixed and we vary $\hat{\gamma}$ 
by changing
$\eta/s$. Hence, throughout this paper, whenever discussing opacity dependencies, we will characterize the opacity via the value of the shear viscosity to entropy density ratio $\eta/s$. Note, however, that these two quantities are inversely proportional.

\begin{table*}
\centering
\begin{tabular}{c|c|c|c|c|c}
      $\d E_\perp^0 / \d\eta$ [$\rm GeV$] & $R$ [${\rm fm}$] & $\hat{\gamma} \times 4\pi \eta/s$ & $\epsilon_2$ & $\epsilon_4$ & $\epsilon_6$ \\\hline
     1280 & 2.78 & 11.3 & 0.416 & 0.210 & 0.0895 
\end{tabular}
\caption{Characteristic properties of the initial condition for the energy density used in this work, corresponding to an average over profiles in the $30-40\%$ centrality class of Pb-Pb collision at $\sqrt{s_{NN}} = 5.02\ {\rm TeV}$ \cite{Borghini:2022iym}, as discussed in Sec.~\ref{sec:init:init}.}
\label{tab:profiles}
\end{table*}

\subsection{Observables}\label{sec:init:obs}

We consider a set of observables which are measured as a function of time $\tau$. Their final state values are taken at finite time, $\tau / R= 4$. These observables are chosen such that they can be easily computed within 
the two frameworks considered in this paper, namely kinetic theory and hydrodynamics.

Specifically, we focus on observables that 
are derived from the energy-momentum tensor, which is the fundamental object of hydrodynamics and can be calculated in kinetic theory as
\begin{align}
    T^{\mu\nu} = \langle p^\mu p^\nu \rangle\;,\label{eq:Tmunu_def}
\end{align}
where angular brackets denote the microscopic average of an observable $O$ with respect to the single-particle distribution function $f$:
\begin{align}
    \langle O \rangle \equiv \int \d P\, f\, O\;,
    \label{eq:avg_momentum}
\end{align}
while $\d P = \nu_{\rm eff} \sqrt{-g} \d^3p / [(2\pi)^3 p_0]$ is the generally-covariant integration measure in momentum space.

We work in the Landau frame, where the local restframe energy density $\epsilon$ and flow velocity $u^\mu$ are given as the timelike eigenvalue and eigenvector of the energy-momentum tensor:
\begin{equation}
    T^{\mu\nu} u_\nu = T^{\mu\nu}_{\rm eq} u_\nu = \epsilon u^\mu,
    \label{eq:Landau}
\end{equation}
where the energy-momentum tensor in thermal equilibrium reads
\begin{equation}
 T^{\mu\nu}_{\rm eq} = (\epsilon + P) u^\mu u^\nu - P g^{\mu\nu}. \label{eq:Tmunueq_def}
\end{equation}

In order to facilitate the comparison between kinetic theory and relativistic hydrodynamics, we use as a substitute for $\d E_\perp/\d \eta = \tau \int_\xT \langle p^\tau p_\perp \rangle$ the integral of the transverse part of the trace of the energy-momentum tensor, $\epsilon_{\rm tr} \equiv T^{\tau\tau} - \tau^2 T^{\eta\eta} = T^{xx} + T^{yy}$, computed as
\begin{align}
 \frac{\d E_{\rm tr}}{d\eta} = \tau \int_{\xT} (T^{xx}+T^{yy}),
 \label{eq:obs_dEtrdeta}
\end{align}
 which is equal to the actual transverse energy per rapidity $\d E_\perp/\d\eta$ whenever the rapidity component of the particle momentum is negligible, $p^\eta \simeq 0$.
Similarly, instead of the flow harmonics $v_n$, we will focus on the ellipticity of the energy flow $\varepsilon_p$, defined in terms of the transverse components of the energy-momentum tensor as
\begin{equation}
 \varepsilon_p e^{2i \Psi_p}= \frac{\int_{\xT} (T^{xx}-T^{yy}+2iT^{xy})}{\int_{\xT} (T^{xx}+T^{yy})}\;,
 \label{eq:obs_epsp}
\end{equation}
where $\Psi_p$ is the symmetry plane angle of the elliptic flow $\varepsilon_p$. 

In order to characterize the expansion rate in the transverse plane, we consider the energy-weighted average of the transverse four-velocity, defined as
\begin{equation}
 \eavg{u_\perp} = \eavg{(u_x^2 + u_y^2)^{1/2}}\;.
 \label{eq:obs_uT}
\end{equation}
The local departure from equilibrium can be characterized in terms of the inverse Reynolds number,
\begin{equation}
 {\rm Re}^{-1}(\tau,\xT) = \left[\frac{6 \pi^{\mu\nu}(\tau,\xT) \pi_{\mu\nu}(\tau,\xT)}{\epsilon^2(\tau,\xT)}\right]^{1/2},\label{eq:Reinv_def}
\end{equation}
where $\pi^{\mu\nu}$ is defined as the non-equilibrium part of the energy-momentum tensor:
\begin{align}
    \pi^{\mu\nu}&= T^{\mu\nu} - T^{\mu\nu}_{\rm eq}\;. \label{eq:pimunu_def}
\end{align}
With the above normalization, ${\rm Re}^{-1} = 1$ when  $T^{\mu\nu}={\rm diag}(\epsilon,\epsilon/2,\epsilon/2,0)$, corresponding to the initial pre-equilibrium free-streaming limit. As a global measure of non-equilibrium effects in the system, we use the energy-weighted average inverse Reynolds number $\eavg{{\rm Re}^{-1}}$.


\section{Evolution Models}\label{sec:evol}

We want to compare the dynamics of several different time evolution frameworks, which first have to be introduced. In Sec.~\ref{sec:evol:RTA}, we discuss the relativistic kinetic model based on the relaxation time approximation (RTA), which is solved using the relativistic lattice Boltzmann approach \cite{Ambrus:2018kug}. Sec.~\ref{sec:opacity_expansion} discusses an analytical approach aimed at approximating the solution of the kinetic theory model for small opacities. Sec.~\ref{sec:evol:vHLLE} summarizes the equations of relativistic hydrodynamics, which are solved using the vHLLE code \cite{Karpenko:2013wva}. Finally, Sec.~\ref{sec:evol:komp} introduces the linear response framework K{\o}MP{\o}ST~\cite{Kurkela:2018wud,Kurkela:2018gitrep}, which was modified to include RTA Green's functions~\cite{Kamata:2020mka}.

\subsection{Kinetic Theory (RTA)}\label{sec:evol:RTA}

As the primary tool to investigate the time evolution of the initial configurations discussed in Sec.~\ref{sec:init:init}, we employ the relativistic Boltzmann equation in the 
Anderson-Witting
relaxation time approximation (RTA) \cite{Anderson:1974,Anderson:1974b,Cercignani:2002,Denicol:2021}:
\begin{equation}
 p^\mu \partial_\mu f = -\frac{p \cdot u}{\tau_R} (f - f_{\rm eq}),
 \label{eq:boltz}
\end{equation}
where $p^\mu = (p^\tau, \mathbf{p}_\perp, p^\eta)$ is the particle four-momentum of massless on-shell particles ($p^2 = 0$) and $\tau_R = 5(\eta / s) / T$ is the relaxation time~\cite{Florkowski:2013lya}. The prefactor is determined by the fact that in conformal RTA, the shear viscosity is given as $\eta = 4\tau_R P / 5$ and the entropy density as $s = 4P / T$. For the remainder of this paper, we will consider that the specific shear viscosity $\eta / s $ is constant.
The restframe velocity $u^\mu$ and energy density $\epsilon = a T^4$ are determined according to Eqs. (\ref{eq:Tmunu_def}),(\ref{eq:Landau}).
As $\tau_R\propto 1/T$, the system obeys conformal symmetry, which simplifies its dynamics. Introducing the reference length scale $\ell_{\rm ref} = R$ and reference energy density $\epsilon_{\rm ref} = \frac{1}{\pi R^3} (\d E_\perp^0 / \d\eta)$, Eq.~\eqref{eq:boltz} can be non-dimensionalized as 
\begin{equation}
 v^\mu \tilde{\partial}_\mu f = - v^\mu u_\mu \hat{\gamma} \widetilde{T} (f - f_{\rm eq}),
 \label{eq:boltz_nondim}
\end{equation}
where $v^\mu = p^\mu / p^\tau$,  $\tilde{\partial}_\mu=\ell_{\rm ref} \partial_\mu$, $\widetilde{T} = T / T_{\rm ref}$ and $T_{\rm ref} = (\epsilon_{\rm ref} / a)^{1/4}$. In this formulation of the equation, it becomes apparent that the time evolution of $f$ parametrically depends only on the opacity $\hat{\gamma}$ introduced in Eq.~\eqref{eq:ghat}. 
The equilibrium distribution appearing on the right-hand side of Eq.~\eqref{eq:boltz} can be identified as the Bose-Einstein distribution
\begin{equation}
 f_{\rm eq} = \frac{1}{\exp\left(p\cdot u(x)/T(x)\right)-1}\; ;
\end{equation}
however, as pointed out in~\cite{Kurkela:2019kip}, the dynamics depend only on the fact that this distribution is isotropic in the local rest frame.
The initial state corresponding to vanishing longitudinal pressure is modeled via
\begin{multline}
 f(\tau_0,\xT,\pT,y-\eta) =\frac{(2\pi)^3}{\nu_{\rm eff}}\frac{\delta(y-\eta)}
 {\tau_0 p_\perp}  \frac{\d N_{0}}{ \d^2\xT\d^2 \pT\d y},
\end{multline}
where $y -\eta = {\rm artanh}(\tau p^\eta / p^\tau)$. The initial particle distribution is assumed to be isotropic with respect to the azimuthal angle $\varphi_p = \arctan(p^y / p^x)$, being connected with the initial transverse-plane energy distribution $\d E^0_\perp / \d \eta \d^2 \xT$ via
\begin{equation}
 \frac{\d E^0_\perp}{\d\eta \d^2\xT} = 2\pi \int_0^\infty \d p_\perp\, p_\perp^2 \frac{\d N_{0}}{ \d^2\xT\d^2 \pT\d y}.
\end{equation}

In this paper, we employ the relativistic lattice Boltzmann (RLB) method \cite{Romatschke:2011hm,Succi:2018,Gabbana:2019ydb} to solve Eq.~\eqref{eq:boltz_nondim}. The full details of the algorithm are given in Sec.~IV.B of Ref.~\cite{Ambrus:2021fej}. The key ideas and simulation parameters are summarized in Appendix~\ref{app:RLB}.  
In the following, we will refer to the numerical solution obtained using the 
lattice Boltzmann algorithm as described above as ``kinetic theory''. 

\subsection{Opacity expansion}\label{sec:opacity_expansion}

For small systems, the dynamical behaviour is expected to be close to free-streaming, with only slight corrections coming from the small but finite number of interactions. In the limit of small opacity, we expand the solution of the Boltzmann equation in opacity up to linear order: $f\approx f^{(0)}+f^{(1)}$. We follow the expansion scheme that was introduced in~\cite{Heiselberg:1998es,Borghini:2010hy}, which has recently also been used in other works examining small systems~\cite{Romatschke:2018wgi,Kurkela:2018ygx,Borghini:2018xum,Kurkela:2021ctp,Bachmann:2022cls}. To zeroth order, there are no interactions, and the time evolution of the phase space distribution is computed in the free-streaming limit
\begin{align}
    p^\mu\partial_\mu f^{(0)} = 0 \label{eq:Boltzmannfreestreaming} \ \ .
\end{align}
Parametrizing the momentum space in terms of $(\pT, y)$, 
$f^{(0)}$ can be related to the distribution at initial time via
\begin{multline}
 f^{(0)}(\tau, \xT, \eta, \pT, y) = f\Bigg(\tau_0, \xT - \vT t(\tau,\tau_0, y-\eta),\\
 y - {\rm arcsinh}\left(\frac{\tau}{\tau_0} \sinh(y - \eta)\right),
 \pT, y\Bigg),
 \label{eq:LO_f0}
\end{multline}
where
\begin{equation}
 t(\tau,\tau_0,y-\eta) = \tau \cosh(y - \eta) - \sqrt{\tau_0^2 +\tau^2 \sinh^2(y - \eta)}.
 \label{eq:LO_fs_t}
\end{equation}

The linear order correction $f^{(1)}$ vanishes at initial time. Its time evolution is given by the scattering rates of the zeroth order solution,
\begin{align}
    p^\mu\partial_\mu f^{(1)} = C[f^{(0)}].\label{eq:Boltzmannfirstorder}
\end{align}
Its explicit expression and properties are presented in Appendix~\ref{app:numerical_LO}.
As parametrically the collision kernel is proportional to the opacity $\hat{\gamma}$, cf. Eq.~\eqref{eq:boltz_F}, we can indeed identify it as the expansion parameter in this scheme.

To enable our scheme to deal with arbitrary input data, the
linear order results have to be computed numerically. The computation requires performing a 6D integral, which in part can be done analytically. The details of the code for linear order results are explained in Appendix~\ref{app:numerical_LO}.

\subsection{Ideal and viscous hydrodynamics}\label{sec:evol:vHLLE}

Relativistic hydrodynamics \cite{Rezzolla:2013} is an effective macroscopic description based on the conservation equations $\nabla_\mu T^{\mu\nu} = 0$ for energy and momentum. After decomposing the energy-momentum tensor $T^{\mu\nu}$ according to Eqs. (\ref{eq:pimunu_def},~\ref{eq:Tmunueq_def}), the equations can be cast in the form
\begin{align}
 \dot{\epsilon} + (\epsilon + P) \theta - \pi^{\mu\nu} \sigma_{\mu\nu} &= 0, \nonumber\\
 (\epsilon + P) \dot{u}^\mu - \nabla^\mu P
 + \Delta^\mu{}_\lambda \partial_\nu \pi^{\lambda\nu} &= 0,
 \label{eq:hydro_cons}
\end{align}
where $\theta = \partial_\mu u^\mu$ is the expansion scalar and $\sigma_{\mu\nu} = \nabla_{\langle \mu} u_{\nu \rangle}$ is the shear tensor, while
$A^{\langle \mu\nu \rangle} = \Delta^{\mu\nu}_{\alpha\beta} A^{\alpha\beta}$, $\Delta^{\mu\nu}_{\alpha\beta} = \frac{1}{2} (\Delta^\mu_\alpha \Delta^\nu_\beta + \Delta^\nu_\alpha \Delta^\mu_\beta) - \frac{1}{3} \Delta^{\mu\nu} \Delta_{\alpha\beta}$ and $\Delta^{\mu\nu} = {g^{\mu\nu} - u^\mu u^\nu}$. 

Equation~\eqref{eq:hydro_cons} provides only four evolution equations, governing the dynamics of $\epsilon$ and $u^\mu$, leaving the dissipative shear-stress $\pi^{\mu\nu}$ as defined in Eqs. (\ref{eq:pimunu_def},~\ref{eq:Tmunueq_def}) unspecified. In ideal hydrodynamics, $\pi^{\mu\nu} = 0$ at all times, such that the system of equations in \eqref{eq:hydro_cons} becomes closed. 

Modeling dissipative effects by means of the Navier-Stokes constitutive equation $\pi^{\mu\nu} \simeq \pi^{\mu\nu}_{\rm NS} = 2\eta \sigma^{\mu\nu}$, where $\eta$ is the shear viscosity, leads to parabolic equations which violate causality and are thus incompatible with special relativity \cite{Hiscock.1983,Hiscock.1985}.
In this paper, we will consider the M\"uller-Israel-Stewart-type theory of second-order hydrodynamics \cite{Muller.1967,Israel.1979}, by which $\pi^{\mu\nu}$ evolves according to the following equation \cite{Denicol:2012cn,Denicol:2021}: 
\begin{multline}
 \tau_\pi \dot{\pi}^{\langle \mu\nu \rangle} + \pi^{\mu\nu} = 2\eta \sigma^{\mu\nu} + 
 2\tau_\pi \pi^{\langle\mu}_\lambda \omega^{\nu\rangle \lambda} \\
 - \delta_{\pi\pi} \pi^{\mu\nu} \theta - 
 \tau_{\pi\pi} \pi^{\lambda\langle \mu}\sigma^{\nu\rangle}_\lambda + \phi_7 \pi_\alpha^{\langle\mu}\pi^{\nu\rangle\alpha},
 \label{eq:hydro_pi}
\end{multline}
where $\omega_{\mu\nu} = \frac{1}{2}[\nabla_\mu u_\nu - \nabla_\nu u_\mu]$ is the vorticity tensor.
The relaxation time $\tau_\pi$, as well as the other coupling coefficients, represent second-order transport coefficients, the values of which are chosen for compatibility with RTA \cite{Jaiswal:2013npa,Molnar:2013lta,Ambrus:2022vif}: 
\begin{align}
 \eta &= \frac{4}{5} \tau_\pi P, &
 \delta_{\pi\pi} &= \frac{4\tau_\pi}{3}, &
 \tau_{\pi\pi} &= \frac{10\tau_\pi}{7}, &
 \phi_7 &= 0,
 \label{eq:hydro_tcoeffs}
\end{align}
while $\tau_\pi = \tau_R$.

Numerical solutions of Eqs.~\eqref{eq:hydro_cons} and \eqref{eq:hydro_pi} reported in this paper are obtained using the open-source viscous HLLE (vHLLE) code~\cite{Karpenko:2013wva},\footnote{Commit number \texttt{efa9e28d24d5115a8d8134852-32fb342b38380f0}.} which we modified to allow the implementation of the initial state considered in this paper (we employed vHLLE also in Ref.~\cite{Ambrus:2021fej} for a similar application). Specifically, we employed the square simulation domain $[-8R,8R] \times [-8R, 8R]$, which we discretized using $401 \times 401$ equidistant points. 
The simulations were performed until the final time $\tau_f = 5R$. The initial state was prepared using insight on the hydrodynamic attractor, as will be discussed in Sec.~\ref{sec:initialization_scaled_hydro}.
In the initial state, a background value of $10^{-7} \times \frac{R}{\tau_0} \epsilon_{\rm ref}$ was added to the energy density to prevent free-streaming artifacts in the system outskirts. The time step $\delta \tau$ was chosen dynamically,
\begin{equation}
 \delta \tau(\tau) = {\rm min} \left[\tau \left(\frac{\delta \tau}{\tau}\right)_{\rm M}, R\left(\frac{\delta \tau}{R}\right)_{\rm M}\right],
\end{equation}
where $(\delta \tau / \tau)_{\rm M} = 0.01$ and $(\delta \tau / R)_{\rm M} = 10^{-3}$.

\subsection{K{\o}MP{\o}ST}\label{sec:evol:komp}

The open-source simulation code K{\o}MP{\o}ST~\cite{Kurkela:2018wud} implements a linearized non-equilibrium time evolution of the energy-momentum tensor $T^{\mu\nu}$ based on the dynamics of a kinetic theory description. It has been developed as a practical tool for describing the early-time far-from-equilibrium dynamics of heavy ion collisions, where the system has not yet hydrodynamized and a non-equilibrium description is required. The original version of K{\o}MP{\o}ST was based specifically on the effective kinetic theory for pure glue QCD~\cite{Arnold:2002zm}. To perform accurate comparisons with the other evolution models used in this paper, a modified version based on the dynamics of RTA was used. For this, we imported the RTA Green's functions calculated in~\cite{Kamata:2020mka}. This version of K{\o}MP{\o}ST is available on Git.

K{\o}MP{\o}ST evolves a given input initial state
from an initial time $\tau_0$ to a final time $\tau$ in a single propagation step. Conceptionally, the output is expected to describe a hydrodynamized system and can be used as input for a subsequent hydrodynamic evolution model. Since the computation of this step involves linearizations in perturbations around a local average value, K{\o}MP{\o}ST has a limited range of applicability in the evolution time.

More specifically, in its default mode with energy perturbations, K{\o}MP{\o}ST propagates the energy momentum tensor in the following way: the values at each point $\bf x$ in the final state are computed from the initial values of $T^{\mu\nu}$ at all causally connected points ${\bf x}'$ in the initial state, meaning points that fulfill $|{\bf{x}}-{\bf x}'|<c(\tau-\tau_0)$. The energy-momentum tensor is divided into a spatial average of the causal past and perturbations around this average:
\begin{align*}
    T_{\bf x}^{\mu\nu}(\tau_0,{\bf x}')=\bar{T}_{\bf x}^{\mu\nu}(\tau_0)+\delta T_{\bf x}^{\mu\nu}(\tau_0,{\bf x}')\;,
\end{align*} where the subscript ${\bf x}$ denotes the fact that the average depends on the position for which the causal past is considered. The average value is evolved according to the laws of Bjorken flow dynamics, assuming local homogeneity in the transverse plane and boost invariance, while the perturbations are propagated in a linear response scheme:
\begin{multline}
    \delta T^{\mu\nu} (\tau,{\bf x})=\int \d^2{\bf x}'\; G^{\mu\nu}_{\alpha\beta}({\bf x},{\bf x}',\tau,\tau_0) \\
    \times\delta T_{\bf x}^{\alpha\beta}(\tau_0,{\bf x}')\frac{\bar{T}_{\bf x}^{\tau\tau}(\tau)}{\bar{T}_{\bf x}^{\tau\tau}(\tau_0)}\;.
\end{multline}
The Green's functions $G^{\mu\nu}_{\alpha\beta}({\bf x},{\bf x}',\tau,\tau')$ have been computed in the respective underlying kinetic theory description and are included in K{\o}MP{\o}ST.

Energy perturbations ($\delta T^{\mu\nu}$) can also be switched off, in which case K{\o}MP{\o}ST propagates only the average energy-momentum tensor taken over the causal past, as discussed above. Some of the phenomenological implications of this mode are discussed below. For all other results in this paper, we employed the modified RTA-K{\o}MP{\o}ST with energy perturbations.

\subsection{Validation of K{\o}MP{\o}ST}\label{sec:validation_of_kompost}

Before employing K{\o}MP{\o}ST to describe 
pre-equilibrium, we first checked to what extent results from modified RTA-K{\o}MP{\o}ST are in agreement with results from full kinetic theory in RTA for our specific initial condition. This comparison was done on the basis of the time evolution of the observables we examined in this paper but also for cross sections through profiles of the energy-momentum tensor after some evolution time.
All K{\o}MP{\o}ST results presented here were 
obtained using 
an initial time of $\tau_0=10^{-6}\;R$.

\begin{figure*}
\begin{tabular}{cc}
    \includegraphics[width=.48\textwidth]{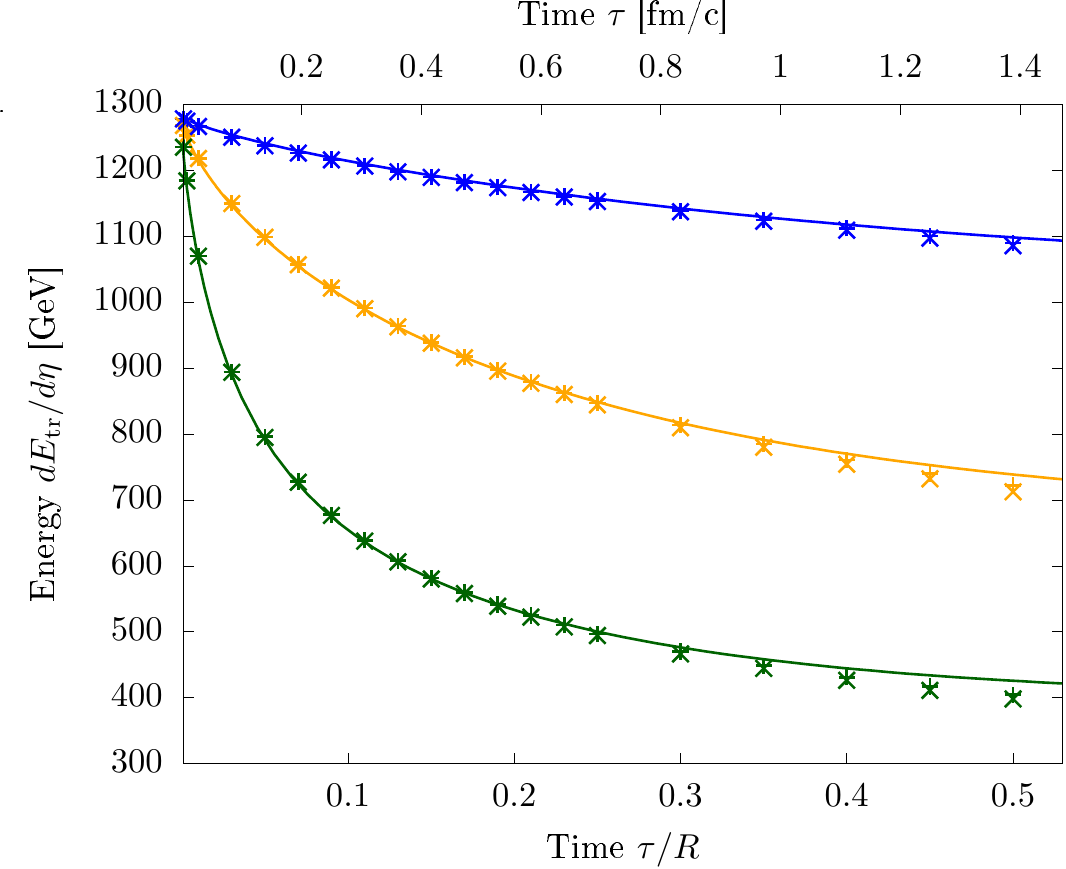} & 
    \includegraphics[width=.48\textwidth]{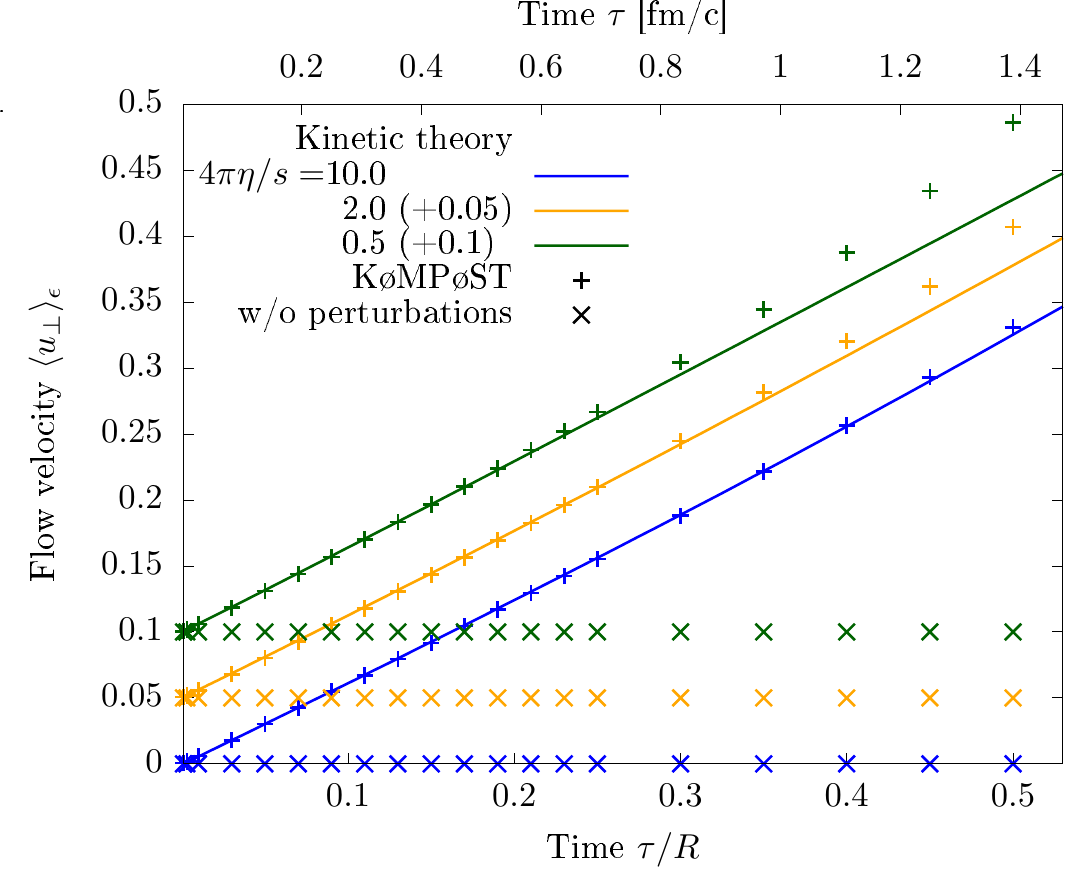} \\
    \includegraphics[width=.48\textwidth]{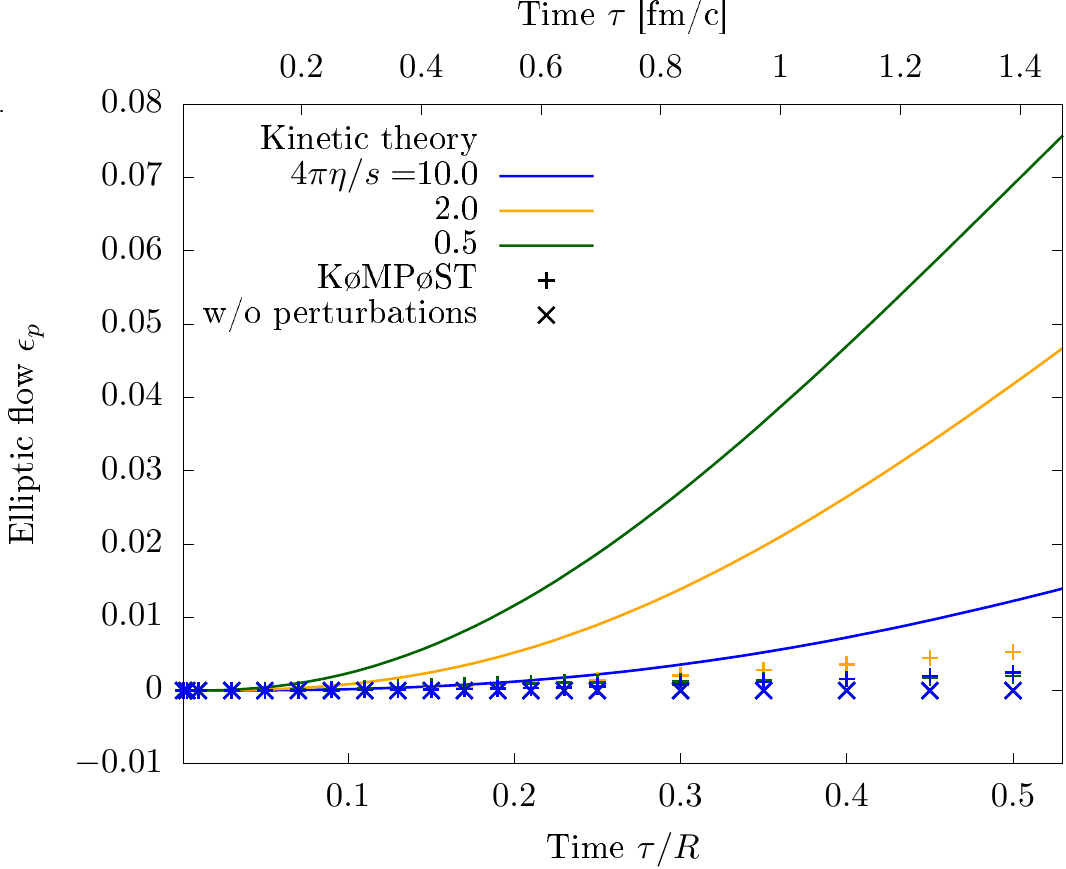} & 
    \includegraphics[width=.48\textwidth]{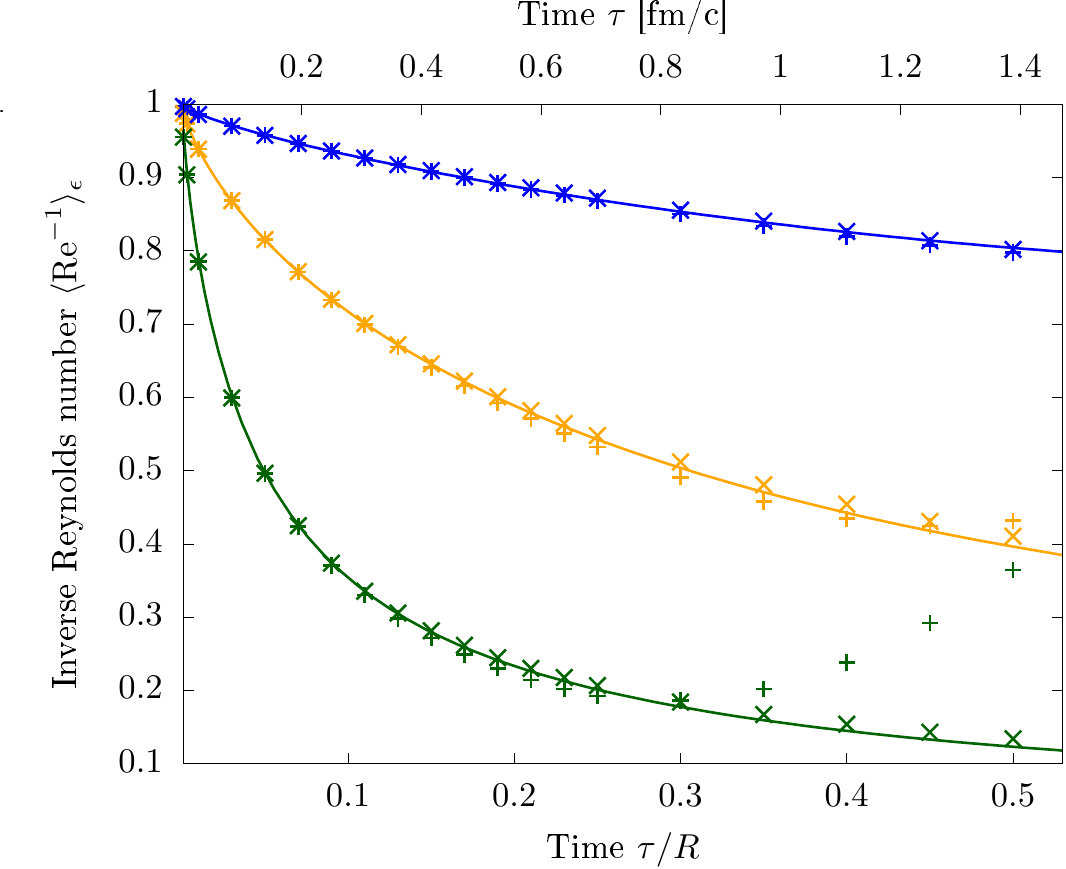}
\end{tabular}
    
    \caption{Time evolution of transverse energy $\d E_{\rm tr} / \d\eta$ [top left, cf. Eq.~\eqref{eq:obs_dEtrdeta}], transverse flow velocity $\langle u_\perp \rangle_\epsilon$ [top right,  cf. Eq.~\eqref{eq:obs_uT}], elliptic flow $\varepsilon_p$ [bottom left, cf. Eq.~\eqref{eq:obs_epsp}] and inverse Reynolds number $\langle \mathrm{Re}^{-1}\rangle_\epsilon$ [bottom right, cf. Eq.~\eqref{eq:Reinv_def}]. Plotted are results from K{\o}MP{\o}ST (RTA) with ($+$ symbols) and without ($\times$ symbols) energy perturbations compared to full kinetic theory results (solid lines) at three different opacities $4\pi\eta/s=0.5$ (green), $2$ (yellow) and $10$ (blue). In the plot of transverse flow velocity, results at different opacities are shifted in value in order to be distinguishable.}
    \label{fig:kompost_comparison_time_evolution}
\end{figure*}

Figure~\ref{fig:kompost_comparison_time_evolution} shows a comparison of the time evolution of four different transverse space integrated observables at three different values of the shear viscousity, namely $4\pi\eta/s=0.5, 2, 10$. The results from K{\o}MP{\o}ST are plotted with symbols "$+$" for the mode with and "$\times$" for the mode without energy perturbations and are benchmarked for times up to $\tau=0.5R$ against the results obtained using a full kinetic theory description, which are plotted with lines. 

The decrease of transverse energy $\d E_{\rm tr} / \d\eta$ is described very well in both modes. As without energy perturbations, the energy-momentum tensor is propagated as if there were no local gradients, it predicts zero transverse flow velocity $\langle u_\perp \rangle_\epsilon$ and elliptic flow $\varepsilon_p$. The mode with energy perturbations can describe the buildup of $\langle u_\perp \rangle_\epsilon$ correctly. On the other hand, while giving nonzero results, it still vastly underestimates the buildup of anisotropic flow $\varepsilon_p$. The inverse Reynolds number $\eavg{\mathrm{Re}^{-1}}$ is well described by both modes at early times, but results from the mode with energy perturbations deviate at very late times.

Generally, the comparison suggests that, for certain observables, K{\o}MP{\o}ST results can be accurate way beyond the timeframe it was intended for, which is on the order of $0.1\ \rm{fm}$. Other observables, in particular those related to anisotropies, are not described correctly.

\begin{figure*}
   \includegraphics[width=.99\textwidth]{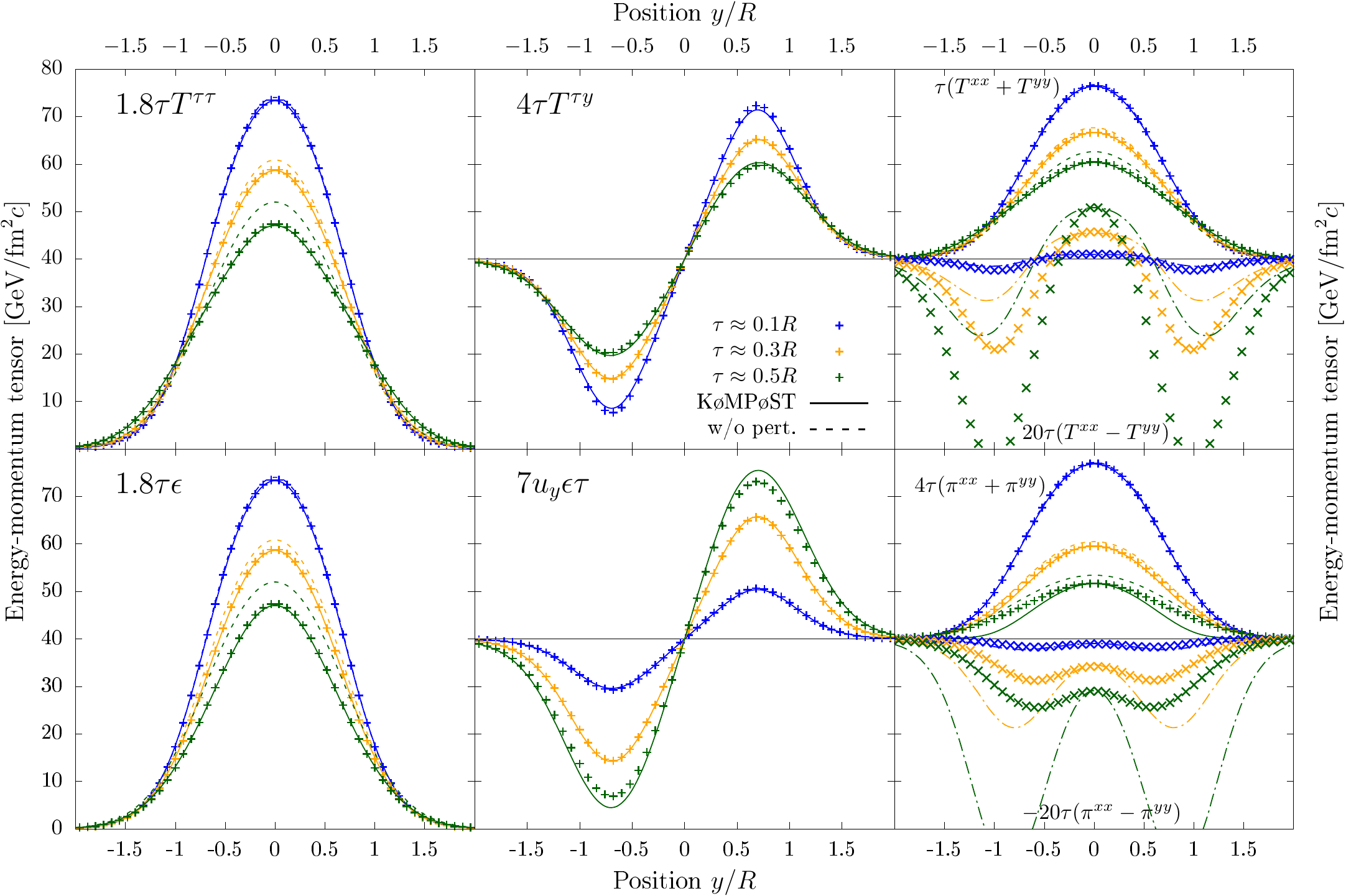}
    \caption{
    Comparison of K{\o}MP{\o}ST (RTA) and full kinetic theory via results for the energy-momentum tensor on the line $x=0$, 
    represented at fixed times
    $\tau / R \simeq 0.1$ in blue, $0.3$ in yellow and $0.5$ in green. 
    The full kinetic theory results are plotted with points ($+$,$\times$), while the K{\o}MP{\o}ST ones obtained with and without energy perturbations are plotted with solid and dashed lines, respectively. Anisotropic observables are nonzero only with energy perturbations and are plotted with point-dashed lines.
    The upper row shows, from left to right, the following components of the energy-momentum tensor: $T^{\tau\tau}$ (left), $T^{\tau y}$ (middle), as well as $T^{xx}+T^{yy}$ and $T^{xx} - T^{yy}$ (right). The lower row shows analogous local rest-frame quantities, namely $\epsilon$ (left), $\epsilon u^y$ (middle), as well as $\pi^{xx}+\pi^{yy}$ and $\pi^{yy}-\pi^{xx}$ (right). Notice the change in sign for the latter when compared to the upper panel. All observables were multiplied with $\tau$
    and rescaled 
    with a constant factor to 
    adjust their magnitudes such that they can be plotted on the same total range of $80\;\mathrm{GeV/fm^2c}$.
    }
    \label{fig:kompost_comparison_profiles}
\end{figure*}

In a further comparison of K{\o}MP{\o}ST to full kinetic theory data, we also investigated profiles of certain components of $T^{\mu\nu}$ at fixed shear viscosity $4\pi\eta/s=2$ and three different fixed times $\tau=0.1R$, $0.3R$ and $0.5R$. The same comparisons were also performed in the local rest frame with analogous quantities that are defined via the variables $\epsilon$, $u^\mu$ and $\pi^{\mu\nu}$. Figure~\ref{fig:kompost_comparison_profiles} illustrates our findings. This time, K{\o}MP{\o}ST results are plotted with lines and full kinetic theory results with symbols.

The results confirm that for energy or energy flow observables like $T^{\tau\tau}$, $T^{\tau y}$ and $T^{xx}+T^{yy}$, K{\o}MP{\o}ST works well even on a local level and in the outskirts of the system for all evolution times that we examined. The only part of the energy-momentum tensor for which K{\o}MP{\o}ST results shows significant deviations are anisotropies in the shear stress, as measured by $T^{xx}-T^{yy}$. While this observable is still correctly reproduced in the central part of the system, it exhibits sizeable deviations of up to a factor of five at a radial distance of $r\gtrsim R$. These deviations also explain the errors in elliptic flow $\varepsilon_p$.

\section{Early-time dynamics of different models}\label{sec:early}

Before the onset of transverse expansion, at times $\tau \ll R$, the system's dynamics is dominated by longitudinal expansion and the effects of transverse expansion can be neglected. Under these conditions, at each point in the transverse plane, the system evolves independently of the transverse neighbourhood and can locally be described by $(0+1)$D longitudinally boost-invariant Bjorken flow. In Bjorken flow, the trajectories of energy, pressure and stress for different initial conditions are known to rapidly converge to a common time evolution curve called the Bjorken flow attractor curve~\cite{Giacalone:2019ldn, Kurkela:2019set}. This means that at late times the system always evolves in the same way. If it is initialized on the attractor, then its entire time evolution is given by the attractor curve. We will describe the features of the attractor scaling solution for both the M\"uller-Israel-Stewart-type second-order hydrodynamics theory and for the conformal kinetic theory in RTA. In Sec.~\ref{sec:early:attractor}, the quantities describing the attractor solutions are introduced. Sec.s~\ref{sec:early:pre-equilibrium} and \ref{sec:early:pre-flow} discuss how the pre-equilibrium evolution impacts the observables of interest, highlighting the possible discrepancies between RTA, viscous hydrodynamics and ideal hydrodynamics. Finally, in Sec.~\ref{sec:initialization_scaled_hydro}, we discuss how viscous and ideal hydrodynamics can be brought in agreement with RTA at late times by scaling the initial conditions.

\subsection{Bjorken attractor}\label{sec:early:attractor}

The $(0+1)$D Bjorken flow can be described in terms of the Bjorken coordinates $(\tau, x, y, \eta)$, with respect to which the velocity becomes $u^\mu \partial_\mu = \partial_\tau$. The energy-momentum tensor takes the diagonal form
\begin{equation}
 T^{\mu\nu} = {\rm diag}(\epsilon, P_T, P_T, \tau^{-2} P_L),
\end{equation}
where $P_T$ and $P_L$ are the transverse and longitudinal pressures, respectively. The shear-stress tensor also becomes diagonal,
\begin{equation}
 \pi^{\mu\nu} ={\rm diag}\left(0, -\frac{1}{2} \pi_d, -\frac{1}{2} \pi_d, \frac{1}{\tau^2} \pi_d\right),
\end{equation}
where $\pi_d$ can be related to $P_T$ and $P_L$ via
\begin{equation}
 P_T = P - \frac{\pi_d}{2}, \quad 
 P_L = P + \pi_d,
\end{equation}
such that $\pi_d = \frac{2}{3}(P_L - P_T)$. The observables of interest for the following section are the inverse Reynolds number defined in Eq.~\eqref{eq:Reinv_def}, and the sum $\epsilon_{\rm tr} = T^{xx} + T^{yy}$, which become
\begin{equation}
 {\rm Re}^{-1} = -\frac{3\pi_d}{\epsilon}, \qquad 
 \epsilon_{\rm tr} = \frac{2\epsilon}{3} - \pi_d = 
 \frac{\epsilon}{3} (2 + {\rm Re}^{-1}).
 \label{eq:bjorken_obs_def}
\end{equation}

The evolution of the energy density $\epsilon$ is governed by the conservation equations $\nabla_\mu T^{\mu\nu}$, where $\nabla_\mu$ is the covariant derivative, which reduces to
\begin{equation}
 \tau \frac{\partial \epsilon}{\partial \tau} + \frac{4}{3} \epsilon + \pi_d = 0.
 \label{eq:bjorken_edot}
\end{equation}
In ideal hydrodynamics, $\pi_d = 0$ and $\tau^{4/3} \epsilon(\tau) = \tau_0^{4/3} \epsilon_0$, where $\epsilon_0$ is the energy density at initial time $\tau_0$. 

In RTA, the dynamics of $\pi_d$ is governed directly by the Boltzmann equation. In viscous hydrodynamics, the evolution of $\pi_d$ can be found from Eq.~\eqref{eq:hydro_pi} and reads:
\begin{equation}
 \tau \frac{\partial \pi_d}{\partial \tau} + \left(\lambda + \frac{4\pi \tilde{w}}{5} + \frac{2 \pi \tilde{w}}{5} \phi_7 \pi_d \right) \pi_d + \frac{16 \epsilon}{45} = 0,
 \label{eq:bjorken_pidot}
\end{equation}
where $\tilde{w}$ is the conformal parameter, 
\begin{equation}
 \tilde{w} = \frac{5\tau}{4\pi \tau_\pi} = \frac{\tau T}{4\pi \eta / s}.
 \label{eq:wt}
\end{equation}
In the above, $s = (\epsilon + P) / T$ is the entropy density for an ultrarelativistic gas at vanishing chemical potential, while $\eta = \frac{4}{5} \tau_\pi P$, as shown in Eq.~\eqref{eq:hydro_tcoeffs}. In Eq.~\eqref{eq:bjorken_pidot}, we introduced the notation 
\begin{equation}
 \lambda = \frac{\delta_{\pi\pi}}{\tau_\pi} + \frac{\tau_{\pi\pi}}{3\tau_\pi},
 \label{eq:bjorken_lambda}
\end{equation}
which evaluates to $38 / 21$ when using the values for the second-order transport coefficients given in Eq.~\eqref{eq:hydro_tcoeffs}. 
We note that in the original MIS theory, $\lambda$ evaluates to $4/3$, while the value $31/ 15$ was advocated in Ref.~\cite{Blaizot:2021cdv} in order to mimic the early time attractor of RTA. 

\begin{figure}[t]
\begin{tabular}{c}
 \includegraphics[width=.48\textwidth]{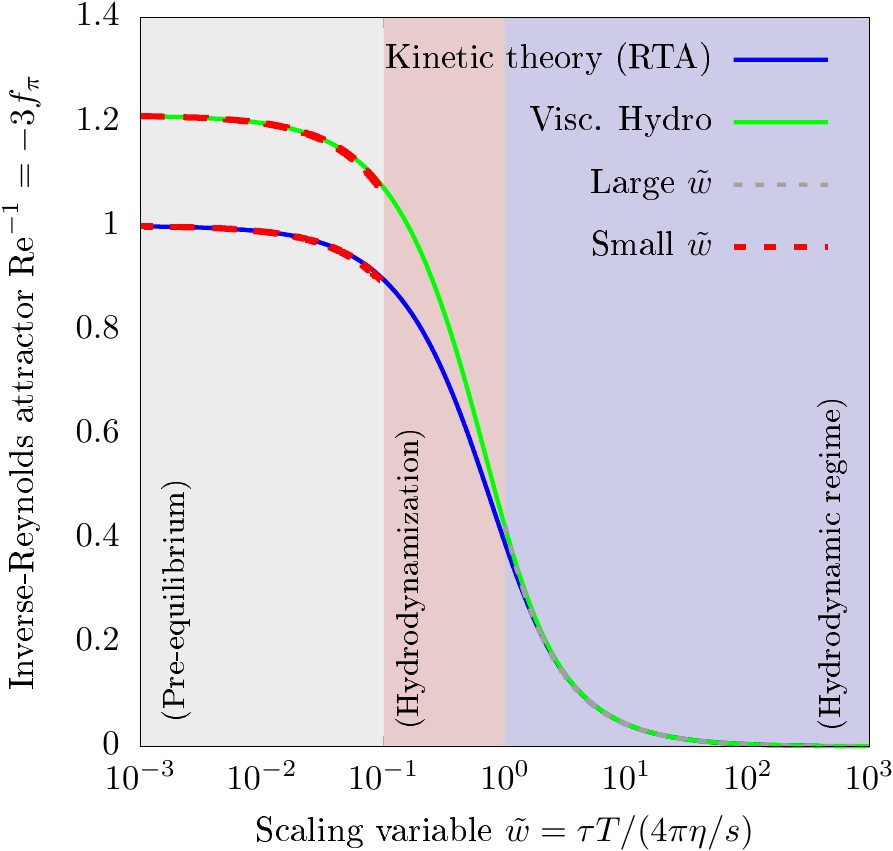} \\
 \includegraphics[width=.48\textwidth]{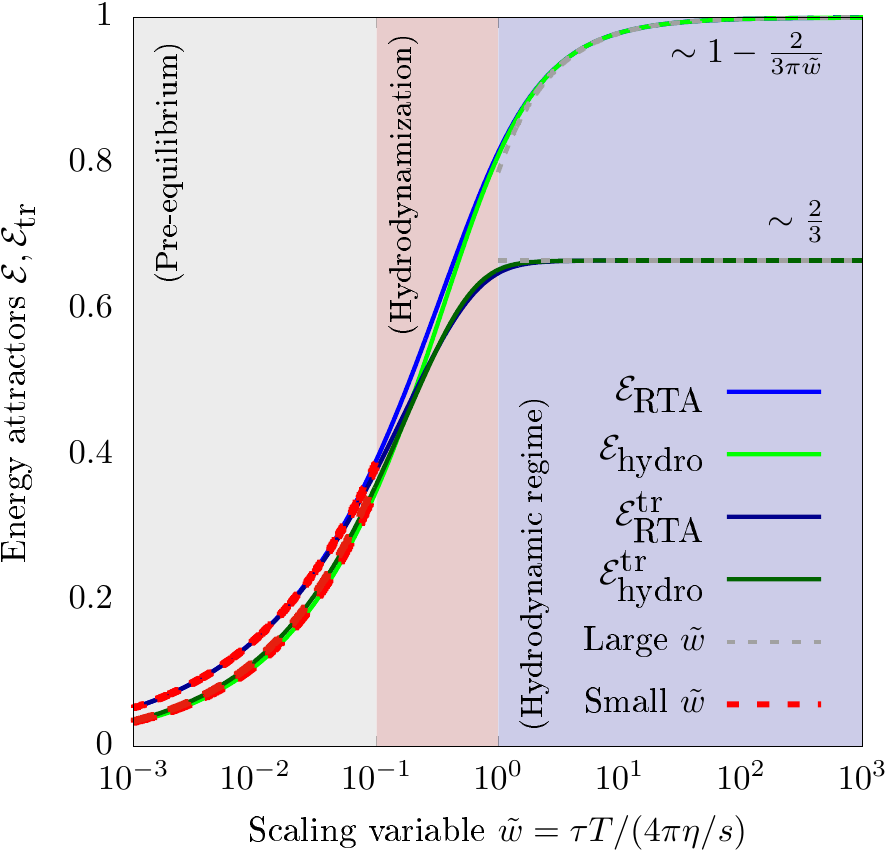} 
\end{tabular}
\caption{Attractor curves for the scaling functions (top) ${\rm Re}^{-1}$ [cf. Eq.~\eqref{eq:bjorken_obs_def}], 
(bottom)
$\mathcal{E}(\tilde{w})$  [upper two curves, light colors, cf. Eq.~\eqref{eq:bjorken_eps_scaling}] and $\mathcal{E}_{\rm tr}(\tilde{w})$ [lower two curves, dark colors, cf. Eq.~\eqref{eq:bjorken_epstr_scaling}] obtained for RTA (blue) and for second-order hydrodynamics (green). 
The large-$\tilde{w}$ asymptotics are shown with dashed gray curves. 
The small-$\tilde{w}$ asymptotics are shown with black and red dashed curves for RTA and hydro, respectively. 
}
\label{fig:attractor_curves}
\end{figure}

Equations \eqref{eq:bjorken_edot} and \eqref{eq:bjorken_pidot} admit scaling solutions with respect to the conformal parameter $\tilde{w}$. To see this, we note that the time derivative of $\tilde{w}$ satisfies
\begin{equation}
 \tau \frac{\d \tilde{w}}{\d \tau} = \tilde{w} \left(\frac{2}{3} - \frac{f_\pi}{4}\right),
 \label{eq:bjorken_wdot}
\end{equation}
where we defined
\begin{equation}
    f_\pi = \frac{\pi_d}{\epsilon}.\label{eq:bjorken_fpi_def}
\end{equation}
Using Eqs.~\eqref{eq:bjorken_edot}, \eqref{eq:bjorken_pidot}, and \eqref{eq:bjorken_wdot}, $f_\pi$ can be shown to satisfy 
\begin{multline}
 \tilde{w} \left(\frac{2}{3} - \frac{f_\pi}{4}\right) \frac{\d f_\pi}{\d\tilde{w}} + \frac{16}{45} \\
 + \left(\lambda - \frac{4}{3} + \frac{4\pi \tilde{w}}{5} + \frac{2 \pi \tilde{w}}{5} \phi_7 \epsilon f_\pi - f_\pi\right) f_\pi = 0,
 \label{eq:bjorken_fprime}
\end{multline}
where $\phi_7 = 0$ for consistency with RTA [see Eq.~\eqref{eq:hydro_tcoeffs}].
Demanding that $f_\pi$ remains finite when $\tilde{w} \rightarrow 0$, its early-time behavior in viscous hydro can be obtained as
\begin{equation}
 f_\pi (\tilde{w} \ll 1) = f_{\pi;0} + f_{\pi;1} \tilde{w} + f_{\pi;2} \tilde{w}^2 + O(\tilde{w}^3),
 \label{eq:bjorken_fpi_early}
\end{equation}
where 
\begin{align}
 f_{\pi;0} &= \frac{1}{2}\left[\lambda - \frac{4}{3} - 
 \sqrt{\left(\lambda - \frac{4}{3}\right)^2 + \frac{64}{45}}\right], \nonumber\\
 f_{\pi;1} &= \frac{\frac{16\pi}{25} f_{\pi;0}^2}{(f_{\pi;0} - \frac{4}{15})^2 + \frac{16}{75}}\left(1 + \frac{1}{2} \phi_7 \epsilon f_{\pi;0}\right),\nonumber\\
 f_{\pi;2} &= \frac{\frac{8\pi}{15} f_{\pi;0} f_{\pi;1}}{\left(f_{\pi;0} - \frac{4}{9}\right)^2 + \frac{16}{405}}  \left(1 - \frac{25 f_{\pi;1}}{16\pi} + f_{\pi;0} \phi_7 \epsilon\right).
 \label{eq:bjorken_fpi_early_coeffs}
\end{align}
When $\lambda = 38/21$, we find $f_{\pi;0} = \frac{1}{105}(25 - 3\sqrt{505}) \simeq -0.404$,
which is different from the limit $-1/3$ in kinetic theory.\footnote{
Expressing $\lambda = f_{\pi;0} + \frac{4}{3} - \frac{16}{45 f_{\pi;0}}$,
it can be seen that $f_{\pi;0} = -1/3$ leads to $\lambda = 31/15$, as pointed out in Ref.~\cite{Blaizot:2021cdv}.} At large values of $\tilde{w}$, $f_\pi(\tilde{w})$ behaves like
\begin{equation}
 f_\pi(\tilde{w} \gg 1) = -\frac{4}{9\pi \tilde{w}} + O(\tilde{w}^{-2}),
 \label{eq:bjorken_fpi_late}
\end{equation}
which is the leading order gradient expansion~\cite{Heller:2015dha} and therefore valid in both viscous hydrodynamics and in RTA. Due to the relations in Eq.~\eqref{eq:bjorken_obs_def}, our observable ${\rm Re}^{-1} = -3f_\pi$ also exhibits attractor behaviour. Its attractor curve is represented as a function of $\tilde{w}$ in the top panel of Figure~\ref{fig:attractor_curves}. Its asymptotic forms at small and large $\tilde{w}$ can be found from Eqs.~\eqref{eq:bjorken_fpi_early} and \eqref{eq:bjorken_fpi_late}, respectively.

We now turn to the energy equation, Eq.~\eqref{eq:bjorken_edot}. On the attractor, when $f_\pi$ depends only on $\tilde{w}$, it is possible to write (cf.~\cite{Giacalone:2019ldn,Du:2022bel})
\begin{equation}
 \tau^{4/3}\epsilon(\tau) = \frac{\tau_0^{4/3} \epsilon_0}{\mathcal{E}(\tilde{w}_0)} \mathcal{E}(\tilde{w}),
 \label{eq:bjorken_eps_scaling}
\end{equation}
where the scaling function $\mathcal{E}(\tilde{w})$ satisfies
\begin{equation}
 \tilde{w} \left(\frac{2}{3} - \frac{f_\pi}{4}\right) \frac{\d \mathcal{E}}{\d \tilde{w}} + f_\pi \mathcal{E} = 0.
 \label{eq:bjorken_Eprime}
\end{equation}
Due to Eq.~\eqref{eq:bjorken_obs_def}, $\epsilon_{\rm tr}$ also admits a  scaling solution,
\begin{align}
 \tau^{4/3} \epsilon_{\rm tr}(\tau) &= 
 \frac{\tau_0^{4/3} \epsilon_0}{\mathcal{E}(\tilde{w}_0)} \mathcal{E}_{\rm tr}(\tilde{w}),
 \nonumber\\
 \mathcal{E}_{\rm tr}(\tilde{w}) &= \left(\frac{2}{3} - f_\pi(\tilde{w})\right) \mathcal{E}(\tilde{w}).\label{eq:bjorken_epstr_scaling}
\end{align}
For $\tilde{w} \ll 1$, $\mathcal{E}(\tilde{w})$ can be obtained as
\begin{equation}
 \mathcal{E}(\tilde{w} \ll 1) = C_{\infty}^{-1} \tilde{w}^{\gamma} (1 + \mathcal{E}_1 \tilde{w} + \dots),
 \label{eq:bjorken_E_early}
\end{equation}
where 
the exponent $\gamma$ and the correction $\mathcal{E}_1$ are given by
\begin{equation}
 \gamma = \frac{12 f_{\pi;0}}{3f_{\pi;0} - 8}, \quad
 \mathcal{E}_1 = -\frac{\frac{32}{3} f_{\pi;1}}{(f_{\pi;0} -\frac{8}{3})^2}.
 \label{eq:bjorken_gamma}
\end{equation}
The constant $C_\infty$ appearing in Eq.~\eqref{eq:bjorken_E_early} is taken such that 
$\lim_{\tilde{w} \rightarrow \infty} \mathcal{E}(\tilde{w}) = 1$, in which case $\mathcal{E}$ has the following late-time asymptotic behavior:
\begin{equation}
 \mathcal{E}(\tilde{w} \gg 1) = 1 - \frac{2}{3\pi \tilde{w}}.
 \label{eq:bjorken_E_late}
\end{equation}
In the case of ideal hydrodynamics, obviously $f_\pi = 0$ (such that $f_{\pi;0} = \gamma = 0$) and $\mathcal{E}(\tilde{w}) = C_\infty = 1$.
The functions $\mathcal{E}(\tilde{w})$ and $\mathcal{E}_{\rm tr}(\tilde{w})$ are shown in the bottom panel of Figure~\ref{fig:attractor_curves} for both viscous hydrodynamics and for  kinetic theory. The normalization factor $C_\infty$ can be obtained in each theory by computing the attractor curve~\cite{Giacalone:2019ldn}. For completeness, we list below the values of $\gamma$ and $C_\infty$ in the relevant theories:
\begin{subequations}\label{eq:bjorken_Cinf}
\begin{align}
    \text{RTA}:& & 
    \gamma &= \frac{4}{9}, & 
    C_\infty &\simeq 0.88,\label{eq:bjorken_Cinf_RTA}\\
    \text{Visc. Hydro}:& & 
    \gamma &=\frac{\sqrt{505}-13}{18}, & 
    C_\infty &\simeq 0.80,\label{eq:bjorken_Cinf_hydro}\\
    \text{Ideal Hydro}:& & 
    \gamma &=0, & 
    C_{\infty} &= 1.\label{eq:bjorken_Cinf_ideal}
\end{align}
\end{subequations}

Due to the normalization $\lim_{\tilde{w} \rightarrow \infty} \mathcal{E}(\tilde{w}) = 1$, the quantities $\tau^{-2/3} \tilde{w}$ and $\tau^{4/3} \epsilon$ can be rewritten as 
\begin{subequations}
\begin{align}
 \tau^{-2/3} \tilde{w} &= (\tau^{-2/3} \tilde{w})_\infty \mathcal{E}^{1/4}(\tilde{w}), \label{eq:bjorken_wtinf_aux}\\
 \tau^{4/3} \epsilon &= (\tau^{4/3} \epsilon)_\infty \mathcal{E}(\tilde{w}),
 \label{eq:bjorken_epsinf_aux}
\end{align}
\end{subequations}
where $(\tau^{-2/3} \tilde{w})_\infty$ and $(\tau^{4/3} \epsilon)_\infty$ represent the corresponding asymptotic, late-time hydrodynamic limits, satisfying
\begin{align}
 (\tau^{-2/3} \tilde{w})_\infty &= \frac{(\tau^{4/3} \epsilon)^{1/4}_\infty}{a^{1/4} 4\pi \eta / s}, &
 (\tau^{4/3} \epsilon)_\infty &= \frac{\tau_0^{4/3} \epsilon_0}{\mathcal{E}(\tilde{w}_0)}.
\label{eq:bjorken_inf_gen}
\end{align}
Taking now the initial time such that $\tilde{w}_0 \ll 1$, Eq.~\eqref{eq:bjorken_E_early} can be used to obtain
\begin{multline}
 (\tau^{4/3} \epsilon)_\infty \simeq C_\infty \left(\frac{4\pi \eta}{s} a^{1/4}\right)^\gamma \\
 \times \left(\tau_0^{(\frac{4}{3} - \gamma) / (1 - \gamma/4)} \epsilon_0\right)^{1 - \gamma/4}.
\label{eq:bjorken_epsinf}
\end{multline}
Equation~\eqref{eq:bjorken_epsinf} tells us that the equilibration dynamics introduce a nontrivial relation between energy densities in equilibrium and in the initial state, as the dependence is nonlinear and the exponents depend on the model description, which was one of the main points of Ref.~\cite{Giacalone:2019ldn}. 

In the pre-equilibrium regime, $\tilde{w}\ll 1$. Under the early-time approximation \eqref{eq:bjorken_E_early}, $\tilde{w}$ can be written in terms of $(\tau^{-2/3} \tilde{w})_\infty$ as
\begin{equation}
 \tilde{w} \simeq 
 \tau^{\frac{2}{3} / (1 - \gamma/4)}
 \left[
 C_\infty^{-1/4}
  (\tau^{-2/3} \tilde{w})_\infty\right]^{1/(1 - \gamma/4)},
\end{equation}
which allows $\epsilon(\tilde{w} \ll 1)$ to be obtained as
\begin{multline}
 \epsilon(\tilde{w} \ll 1) \simeq 
 \tau^{(\gamma - \frac{4}{3}) / (1 - \gamma / 4)} \\\times
 \left[C_\infty^{-1} \left(\frac{4\pi \eta}{s} a^{1/4} \right)^{-\gamma} 
 (\tau^{4/3} \epsilon)_\infty\right]^{1 / (1 - \gamma / 4)}.
\end{multline}
Substituting the expression~\eqref{eq:bjorken_epsinf} for $(\tau^{4/3} \epsilon)_\infty$ 
manifestly shows that $\tau^{(\frac{4}{3} - \gamma) / (1 - \gamma/4)} \epsilon$ becomes independent of $\tau$ as $\tau \rightarrow 0$:
\begin{equation}
 \epsilon(\tilde{w} \ll 1) \simeq \left(\frac{\tau_0}{\tau}\right)^{(\frac{4}{3} - \gamma) / (1 - \frac{\gamma}{4})} \epsilon_0.
 \label{eq:bjorken_eps_early}
\end{equation}

\subsection{Pre-equilibrium evolution}\label{sec:early:pre-equilibrium}

\begin{figure}[b]
   
   \centerline{ 
    \includegraphics[width=.99\columnwidth]{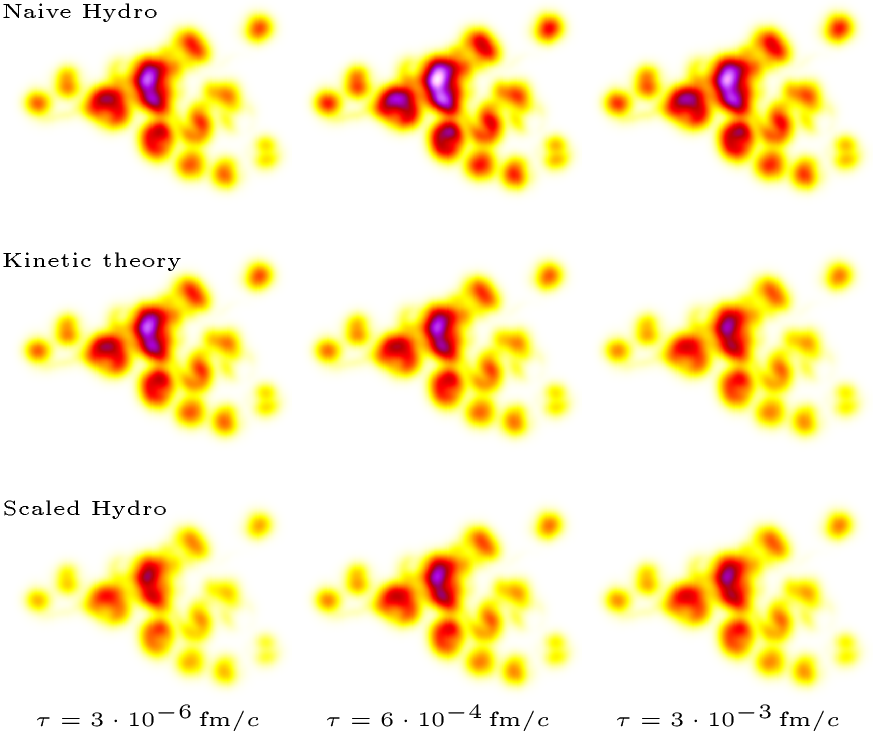}
    }
    
    \caption{Early time evolution of the transverse profile of the restframe energy density $\tau\epsilon$ for an example event in the $30-40\%$ centrality class of Pb-Pb collisions in naive viscous hydrodynamics (top), kinetic theory (middle) and scaled viscous hydrodynamics (bottom) at an opacity $4\pi\eta/s=0.05$.}
    \label{fig:pre-eq_profiles}
\end{figure}

We now consider a system which is no longer homogeneous in the transverse plane, such that the energy density becomes a function of both $\tau$ and $\xT$, $\epsilon \equiv \epsilon(\tau, \xT)$.
At early times $\tau \ll R$ we can neglect transverse dynamics and describe the dynamics locally by Bjorken flow (we will discuss early-time transverse expansion effects on the build-up of flow in the Sec.~\ref{sec:early:pre-flow}). Under this approximation, at each point $\xT$ of the transverse plane, we can assume that $\epsilon(\tau, \xT)$ follows an evolution along the attractor curve according to the local value of the conformal variable, $\tilde{w} \equiv \tilde{w}(\tau, \xT)$.
Moreover, we consider that $\tilde{w}_0(\xT) \ll 1$ throughout the system, such that the full pre-equilibrium evolution is captured during the system's evolution.

Neglecting the dynamics in the transverse plane, such that $T^{xx} = T^{yy} = \frac{1}{2} \epsilon_{\rm tr}$, $\d E_{\rm tr} / \d\eta$ defined in Eq.~\eqref{eq:obs_dEtrdeta} can be written as
\begin{equation}
 \frac{\d E_{\rm tr}}{\d\eta} = \tau \int_{\xT} \left(\frac{2}{3} - f_\pi\right) \epsilon,
 \label{eq:eq_dEtrdeta}
\end{equation}
where Eq.~\eqref{eq:bjorken_obs_def} was employed to replace $\epsilon_{\rm tr}$ and $f_\pi = -{\rm Re}^{-1} / 3$. Using now Eqs.~\eqref{eq:bjorken_epsinf_aux} and \eqref{eq:bjorken_epsinf} to replace $\epsilon$, we arrive at
\begin{multline}
 \frac{\d E_{\rm tr}}{\d\eta} = \tau^{-1/3} a \left(\frac{4\pi \eta}{s}\right)^\gamma 
 \int_{\xT} \left(\tau_0^{(\frac{4}{3} - \gamma) / (1 - \gamma/4)} \frac{\epsilon_0}{a}\right)^{1 - \frac{\gamma}{4}} \\\times
 \left(\frac{2}{3} - f_\pi\right) C_\infty \mathcal{E}.
 \label{eq:eq_dEtrdeta_full}
\end{multline}
The above equation can be employed to estimate the evolution of $\d E_{\rm tr} / \d\eta$ due solely to longitudinal expansion over the whole range of $\tau$. 

At a fixed value of $\tau$, the conformal parameter $\tilde{w}$ spans the interval $0$ (reached at infinitely large distances from the system's center of mass) up to the value $\tilde{w}_{\rm max}$ corresponding to the maximum value of the temperature. For sufficiently small values of $\tau$, $\tilde{w}_{\rm max} \ll 1$ and Eqs.~\eqref{eq:bjorken_fpi_early}, \eqref{eq:bjorken_eps_early} can be used to approximate $f_\pi$ and $\epsilon$, leading to 
\begin{equation}
 \frac{\d E_{\rm tr}}{\d\eta} \simeq
 \left(\frac{\tau_0}{\tau} \right)^{\frac{1}{3}(1 - 9\gamma/4) / (1 - \gamma/4)} 
 \frac{\d E^0_{\rm tr}}{\d\eta}.
 \label{eq:eq_dEtrdeta_early}
\end{equation}
The above relation shows that in RTA ($\gamma = 4/9$), $\d E_{\rm tr} / \d\eta$ remains constant during pre-equilibrium. Conversely, in viscous hydrodynamics, $\gamma > 4/9$ and consequently $\d E_{\rm tr} / \d\eta$ increases with time. As expected, in ideal hydrodynamics, $\d E_{\rm tr} / \d\eta$ decreases as $\tau^{-1/3}$.

In the limit $\tilde{w} \gg 1$, $f_\pi \sim \tilde{w}^{-1}$ and $\mathcal{E} \simeq 1$, as shown in Eqs.~\eqref{eq:bjorken_fpi_late} and \eqref{eq:bjorken_E_late}, such that $\tau^{4/3} \epsilon$ can be approximated by $(\tau^{4/3} \epsilon)_\infty$ by virtue of Eq.~\eqref{eq:bjorken_epsinf_aux}. Using Eq.~\eqref{eq:bjorken_epsinf}, $\d E_{\rm tr} / \d\eta$ reduces to
\begin{align}
 \frac{\d E_{\rm tr}}{\d\eta} &\simeq 
 \frac{2\tau^{-1/3}}{3} C_\infty \left(\frac{4\pi \eta}{s} a^{1/4} \right)^\gamma 
 \tau_0^{\frac{4}{3} - \gamma}
 \int_{\xT} \epsilon_0^{1-\gamma/4}.
 \label{eq:eq_dEtrdeta_late}
\end{align}
The above equation shows that at late times, $\d E_{\rm tr} / \d\eta$ decrease as $\tau^{-1/3}$. The amount of energy available at a given time $\tau$ depends explicitly on the dynamical theory (ideal and viscous hydrodynamics, RTA).

We now consider another important effect arising due to the pre-equilibrium evolution, namely inhomogeneous cooling.
During pre-equilibrium, neighbouring points in the transverse plane undergo cooling at differing rates according to their local attractors. 
As pointed out in Refs.~\cite{Giacalone:2019ldn,Ambrus:2021fej}, the characteristics of the inhomogeneities in the transverse plane change during pre-equilibrium, as can be seen by looking at the eccentricity $\epsilon_n$, defined as
\begin{equation}
 \epsilon_n = -
 \frac{\displaystyle \int_{\xT} x_\perp^n \epsilon \cos[n(\phi_x - \Psi_n)]}
 {\displaystyle \int_{\xT} x_\perp^n \epsilon} 
 \label{eq:eq_epsn}
\end{equation}
When $\tilde{w} \ll 1$, Eq.~\eqref{eq:bjorken_eps_early} can be employed to show that $\epsilon_n(\tau) \simeq \epsilon_n(\tau_0)$ and the eccentricities $\epsilon_n$ remain constant during pre-equilibrium. When $\tilde{w} \gg 1$, $\epsilon_n$ is modified to
\begin{equation}
 \epsilon_n \simeq -\frac{\displaystyle \int_{\xT} x_\perp^n \epsilon_0^{1-\gamma/4} \cos[n(\phi_x - \Psi_n)]}
 {\displaystyle \int_{\xT} x_\perp^n \epsilon_0^{1-\gamma/4}}.
 \label{eq:eq_epsn_late}
\end{equation}
The above relation shows that inhomogeneous cooling leads to modifications of all eccentricities of the initial profile, except in the case of ideal hydrodynamics ($\gamma = 0$).

The effects of the different behaviour for global and inhomogeneous cooling in different model descriptions are illustrated in Figure~\ref{fig:pre-eq_profiles}. It shows the pre-equilibrium evolution of the energy density profile 
multiplied by the Bjorken time, $\tau \epsilon$,
for an example event in the $30-40\%$ centrality class of Pb-Pb collisions in kinetic theory and viscous hydrodynamics with either the same or a scaled initial condition. At very early times, this quantity is constant in kinetic theory, but later it decreases slightly due to equilibration. Meanwhile, in hydrodynamics it increases first before transitioning to a decreasing trend. The speed of these transitions in both cases depends on the local temperature, meaning that e.g. the peak values will start decreasing earlier than the values in the outskirts of the system, i.e. the system cools inhomogeneously. After equilibration, the time evolution will uniformly follow the same power law in both models, but the differences due to the different pre-equilibrium evolution will persist. But the knowledge of the local attractor scaling behaviour allows to anticipate the differences between kinetic theory and hydrodynamics and apply a corresponding local scaling prescription to the initial condition of hydro. It then initially takes smaller values than in kinetic theory but dynamically approaches it during pre-equilibrium and reaches agreement after equilibration. This initialization scheme is explained in more detail in Sec~\ref{sec:initialization_scaled_hydro}.

\subsection{Pre-flow estimation}\label{sec:early:pre-flow}

We now estimate the buildup of flow during the pre-equilibrium evolution, which we quantify via the observable $\eavg{u_\perp}$ defined in Eq.~\eqref{eq:obs_uT}. The basis of our analysis is to consider that the transverse dynamics represent a small perturbation on top of the purely-longitudinal dynamics discussed in Secs.~\ref{sec:early:attractor} and \ref{sec:early:pre-equilibrium}, which we consider to be dominant. The idea of this calculation is similar to the one presented in Ref.~\cite{Vredevoogd:2008id}.

At early times $\tau \ll R$, when the transverse flow is negligible, we can write $T^{\mu\nu} = T^{\mu\nu}_B + \delta T^{\mu\nu}$, where $T^{\mu\nu}_B = {\rm diag}(\epsilon_B, P_{B;T}, P_{B;T}, 
 \tau^{-2}
P_{B;L})$ is the background (Bjorken) solution of the local, equivalent $(0+1)$D system (we also consider that at initial time, $\tilde{w}_0 \ll 1$ throughout the transverse plane).
Further assuming that $\delta T^{\mu\nu}$ is small and imposing the Landau frame condition, $T^\mu_\nu u^\nu = \epsilon u^\mu$, we write $u^\mu = u^\mu_B + \delta u^\mu$ and $\epsilon = \epsilon_B + \delta \epsilon$ and find
\begin{equation}
 \delta \epsilon = \delta T^{\tau\tau}, \qquad 
 \delta u^i = \frac{\delta T^{\tau i}}{\epsilon_B + P_{B;T}},
 \label{eq:flow_Landau}
\end{equation}
while $\delta u^\tau = 0$ as required by $u^\mu_B \delta u_\mu = 0$. Thus, the flow buildup can be estimated from the build-up of $\delta T^{\tau i}$.

We can now derive a dynamical equation for $T^{\tau i}$ via the conservation equations $\nabla_\mu T^{\mu\nu}=0$, which in a general coordinate system reads
\begin{align}
 \nabla_\mu T^{\mu\nu} &= \partial_\mu T^{\mu\nu} + \Gamma^\mu{}_{\lambda\mu} T^{\lambda\nu} + \Gamma^\nu{}_{\lambda\mu} T^{\mu\lambda}\;,
\end{align}
where $\Gamma^\lambda{}_{\mu\nu} = \frac{1}{2} g^{\lambda\rho}(\partial_\nu g_{\rho\mu} + \partial_\mu g_{\rho \nu} - \partial_\rho g_{\mu\nu})$ are the Christoffel symbols.
In the Bjorken coordinate system $(\tau, x, y, \eta)$, the only non-vanishing Christoffel symbols are $\Gamma^\tau{}_{\eta\eta} = \tau$ and $ \Gamma^{\eta}{}_{\tau\eta} = \Gamma^{\eta}{}_{\eta\tau} = \tau^{-1}$, such that the equation for $\nu = i$ becomes
\begin{align}
 \frac{1}{\tau} \frac{\partial(\tau T^{\tau i})}{\partial \tau} + \partial_j T^{ij} &= 0\;.
 \label{eq:flow_cons}
\end{align}
Splitting the energy-momentum tensor into a local Bjorken flow part and a small perturbation as discussed above, we find:
\begin{align}
 \frac{1}{\tau} \frac{\partial(\tau \delta T^{\tau i})}{\partial \tau} + \partial_i P_{B;T} + \partial_j \delta T^{ij} &= 0\;.\label{eq:flow_cons_aux}
\end{align}
Noting that $\delta T^{ij}$ represents a higher-order correction, 
the leading-order contribution to $\delta T^{\tau i}$ can be obtained by solving 
\begin{equation}
 \frac{\partial (\tau \delta T^{\tau i})}{\partial \tau} \simeq -\tau \partial_i P_{B;T}.
 \label{eq:flow_dTtaui}
\end{equation}

In the above, $P_{B;T}$ evolves according to the local Bjorken attractor, such that $P_{B;T} \simeq \epsilon_B \left(\frac{1}{3} - \frac{1}{2} f_{\pi;B}\right)$. Using Eq.~\eqref{eq:bjorken_epsinf_aux} to replace $\epsilon_B$, the spatial gradient of $P_{B;T}$ can be obtained as:
\begin{equation}
 \frac{\partial_i P_T}{P_T} = \frac{\partial_i (\tau^{4/3} \epsilon)_\infty}{(\tau^{4/3} \epsilon)_\infty} + 
 \left(\frac{\mathcal{E}'}{\mathcal{E}} - 
 \frac{\frac{1}{2} f_\pi'}{\frac{1}{3} - \frac{1}{2} f_\pi} \right) \partial_i \tilde{w},
 \label{eq:flow_gradPT_aux}
\end{equation}
where the prime denotes differentiation with respect to $\tilde{w}$. 
Here and henceforth, we will drop the $B$ subscript for brevity, keeping in mind that all instances of $P_T$, $\epsilon$, $f_\pi$ and the corresponding conformal variable $\tilde{w}$ are evaluated according to the background $(0+1)$D Bjorken attractor.

Since $(\tau^{4/3} \epsilon)_\infty$ depends on the transverse coordinates only through the initial profile [see Eq.~\eqref{eq:bjorken_epsinf}], the first term on the right-hand side of the above relation evaluates in the limit $\tilde{w}_0 \ll 1$ to
\begin{equation}
 \frac{\partial_i (\tau^{4/3} \epsilon)_\infty}{(\tau^{4/3} \epsilon)_\infty} = \left(1 - \frac{\gamma}{4}\right) \frac{\partial_i \epsilon_0}{\epsilon_0}.
 \label{eq:flow_gradepsinf}
\end{equation}
The gradient of $\tilde{w}$ appearing in Eq.~\eqref{eq:flow_gradPT_aux} can be written in terms of that of $(\tau^{-2/3} \tilde{w})_\infty$ starting from Eq.~\eqref{eq:bjorken_wtinf_aux},
\begin{align}
 \frac{\partial_i \tilde{w}}{\tilde{w}} &= \left(1 - \frac{\tilde{w} \mathcal{E}'}{4\mathcal{E}}\right)^{-1} \frac{\partial_i (\tau^{-2/3} \tilde{w})_\infty}{(\tau^{-2/3} \tilde{w})_\infty}\nonumber\\
 &= \frac{1}{4}\left(1 - \frac{\gamma}{4}\right) \left(1 - \frac{\tilde{w} \mathcal{E}'}{4\mathcal{E}}\right)^{-1}
 \frac{\partial_i \epsilon_0}{\epsilon_0},
 \label{eq:flow_gradwt}
\end{align}
where the equality on the second line is established using 
the relations \eqref{eq:bjorken_inf_gen} and \eqref{eq:flow_gradepsinf}.
Substituting Eqs.~\eqref{eq:flow_gradepsinf} and  \eqref{eq:flow_gradwt} into Eq.~\eqref{eq:flow_gradPT_aux} gives
\begin{equation}
 \frac{\partial_i P_T}{P_T} =
 \frac{1 - \gamma/4}{1 - \frac{\tilde{w} \mathcal{E}'}{4\mathcal{E}}}
 \left(1 - \frac{\tilde{w} f_\pi'}{\frac{8}{3} - 4 f_\pi}\right)
 \frac{\partial_i \epsilon_0}{\epsilon_0}.
 \label{eq:flow_gradPT}
\end{equation}
Substituting Eq.~\eqref{eq:flow_gradPT} in Eq.~\eqref{eq:flow_dTtaui} and integrating with respect to $\tau$, we arrive at
\begin{equation}
\delta T^{\tau i} =- \frac{1}{\tau}\left(1 - \frac{\gamma}{4}\right) 
 \frac{\partial_i  \epsilon_0}{\epsilon_0}
 \int_{\tau_0}^\tau d\tau
 \frac{\frac{1}{3} - \frac{1}{2} f_\pi - \frac{\tilde{w}}{8} f_\pi'}{1 - \frac{\tilde{w}}{4\mathcal{E}} \mathcal{E}'} \tau \epsilon\;.\label{eq:Ttaui_time_int}
\end{equation}

Considering now that $\tilde{w} \ll 1$ throughout the system, we can use Eqs.~\eqref{eq:bjorken_E_early},~\eqref{eq:bjorken_fpi_early} and~\eqref{eq:bjorken_eps_early} to approximate $f_{\pi} \simeq f_{\pi;0} = -(2\gamma / 3) / (1 - \gamma / 4)$, $\mathcal{E} \simeq C_\infty^{-1} \tilde{w}^\gamma$ and $\epsilon = (\tau_0/\tau)^{2-\alpha}\epsilon_0$, where $\alpha = (\gamma + 4/3)/[2(1 - \gamma/4)]$, which reduces to $\alpha= 2/3$, $1$ and $1.071$ in ideal hydro, RTA and viscous hydro, respectively. To leading order, we find
\begin{equation}
 \tau^{2-\alpha} \delta T^{\tau i} = -
 \frac{\tau}{2} \left[1 -\left(\frac{\tau_0}{\tau}\right)^\alpha\right] \partial_i (\tau_0^{2-\alpha} \epsilon_0),
\end{equation}
which allows the macroscopic velocity to be estimated as
\begin{equation}
 \delta u^i(\tilde{w} \ll 1) \simeq -\frac{3\tau}{8} \left(1 - \frac{\gamma}{4}\right)
 \left[1 -\left(\frac{\tau_0}{\tau}\right)^\alpha\right]
 \frac{\partial_i \epsilon_0}{\epsilon_0}. 
 \label{eq:flow_du}
\end{equation}
As expected, the flow velocity is driven by the gradients of the initial energy density profile. In addition, when $\tau \gg \tau_0$, $\delta u^i$ exhibits a linear increase with $\tau$, independently of the value of $\gamma$. The prefactor governing the overall amplitude of $\delta u^i$ is however $\gamma$-dependent.
We can now estimate the early-time evolution of $\eavg{u_\perp}$, defined in Eq.~\eqref{eq:obs_uT}, as follows:
\begin{multline}
 \langle u_\perp \rangle_{\epsilon,{\rm early}} \simeq 
 \frac{3\tau}{8} \left(1 - \frac{\gamma}{4}\right) \left[1 -\left(\frac{\tau_0}{\tau}\right)^\alpha\right] \\\times \left(\int_\xT \epsilon_0\right)^{-1}
 \int_\xT |\nabla_\perp \epsilon_0|,
 \label{eq:flow_duT_early}
\end{multline}
where $|\nabla_\perp \epsilon_0| = [(\partial_x \epsilon_0)^2 + (\partial_y \epsilon_0)^2]^{1/2}$.

In general, the time dependence of the integrand 
in Eq. \eqref{eq:Ttaui_time_int} is too complicated to integrate analytically. But it again takes a simple form in the Bjorken flow equilibrium stage, where $\tau^{4/3}P_T \simeq \frac{1}{3}(\tau^{4/3}\epsilon)_\infty$. At late times, when the duration of pre-equilibrium is small compared to the elapsed time, its contribution in the time integration is negligible and $\delta T^{\tau i}$ and $\delta u^i$ asymptote to
\begin{subequations}
\begin{align}
 \delta T^{\tau i}(\tilde{w}\gg 1) &\simeq -\frac{1}{2 \tau^{1/3}} \left[1 - \left(\frac{\tau_0}{\tau}\right)^{2/3}\right] \partial_i(\tau^{4/3} \epsilon)_\infty,
 \label{eq:flow_dTtaui_late}\\
 \delta u^i(\tilde{w}\gg 1) &\simeq -\frac{3\tau}{8} \left[1 - \left(\frac{\tau_0}{\tau}\right)^{2/3}\right]  \frac{\partial_i (\tau^{4/3} \epsilon)_\infty}{(\tau^{4/3} \epsilon)_\infty},
 \label{eq:flow_du_late}
\end{align}
\end{subequations}
such that $\eavg{u_\perp}$ becomes
\begin{equation}
 \langle u_\perp \rangle_{\epsilon,\rm late} \simeq \frac{3\tau}{8} \left[1 - \left(\frac{\tau_0}{\tau}\right)^{2/3}\right]
 \frac{\int_\xT |\nabla_\perp \epsilon_0^{1 - \gamma/4}|}{\int_\xT \epsilon_0^{1-\gamma/4}}.
 \label{eq:flow_duT_late}
\end{equation}
Note that the above equation was derived under the assumption that $\delta u^i$ is small and thus holds only when the system hydrodynamizes before transverse expansion sets in.

\begin{table}
\begin{tabular}{c||l|ll|ll}
 & Kinetic & \multicolumn{2}{c}{Naive hydro} & \multicolumn{2}{|c}{Scaled hydro} \\\cline{3-6}
 & theory & Ideal & Viscous & Ideal & Viscous\\\hline\hline
 $\gamma$ & $4/9$ & $0$ & $0.526$ & $0$ & $0.526$\\
 $\alpha$ & $1$ & $2/3$ & $1.071$ & $2/3$ & $1.071$ \\\hline
 ${\displaystyle \frac{R}{\tau}
 \frac{\langle u_\perp \rangle_{\epsilon,{\rm early}}}{1 - (\tau_0/\tau)^\alpha}}$ & $0.614$ & $0.691$ & $0.600$ & $0.658$ & $0.606$ \\\hline
 ${\displaystyle \frac{R}{\tau}
 \frac{\langle u_\perp \rangle_{\epsilon,{\rm late}}}{1 - (\tau_0/\tau)^{2/3}}}$ &
 $0.658$ & $0.691$ & $0.652$ & $0.658$ & $0.658$
\end{tabular}
\caption{Estimates for the pre-flow generated in kinetic theory, ideal hydrodynamics and viscous hydrodynamics (see Sec.~\ref{sec:initialization_scaled_hydro} for details regarding the naive and scaled hydrodynamics setups).
\label{tbl:pre-flow}}
\end{table}

The right-hand side of Eqs.~\eqref{eq:flow_duT_early} and \eqref{eq:flow_duT_late} can be evaluated numerically for the $30-40\%$ centrality profile that we are considering in this paper. The results for the different theories (kinetic theory, ideal hydrodynamics and viscous hydrodynamics) are shown in Table~\ref{tbl:pre-flow}. Here, we contrast the ``naive'' and ``scaled'' initial conditions for hydrodynamics, which will be discussed in detail in the following subsection. In the early-time regime, it can be seen that kinetic theory leads to more flow than viscous hydrodynamics ($2\%$ and $1\%$ more for the naive and scaled initialization, respectively), while ideal hydrodynamics leads to more flow than kinetic theory ($13\%$ and $7\%$ more for the naive and scaled initializations, respectively). In the late-time limit, both ideal and viscous hydrodynamics are brought in agreement with kinetic theory when the scaled initialization is employed. In the case of the naive initialization, ideal hydrodynamics gives about $5\%$ more flow, while viscous hydrodynamics underestimates the flow by less than $1\%$.

\subsection{Setting initial conditions}\label{sec:initialization_scaled_hydro}

From the discussion in the previous subsection, it becomes clear that the pre-equilibrium evolution of the fluid depends on the theory employed to describe it. We take as the ``correct'' evolution that described by kinetic theory, when $\d E_{\rm tr} / \d\eta$ remains constant during the free-streaming stage of pre-equilibrium. This can be seen by setting $\gamma = 4/9$ in Eq.~\eqref{eq:eq_dEtrdeta_early}. Since in viscous hydrodynamics, $\gamma \simeq 0.526 > 4/9$, $\d E_{\rm tr} / \d\eta$ will actually increase during pre-equilibrium, thus leading for the same initial energy profile to an unphysically higher transverse plane energy at late times, as illustrated in Figure~\ref{fig:pre-eq_profiles}. Similarly, the change in eccentricity due to the pre-equilibrium evolution will be different compared to kinetic theory. We will now discuss how these phenomena specifically affect the pre-equilibrium evolution of our initial state as given in Sec.~\ref{sec:init:init} and how they are counteracted by locally scaling the initial condition. We will then give the quantitative details of the scaling prescription.

\begin{figure}
    \centering
    \includegraphics[width=0.999\linewidth]{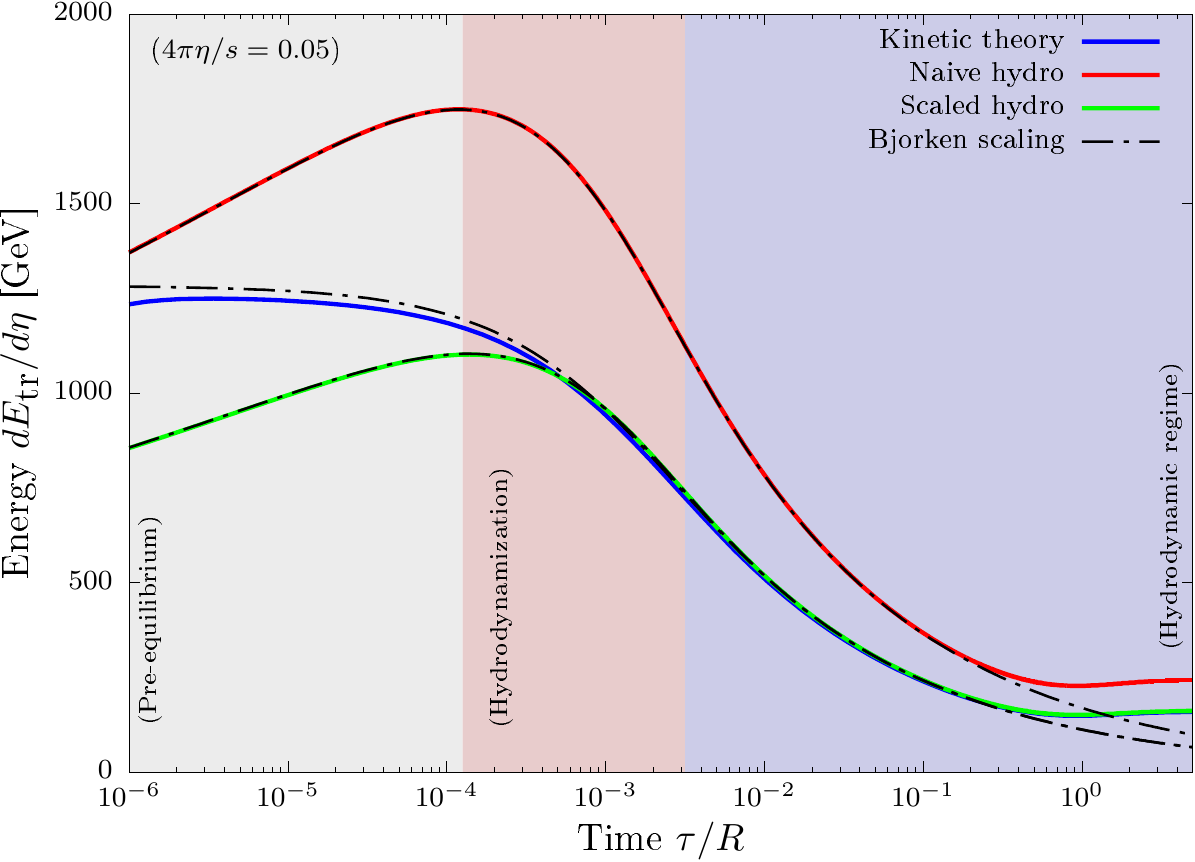} \\
    \includegraphics[width=0.999\linewidth]{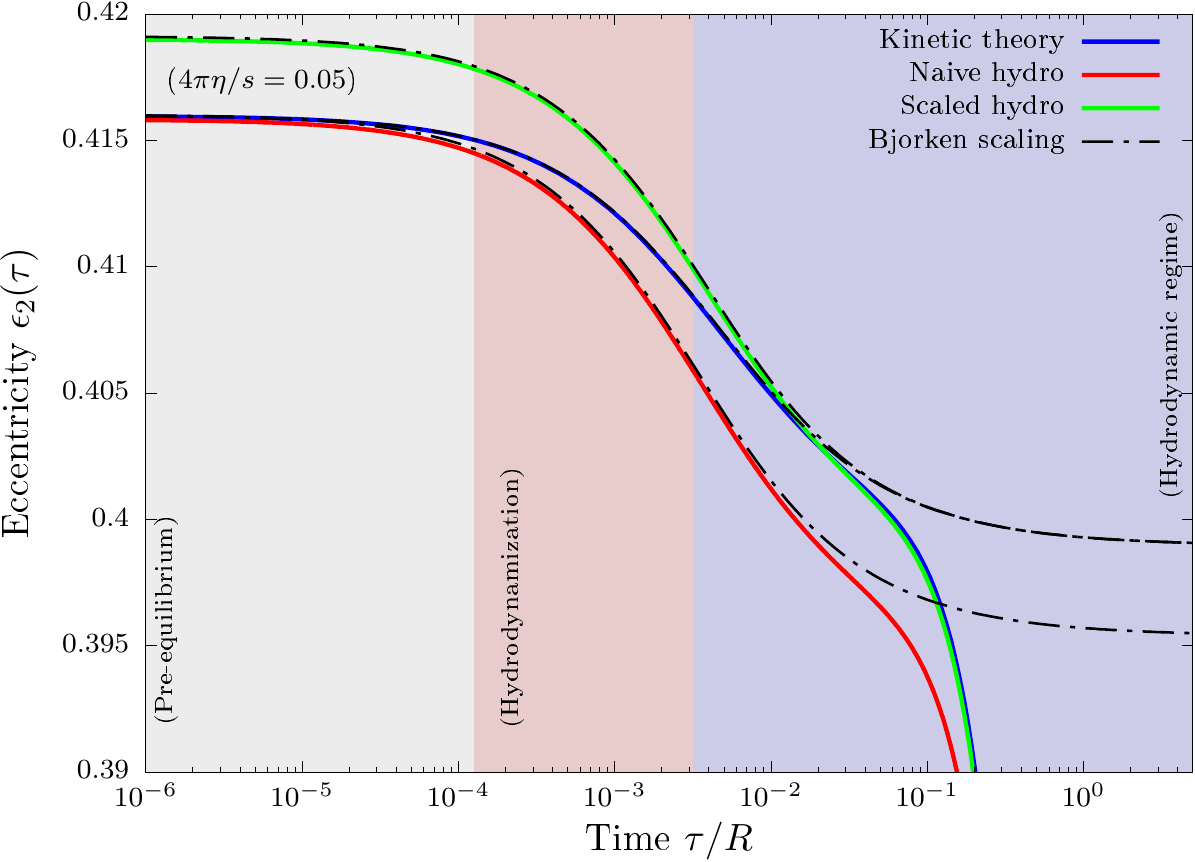}
    \caption{Early time evolution of transverse energy $\d E_{\rm tr} / \d\eta$ [top, cf. Eq.~\eqref{eq:obs_dEtrdeta}] and ellipticity $\epsilon_2$ [bottom, cf. Eq.~\eqref{eq:obs_eccn}] in kinetic theory (blue), naive hydrodynamics (red) and scaled hydrodynamics (green). Hydrodynamics behaves differently in pre-equilibrium, such that differences to a kinetic theory description build up. This can be counteracted by scaling the initial condition.}
    \label{fig:eccentricity_decrease}
\end{figure}

Figure~\ref{fig:eccentricity_decrease} illustrates the size of the effect on transverse energy $\d E_{\rm tr}/\d\eta$ in the top panel and ellipticity $\epsilon_2$ in the bottom panel. In naive hydrodynamics using the same initial condition for the energy density as kinetic theory and initial pressure determined by the hydrodynamic attractor, $\d E_{\rm tr}/\d\eta$ rises to a value which is about $1.5$ times larger than in kinetic theory at the onset of equilibration and will remain in disagreement throughout the rest of the evolution. The dashed lines show predictions of the behaviour in the local Bjorken flow scaling approximation according to Eq.~\eqref{eq:eq_dEtrdeta_full}. In our proposed scheme the initial value of $\d E_{\rm tr}/\d\eta$ is scaled down in such a way that it dynamically reaches agreement with kinetic theory. Similarly, we find that the ellipticity decreases in both kinetic theory and in hydro, but more so in the latter case. This means that in naive hydro the eccentricity will have a smaller value at the onset of the buildup of transverse flow than in kinetic theory, which will result in smaller final values of elliptic flow. With the scaling scheme, the initial ellipticity is scaled up in hydrodynamics and will come into agreement with kinetic theory after equilibration. 

As the local scaling factor for the hydrodynamic initial condition is computed in the local Bjorken flow approximation, it assumes that the system will fully equilibrate before the onset of transverse expansion. How well this works in practice will be discussed in Sec.~\ref{sec:time_evolution_kinetic_theory_scaled_hydro}.

We now move to the quantitative analysis of the pre-equilibrium behaviour in the two hydro schemes. In the first one, dubbed ``naive hydrodynamics'', we will impose the same energy density $\epsilon_0$ at initial time $\tau_0$ as in kinetic theory.  
We first note that the RTA initial conditions given in Eq.~\eqref{eq:init_Tmunu} are not compatible with the hydrodynamic attractor. 

Indeed, noting the relations $P_T = \epsilon(\frac{1}{3} - \frac{f_\pi}{2})$ and $P_L = \epsilon(\frac{1}{3} + f_\pi)$, the early-time expression for $T^{\mu\nu}$ reads
\begin{equation}
 T^{\mu\nu}_{0} = \epsilon_0 \times {\rm diag}\left(1,\frac{1}{3} - \frac{f_\pi}{2}, \frac{1}{3} - \frac{f_\pi}{2}, \frac{1}{3 \tau^2} + \frac{f_\pi}{\tau^2} \right),
 \label{eq:init_Tmunu_hydro}
\end{equation}
where $f_\pi \equiv f_\pi(\tilde{w}_0)$ depends on the local value of the conformal variable, $\tilde{w}_0 \equiv \tilde{w}_0(\xT) = \tau_0 T_0(\xT) / (4\pi \eta / s)$, with $T_0(\xT) = [\epsilon_0(\xT) / a]^{1/4}$. In order to evaluate $f_\pi(\tilde{w}_0)$, we follow Ref.~\cite{Romatschke:2017vte} and employ a simple Pad\'e approximation interpolating between the $\tilde{w} \ll 1$ and $\tilde{w} \gg 1$ limits given in Eqs.~\eqref{eq:bjorken_fpi_early} and \eqref{eq:bjorken_fpi_late}:
\begin{equation}
 f_\pi(\tilde{w}) \simeq \frac{c_0 + c_1\tilde{w}}{d_0 + d_1 \tilde{w} + d_2\tilde{w}^2},
 \label{eq:init_Tmunu_hydro_fpi_pade}
\end{equation}
where the coefficients $c_0$, $c_1$, $d_0$, $d_1$, and $d_2$ are computed to ensure second order accuracy at small $\tilde{w}$ and first order accuracy at large $\tilde{w}$:
\begin{gather}
 d_0 = \frac{4 f_{\pi; 1}}{9\pi} - f_{\pi;0}^2, \qquad 
 d_1 = f_{\pi;0} f_{\pi;1} - \frac{4 f_{\pi;2}}{9\pi}, \nonumber\\
 d_2 = f_{\pi;0} f_{\pi;2} - f_{\pi;1}^2, \quad
 c_0 = d_0 f_{\pi; 0}, \quad 
 c_1 = -\frac{4d_2}{9 \pi}.
 \label{eq:init_Tmunu_hydro_fpi_pade_coeffs}
\end{gather}
The coefficients $f_{\pi;0}$, $f_{\pi;1}$, and $f_{\pi;2}$ are given in Eq.~\eqref{eq:bjorken_fpi_early_coeffs}.
In the limit $\tilde{w} \rightarrow 0$, when $f_\pi \rightarrow f_{\pi;0}  = -\frac{2\gamma}{3} / (1 - \gamma / 4)$, Eq.~\eqref{eq:init_Tmunu_hydro} reduces to
\begin{equation}
 T^{\mu\nu}_{0} = \frac{\epsilon_0}{1 - \gamma/4} {\rm diag}\left(1 - \frac{\gamma}{4}, \frac{1}{3} +\frac{\gamma}{4}, \frac{1}{3}+ \frac{\gamma}{4}, \frac{1}{3\tau^2} - \frac{3\gamma}{4\tau^2}\right),
 \label{eq:init_Tmunu_hydro_w0}
\end{equation}
which coincides with the initialization employed for RTA [shown in Eq.~\eqref{eq:init_Tmunu}] in the case when $\gamma = 4/9$.
Since in hydrodynamics, $\gamma > 4/9$, the initial transverse-plane energy when $\tilde{w}_0 \ll 1$ will be larger than in RTA:
\begin{equation}
 \frac{\d E_{{\rm tr};\gamma}^{0}}{\d\eta} = \frac{2}{3} \frac{1 + 3\gamma/4}{1 - \gamma/4}  
 \frac{\d E_{{\rm tr};{\rm RTA}}^0}{\d\eta}.
 \label{eq:dEtrdeta_hydro_vs_RTA}
\end{equation}
This explains why at initial time the naive hydro curve in Figure~\ref{fig:eccentricity_decrease} starts above the kinetic theory one.

Acknowledging that viscous hydrodynamics does not capture correctly the pre-equilibrium evolution of the fluid, we propose to change the initialization of hydrodynamics in such a way that the energy density $\epsilon$ locally agrees with the kinetic theory prediction at late times. In principle, this works only when the pre-equilibrium evolution ends before the onset of transverse expansion. Taking $a$ and $\eta / s$ to be identical in the two theories and demanding that they both reach the same $(\tau^{4/3} \epsilon)_\infty$ value when $\tau \rightarrow \infty$,  Eq.~\eqref{eq:bjorken_epsinf} shows that the local modification of the initial energy density in hydrodynamics (denoted $\epsilon_{0,\gamma}$) is
\begin{equation}
 \epsilon_{0,\gamma} = \left[ \left(\frac{4\pi\eta/s}{\tau_0} a^{1/4} 
 \right)^{\frac{1}{2} - \frac{9\gamma}{8}} 
 \left(\frac{C_\infty^{\rm RTA}}{C_\infty^\gamma}\right)^{9/8}
 \epsilon_{0,{\rm RTA}}\right]^{\frac{8/9}{1 - \gamma/4}},
 \label{eq:scaling}
\end{equation}
where the specific shear viscosity $\eta / s$ is considered to have the same value in viscous hydrodynamics and in kinetic theory.
Using the above energy profile in  Eqs.~\eqref{eq:eq_dEtrdeta_late}, \eqref{eq:eq_epsn_late} and \eqref{eq:flow_duT_late} shows that after pre-equilibrium (i.e., at large $\tilde{w}$), $\d E_{\rm tr} / \d\eta$, the eccentricities $\epsilon_n$ and the average flow velocity $\eavg{u_\perp}$ will reach the corresponding RTA limits, irrespective of the value of $\gamma$. We note, however, that the pre-equilibrium behaviour of all of the above observables will still be different from that in RTA. 

Before ending this section, we emphasize that the rescaling of the initial conditions shown in Eq.~\eqref{eq:scaling} is not only possible, but also mandatory for ideal hydrodynamics simulations, when $\gamma = 0$ and $C_{\infty} = 1$. While when applying the scaling procedure to viscous hydrodynamics, $\eta / s$ was considered as an invariant physical parameter, in ideal hydrodynamics (when $\eta = 0$), this is no longer the case. Instead, the factor $\eta/s$ helps rescale the initial energy density such that at late times, $\tau^{4/3} \epsilon$ obtained in ideal hydrodynamics would match the one in a hypothetical RTA system with that given value of $\eta/ s$. The agreement between ideal hydro and RTA can be expected of course only in the limit $\eta / s \rightarrow 0$. Specifically, Eq.~\eqref{eq:scaling} reduces in the case of ideal hydro to
\begin{equation}
 \epsilon_{0,{\rm id}} = \left(\frac{4\pi \eta}{s} \right)_{\rm RTA}^{4/9} C_\infty^{\rm RTA} \frac{a^{1/9}}{\tau_0^{4/9}} \epsilon_{0,{\rm RTA}}^{8/9}.
 \label{eq:scaling_id}
\end{equation}
 When comparing the ideal hydro result to kinetic theory calculations, we employ the above formula with $4\pi\eta/s = 1$, and for $\d E_{\rm tr} / \d\eta$ we rescale the final results with $(4\pi \eta/s)^{4/9}$ according to Eq.~(\ref{eq:scaling_id}) when comparing to kinetic theory at other values of $4\pi \eta/s$.

\section{Space-time evolution at different opacities and in different setups}\label{sec:space-time_evolution}

The different behaviour of hydrodynamics compared to kinetic theory in pre-equilibrium can best be assessed via the time dependence of the studied observables. This also allows to study the 
behaviour during
different stages of the collision. In Sec.~\ref{sec:time_evolution_profiles}, we discuss the time evolution of transverse profiles of the system in kinetic theory. Sec.~\ref{sec:time_evolution_kinetic_theory_scaled_hydro} compares the time evolution of the tracked observables in kinetic theory and scaled viscous hydrodynamics. These are then used as the basis for a discussion of the time evolution in hybrid simulation schemes in Sec.~\ref{sec:time_evolution_hybrid}.

\subsection{Evolution of transverse profiles in kinetic theory}\label{sec:time_evolution_profiles}

\begin{figure*}
    \centering
    \begin{tabular}{ccc}
          \small $4\pi\eta/s=0.5$ & \small $4\pi\eta/s=3$ & \small $4\pi\eta/s=10$  \\
    \includegraphics[width=.3\textwidth]{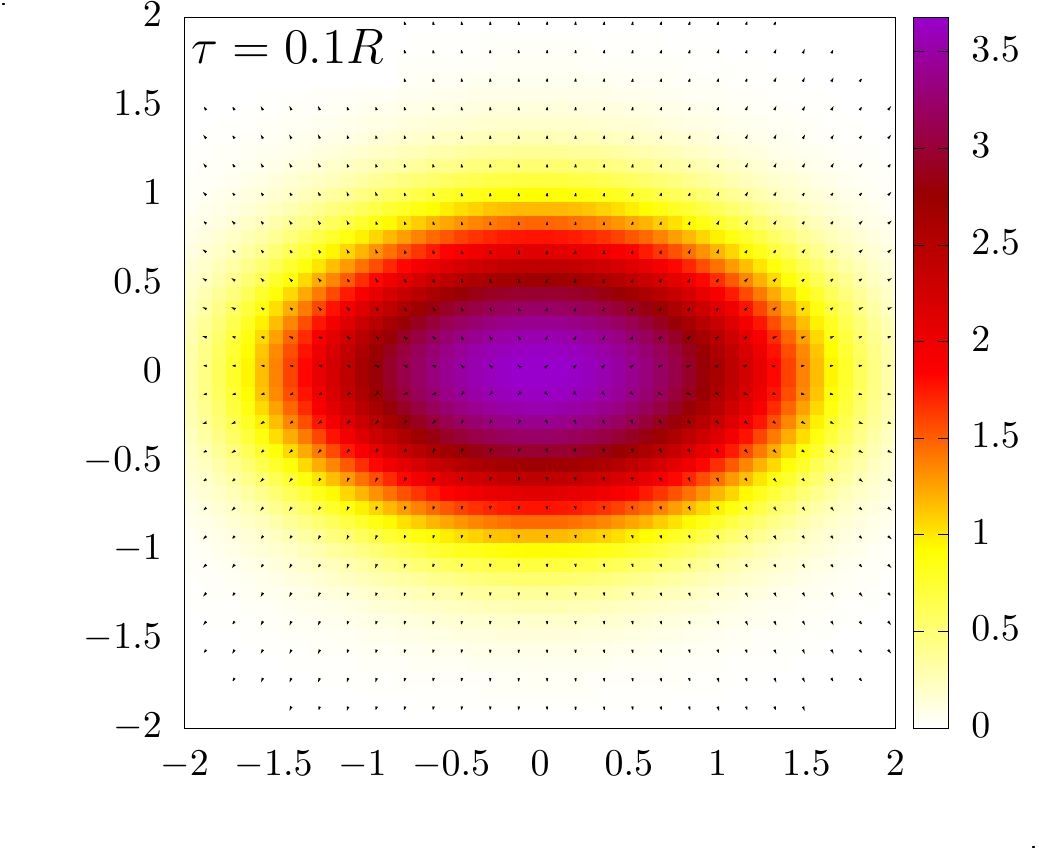} & 
    \includegraphics[width=.3\textwidth]{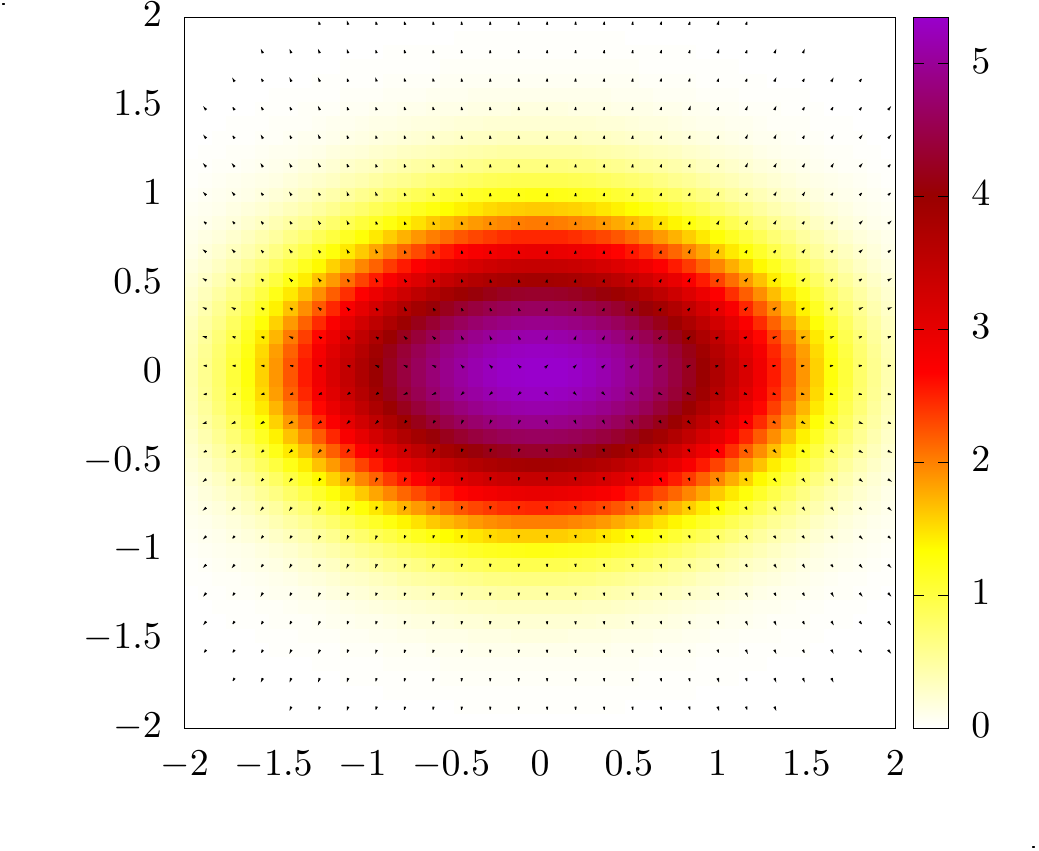} & 
    \includegraphics[width=.3\textwidth]{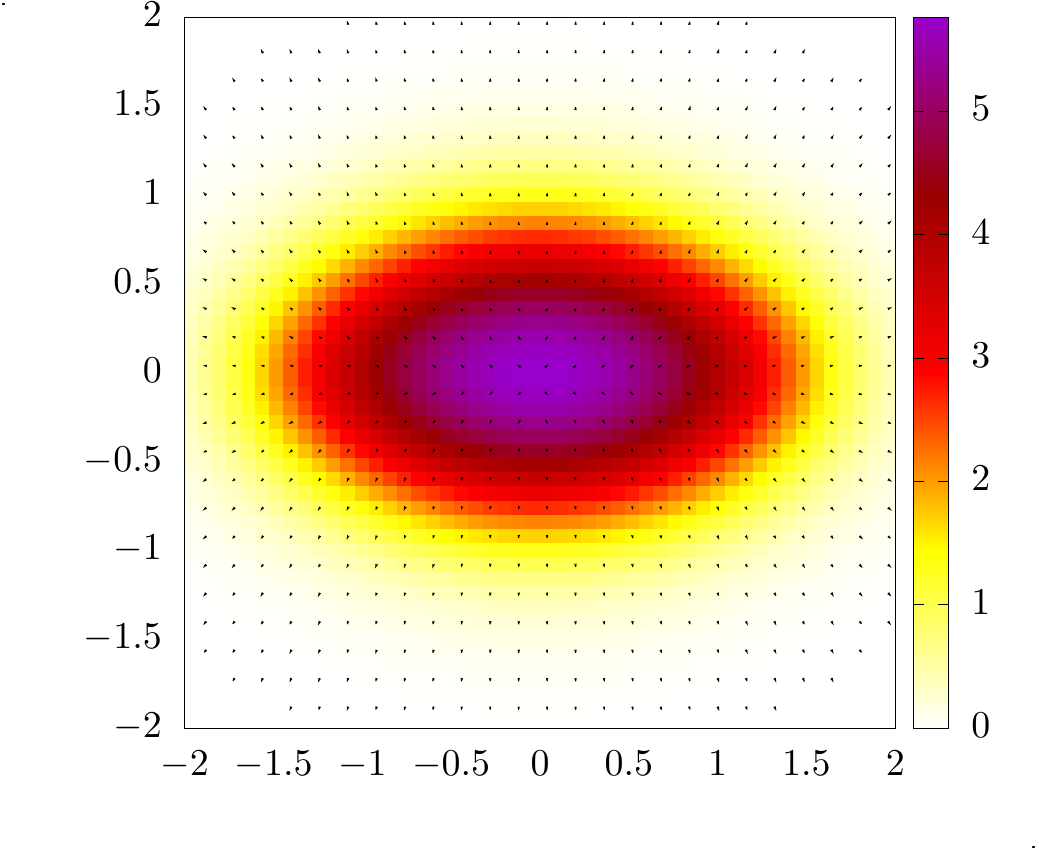} \\
    \includegraphics[width=.3\textwidth]{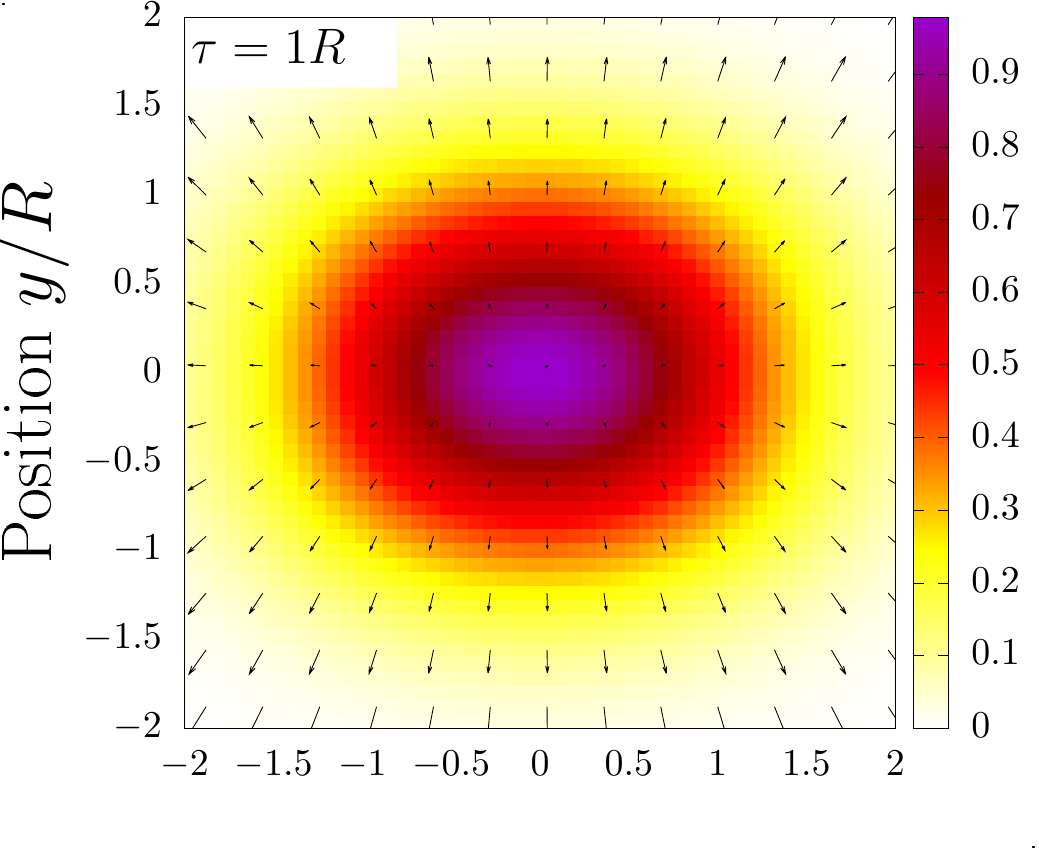} &
    \includegraphics[width=.3\textwidth]{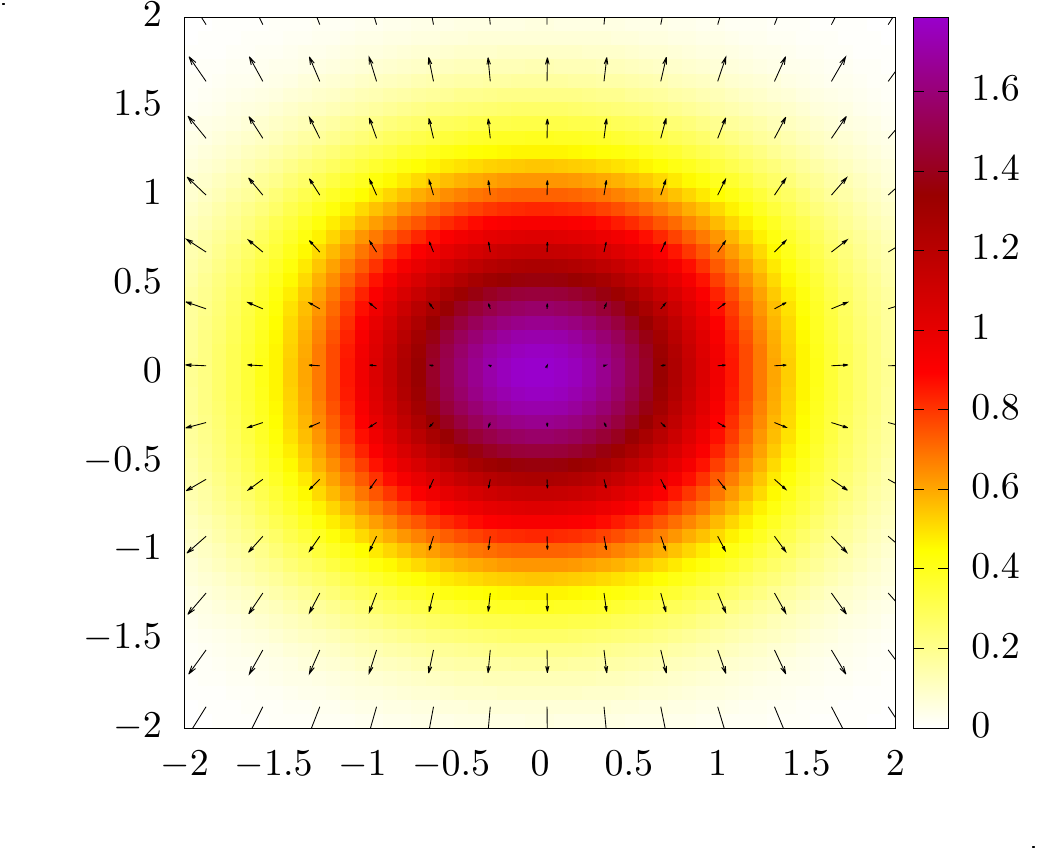} & 
    \includegraphics[width=.3\textwidth]{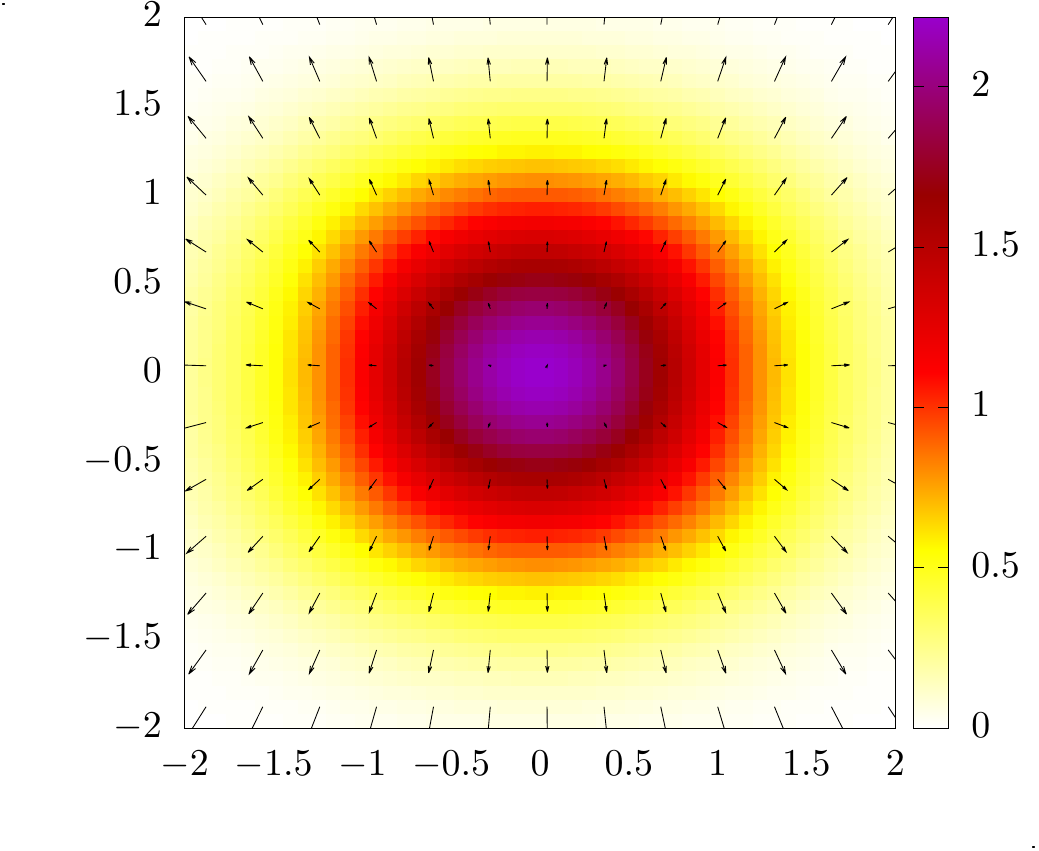} \\
     \includegraphics[width=.3\textwidth]{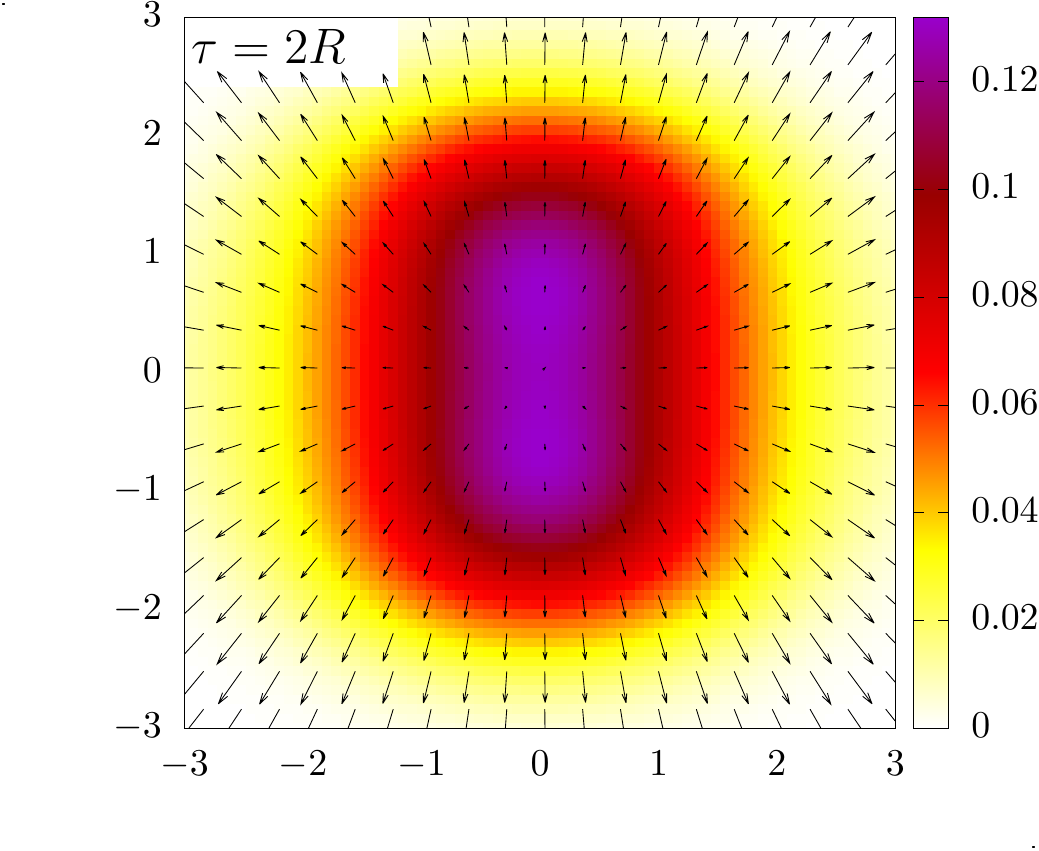} & 
     \includegraphics[width=.3\textwidth]{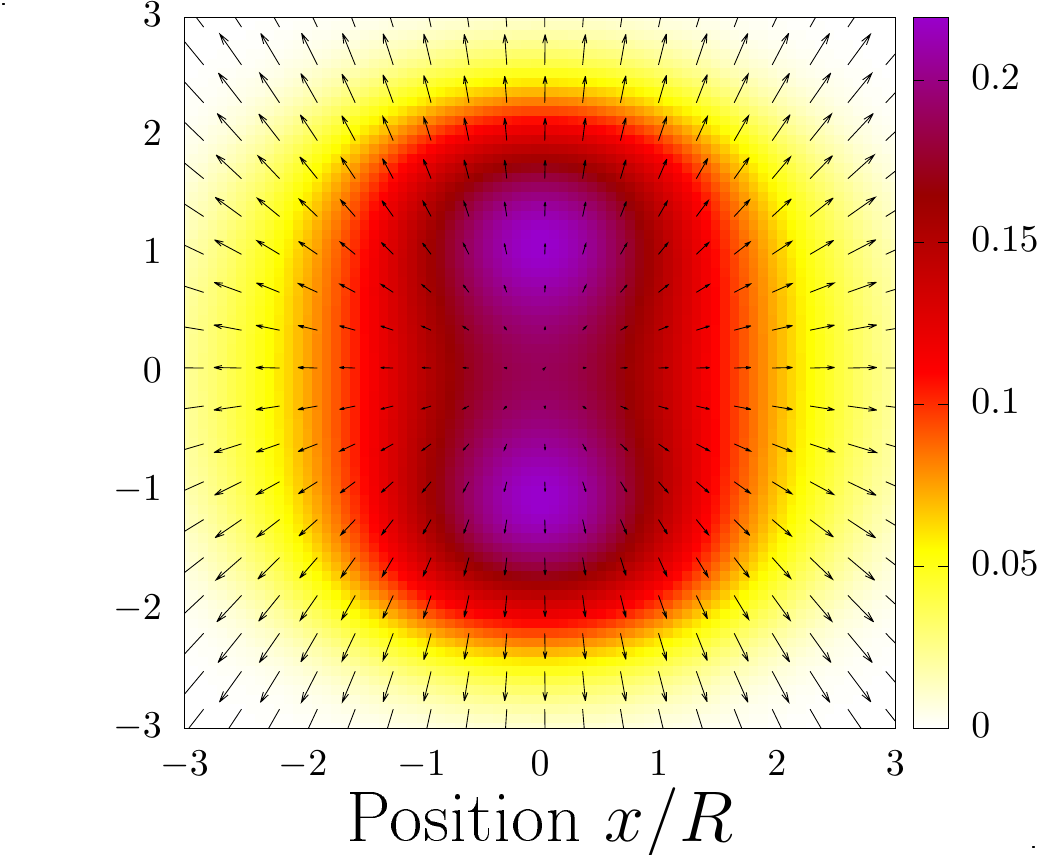} & 
     \includegraphics[width=.3\textwidth]{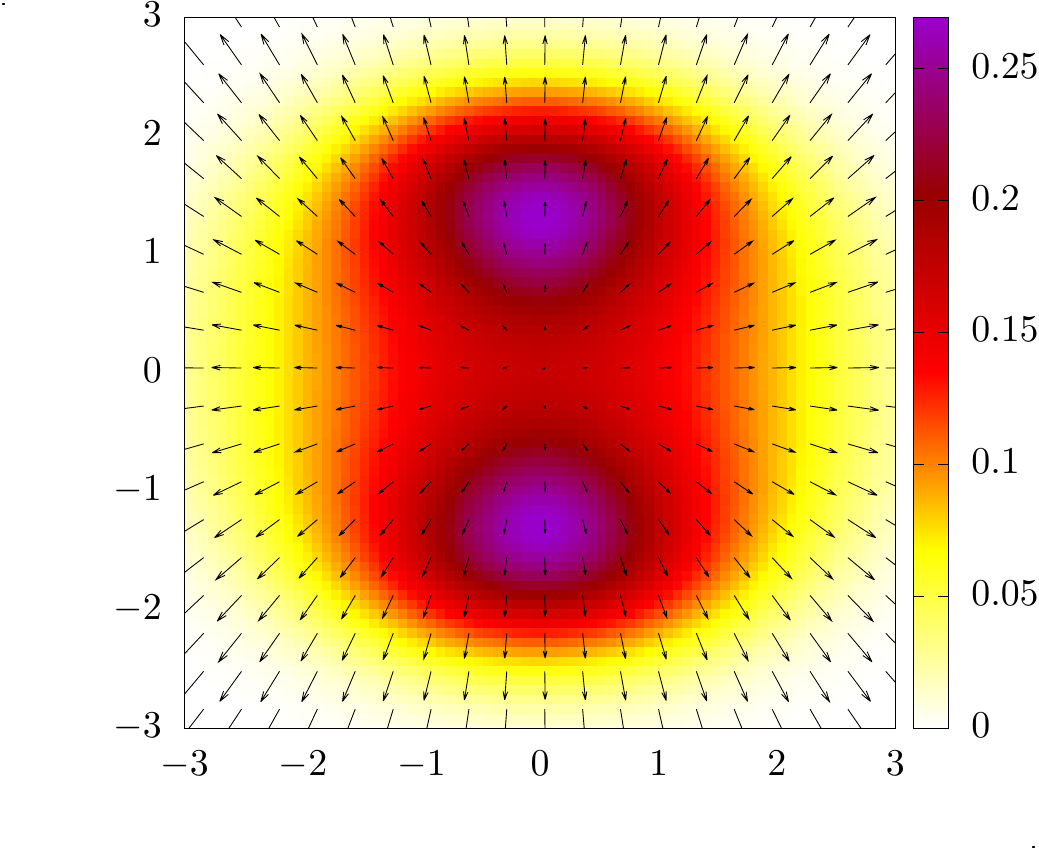}
    \end{tabular}
    \caption{Time evolution of transverse profiles of the restframe energy density $\tau\epsilon$ in a heatmap plot together with transverse components of the flow velocity $(u^x,u^y)$ as a vector field plot for the averaged initial condition used in this work at different opacities $4\pi\eta/s=0.5$ (left), $3$ (middle) and $10$ (right). The snapshot times $\tau=0.1R$ (top), $\tau=1R$ (middle) and $\tau=2R$ (bottom) were chosen as the beginning, peak and end of the buildup of elliptic flow $\varepsilon_p$.
    }
    \label{fig:transverse_e}
\end{figure*}
 
We now want to discuss the system's time evolution at different opacities resolved in transverse space. This is illustrated in Figure~\ref{fig:transverse_e} via heat map plots of the time scaled local rest frame energy density $\tau\epsilon$ together with a vector plot of the transverse components of the flow velocity $u^\mu$ at three different values of the shear viscosity, $4\pi\eta/s=0.5, 3, 10$, which are representative of the regimes of hydrodynamic behaviour, close-to-free-streaming behaviour and the intermediate transitioning regime. The time evolution of these profiles is sampled at three different times, $\tau=0.1R$, $1R$ and $2R$, which mark the beginning, peak and end of the buildup of elliptic flow $\varepsilon_p$, as 
will be
discussed in Section~\ref{sec:time_evolution_kinetic_theory_scaled_hydro}.
 
At the earliest time, $\tau=0.1R$, transverse dynamics have not had a large effect yet: flow velocities are negligible and the main geometric properties of the profile remain unchanged. The only obvious difference is the overall scale. 
At smaller 
$\eta/s$, the system starts cooling sooner, performing more work against the longitudinal expansion, resulting in significantly smaller energy densities when compared to 
larger $\eta/s$.

$\tau=1R$ marks the characteristic time where  
transverse expansion effects become significant.
Here, we see the profile taking on a more circular shape. We also see significant flow velocities, which rise in magnitude with 
the distance from the center. For smaller shear viscosity $\eta/s$, meaning larger interaction rates, the system tends to lump together more, resulting in a smaller spatial extent and smaller flow velocities compared to larger $\eta/s$.

At the largest selected time, $\tau=2R$, the interaction rate in the system has significantly decreased due to 
the dilution caused by
the transverse expansion. Over time, the dynamics will approach a free-streaming expansion in all directions. It is apparent in all three cases that the system has expanded mainly in the directions of larger gradients in the initial state. 
For small shear viscosity $\eta/s$, the system's energy density is still peaked in the center due to stronger collective behaviour.
On the other hand, at large $\eta/s$, the system evolution is closer to a
free-streaming propagation of the initial state, resulting in two high-density areas at distances $r\approx\tau$ from the center. Though the difference is barely visible, the built-up flow velocities are larger for larger $\eta/s$.

\begin{figure*}
    \centering
    \begin{tabular}{ccc}
          $4\pi\eta/s=0.5$ & $4\pi\eta/s=3$ & $4\pi\eta/s=10$  \\
   
    \includegraphics[width=.3\textwidth]{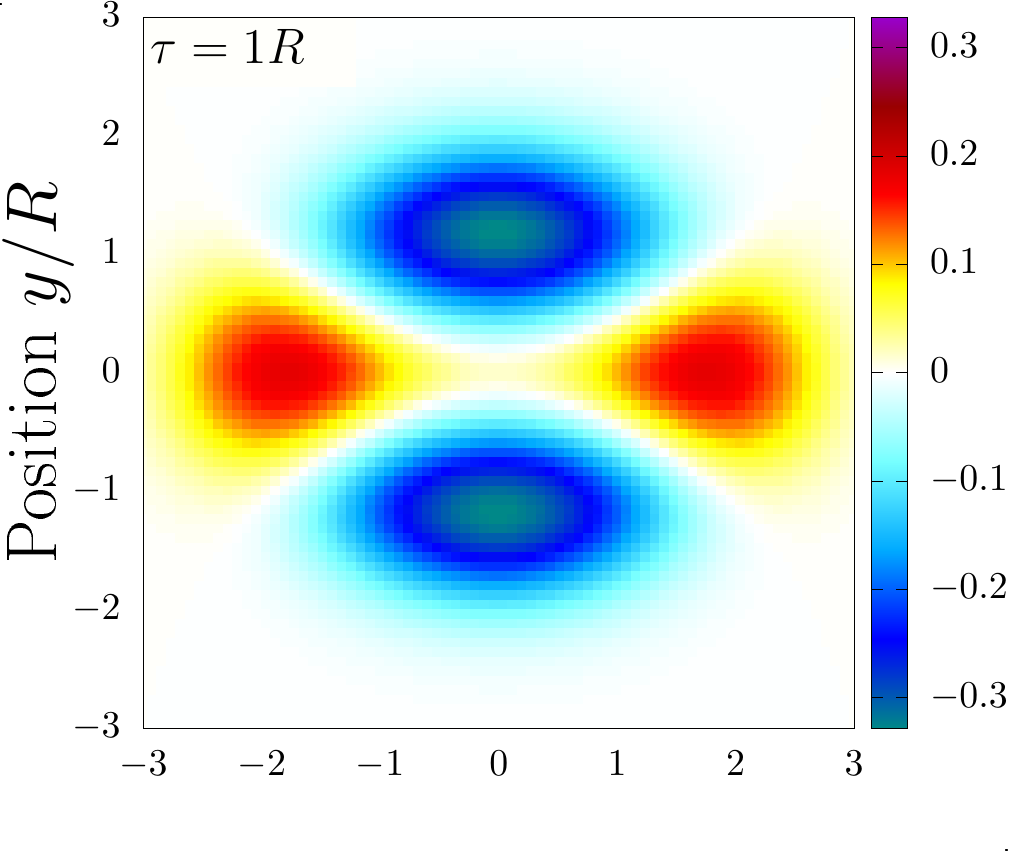} & \includegraphics[width=.3\textwidth]{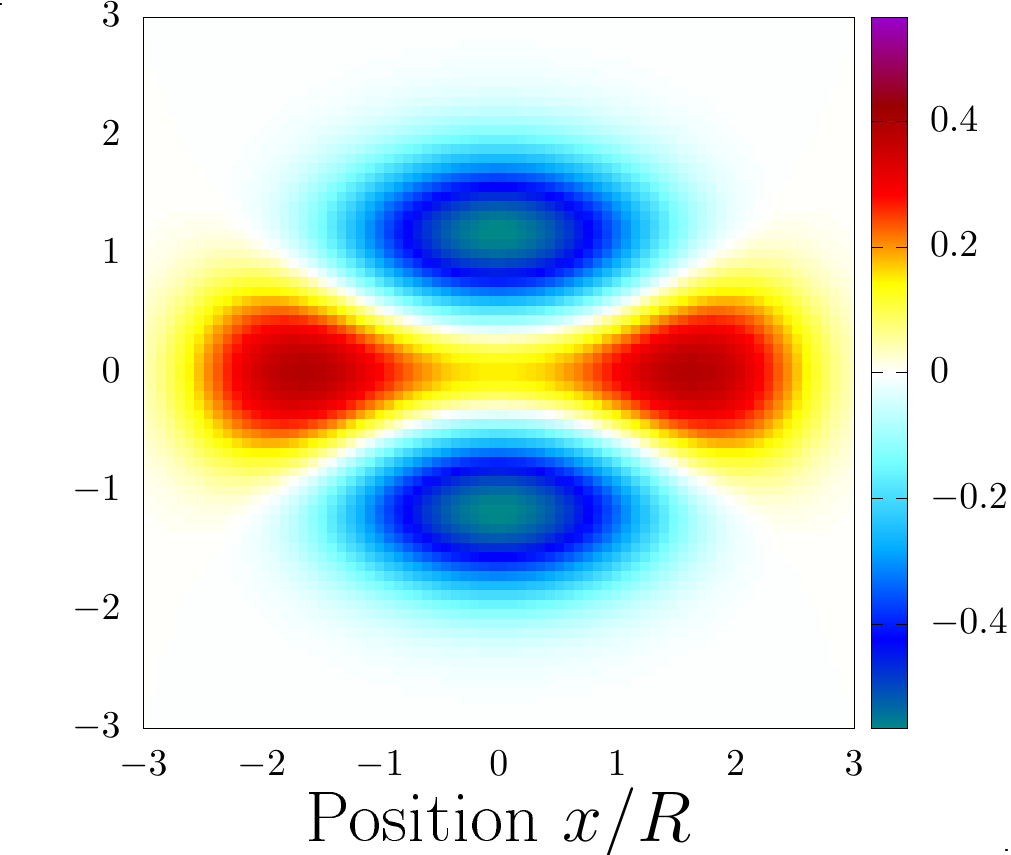} & \includegraphics[width=.3\textwidth]{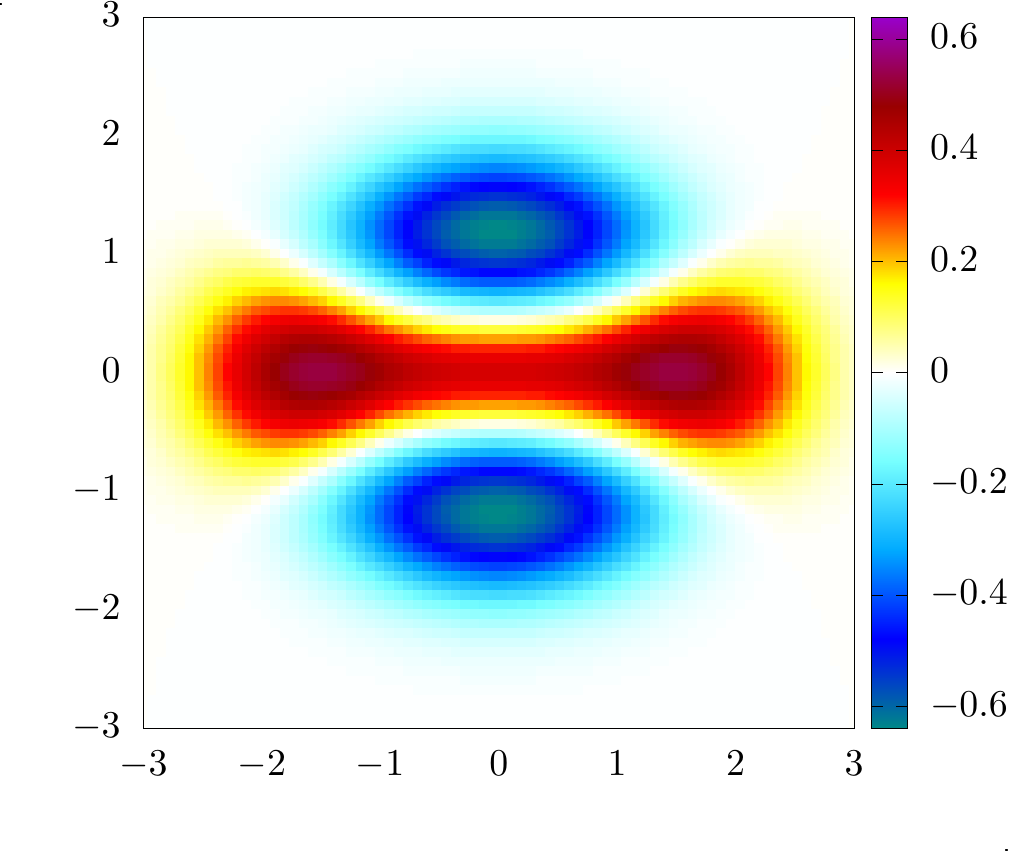}
    
    \end{tabular}
    \caption{Transverse profiles of the transverse anisotropy $\tau(T^{xx}-T^{yy})$ 
    in kinetic theory at time $\tau=1R$ for different opacities $4\pi\eta/s=0.5$ (left), $3$ (middle), $10$ (right). 
    }
    \label{fig:transverse_Txx-Tyy}
\end{figure*}

We can discern additional spatially resolved information on the opacity dependence of the system's evolution by also comparing profile plots of the anisotropic stress, $T^{xx}-T^{yy}$, which are presented in Figure~\ref{fig:transverse_Txx-Tyy}. Per definition in Eq.~\eqref{eq:obs_epsp}, the transverse integral of this quantity is proportional to elliptic flow $\varepsilon_p$, which builds up more at smaller values of $\eta/s$. Note that the symmetry-plane 
phase factor takes the value 
$e^{2i \Psi_p} = -1$
in this case, such that a negative integral results in positive $\varepsilon_p$. The plots show that the transverse plane separates into regions with different sign of the anisotropic stress. The behaviour in the outskirts is dictated by transverse expansion, resulting in positive values in $\pm x$-direction and negative values in $\pm y$-direction. The buildup of elliptic flow seems to proceed mainly via the positive parts decaying more than negative ones. At small opacities in the right panel, particles propagate with few interactions. Due to the initial almond shape, most of the particles in the center propagate in $\pm x$-direction, resulting in a larger $T^{xx}$ than $T^{yy}$. At large opacities in the left panel, the system is hydrodynamized and the anisotropic stress comes mostly from the direction of flow. Since the flow components $u_x$ and $u_y$ are zero in the center of the system, the anisotropic stress vanishes there.

\subsection{Time evolution of observables in kinetic theory and hydrodynamics} \label{sec:time_evolution_kinetic_theory_scaled_hydro}

\begin{figure*}
    \centering
    \begin{tabular}{ccc}
    \small $4\pi\eta/s=0.5$&\small $4\pi\eta/s=3$&\small $4\pi\eta/s=10$\\
    \includegraphics[width=.18\textwidth]{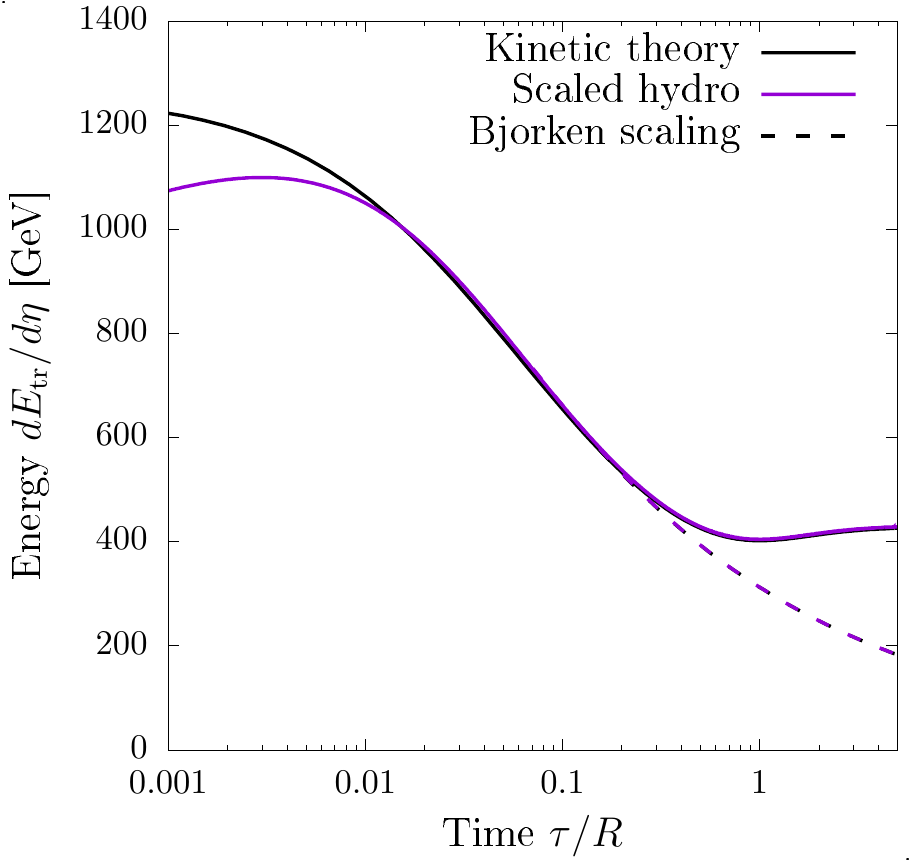}  & 
    \includegraphics[width=.18\textwidth]{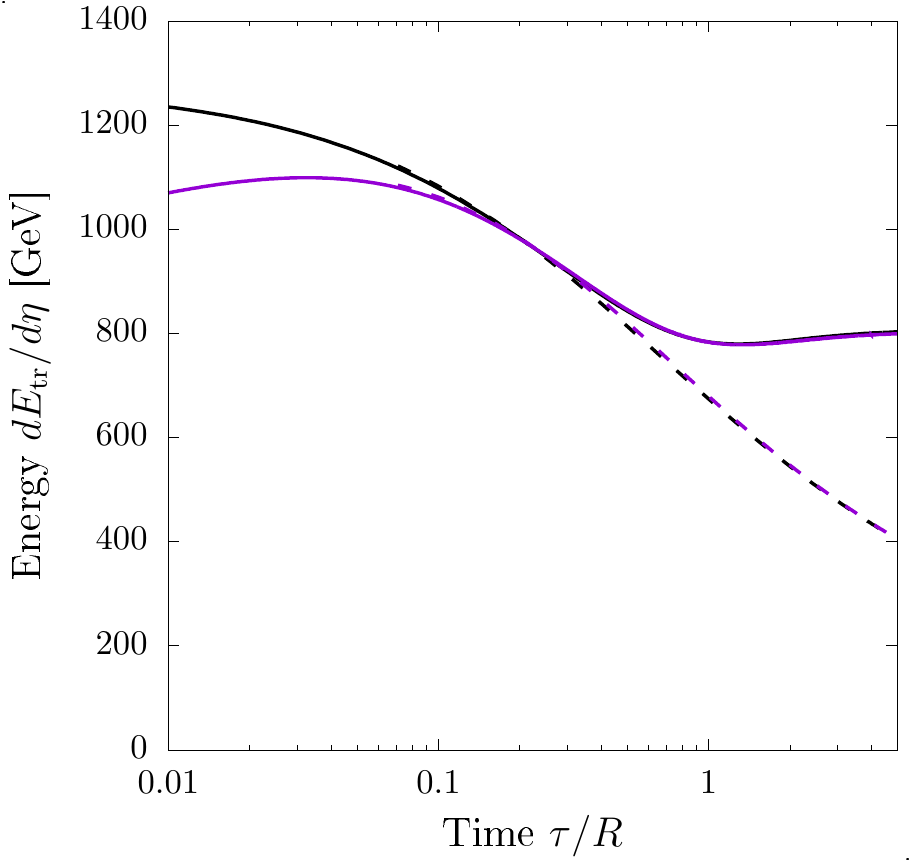}&
    \includegraphics[width=.18\textwidth]{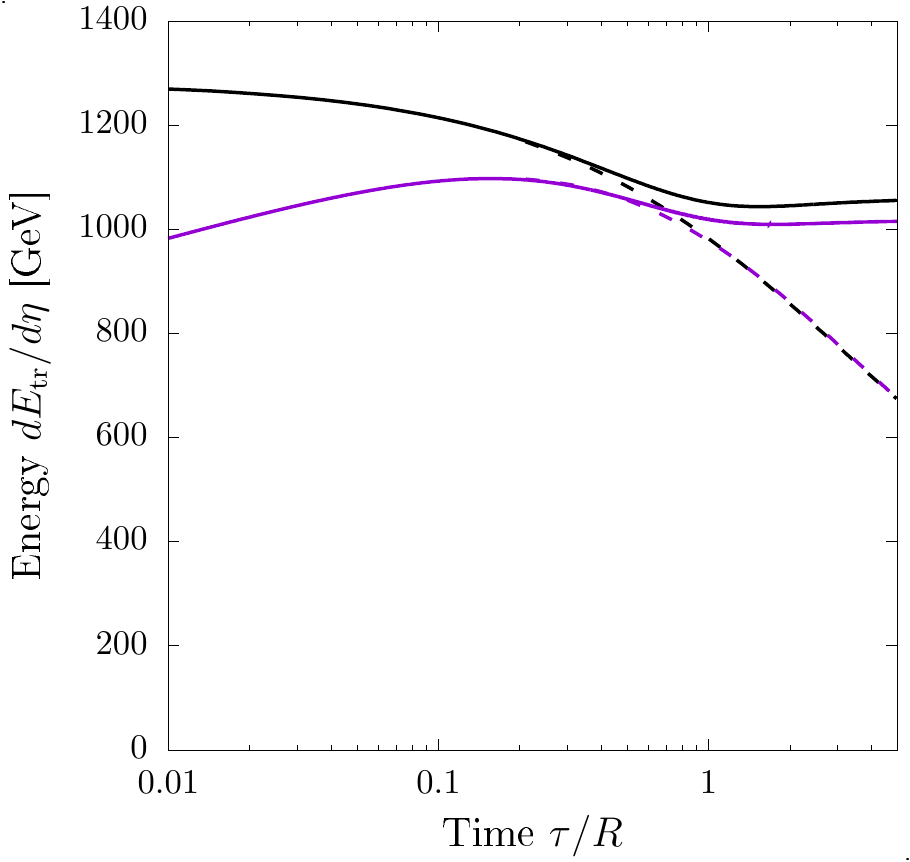}\\
    \includegraphics[width=.18\textwidth]{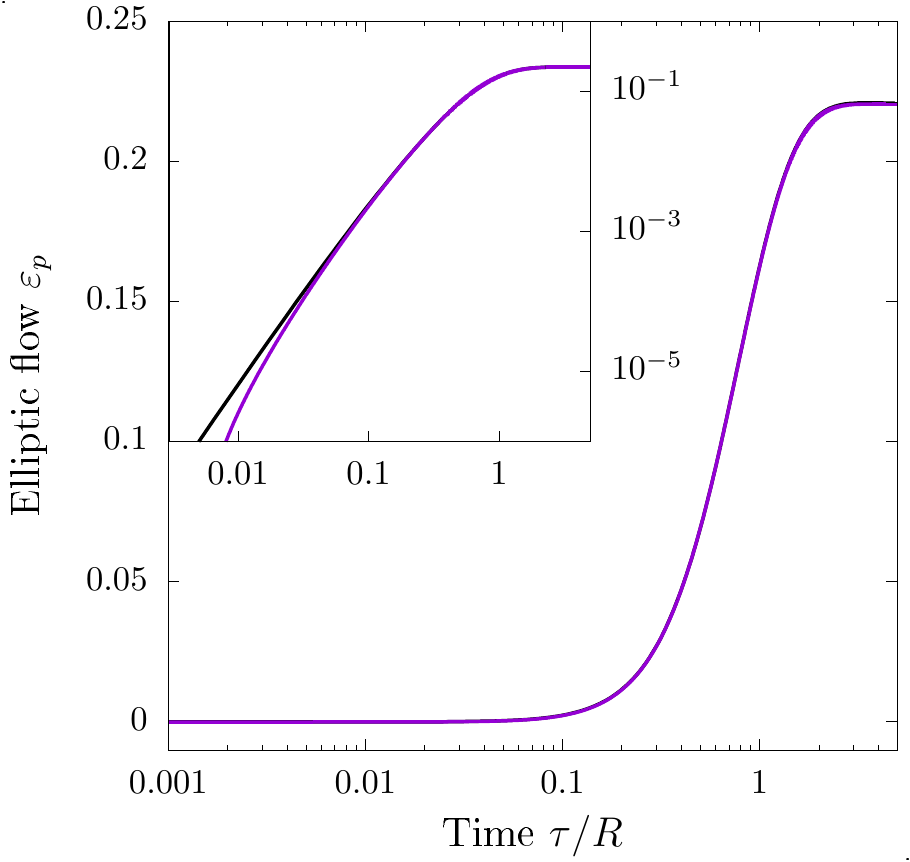}  &
    \includegraphics[width=.18\textwidth]{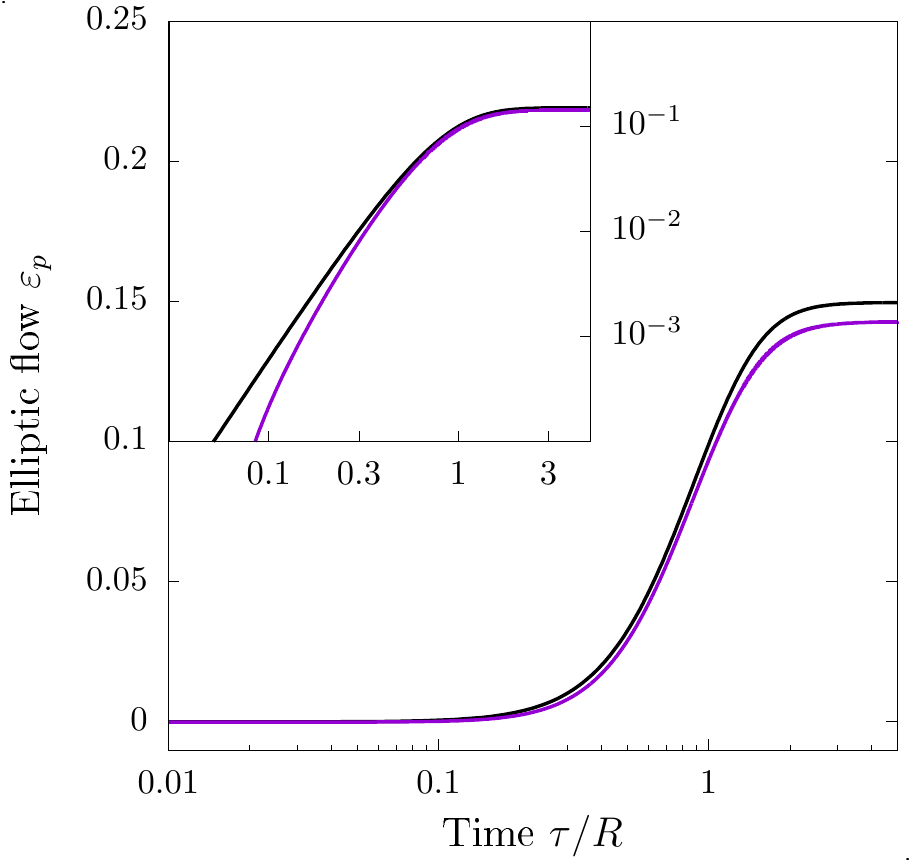}&
    \includegraphics[width=.18\textwidth]{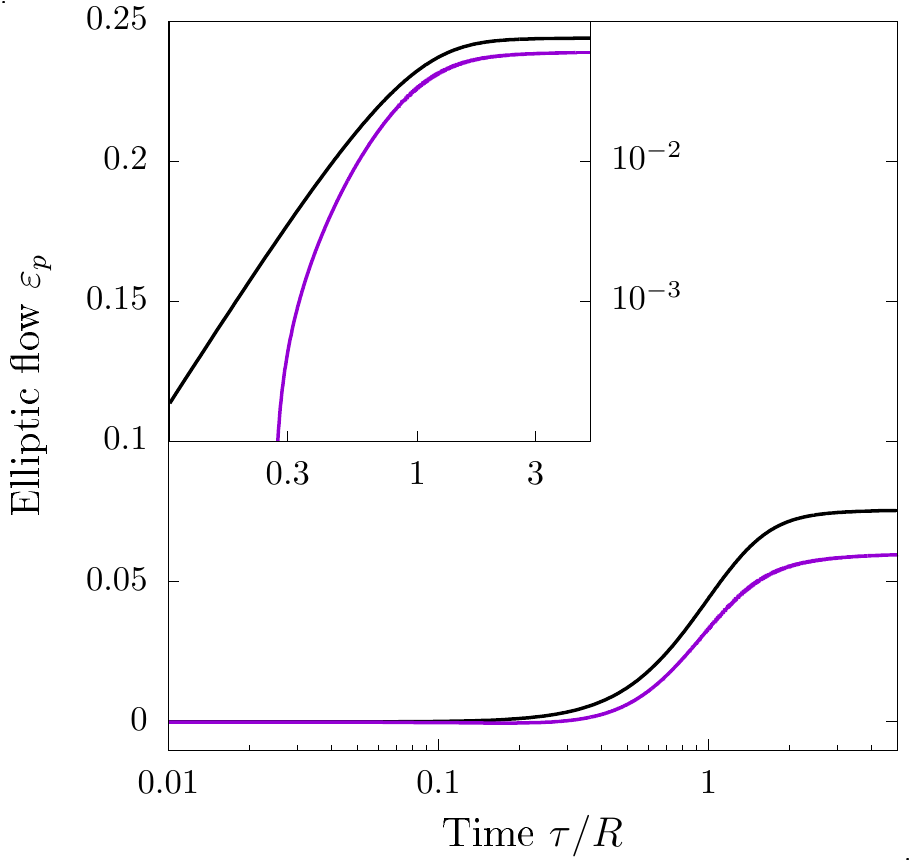}\\
    \includegraphics[width=.18\textwidth]{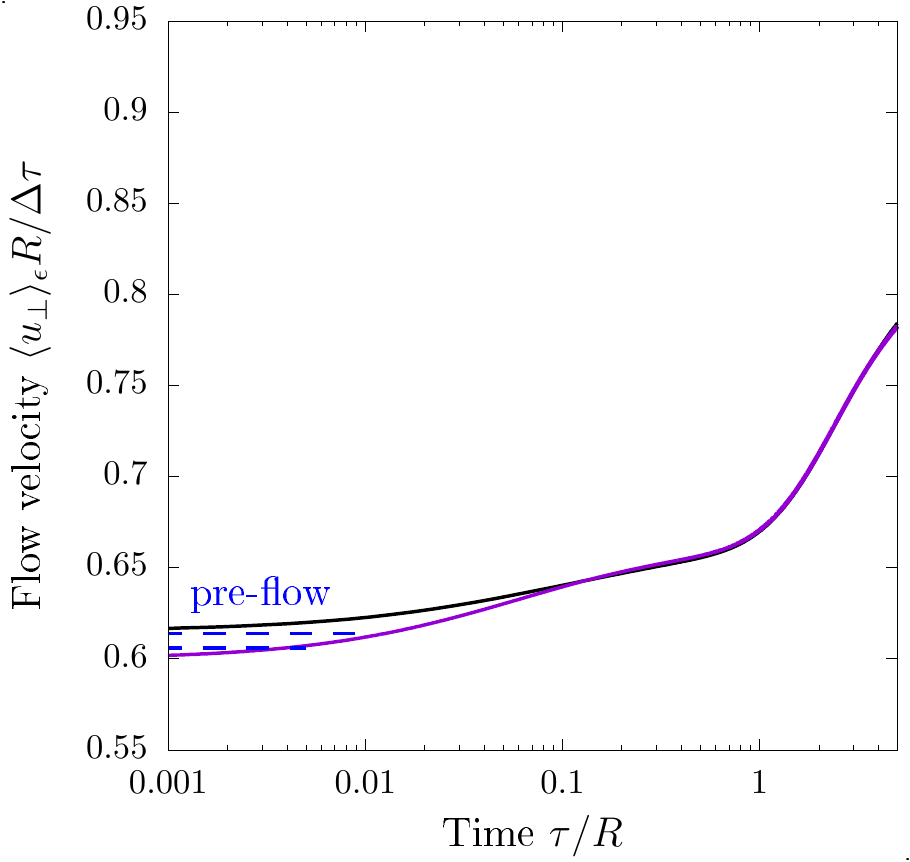}  & 
    \includegraphics[width=.18\textwidth]{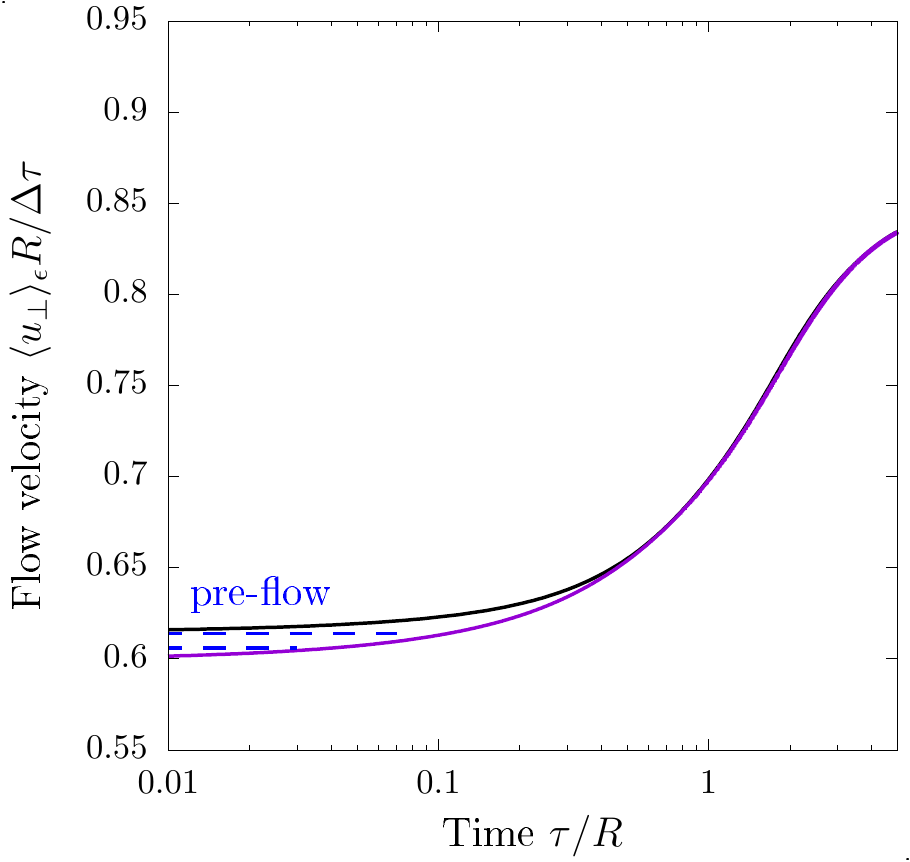}&
    \includegraphics[width=.18\textwidth]{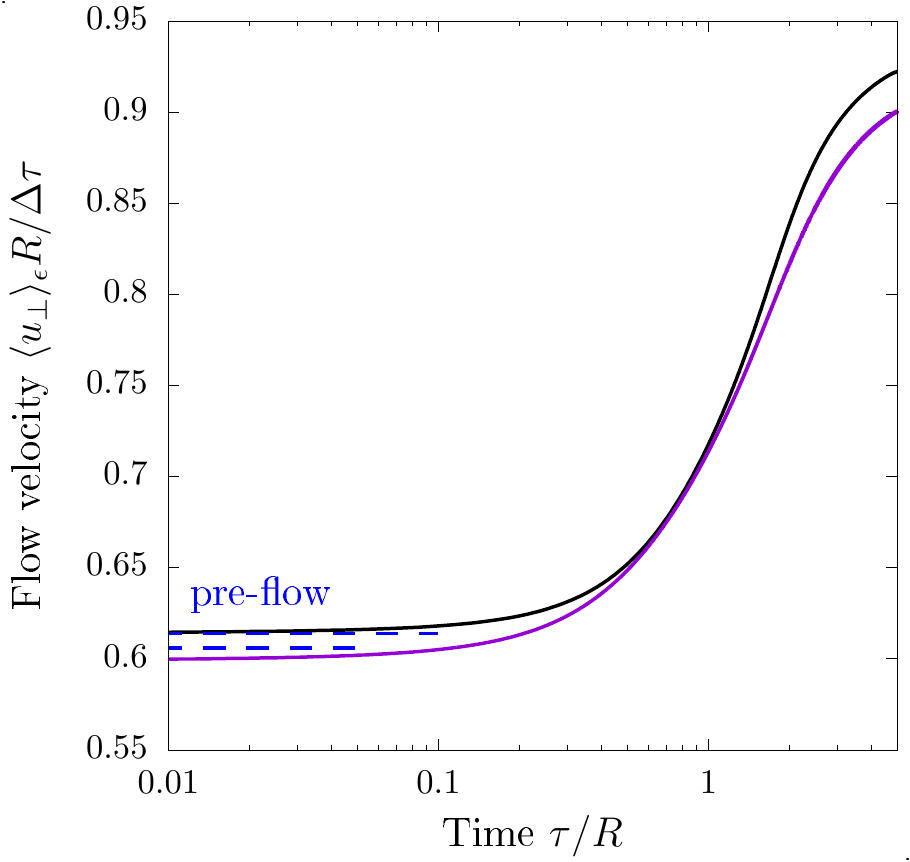}\\
    \includegraphics[width=.18\textwidth]{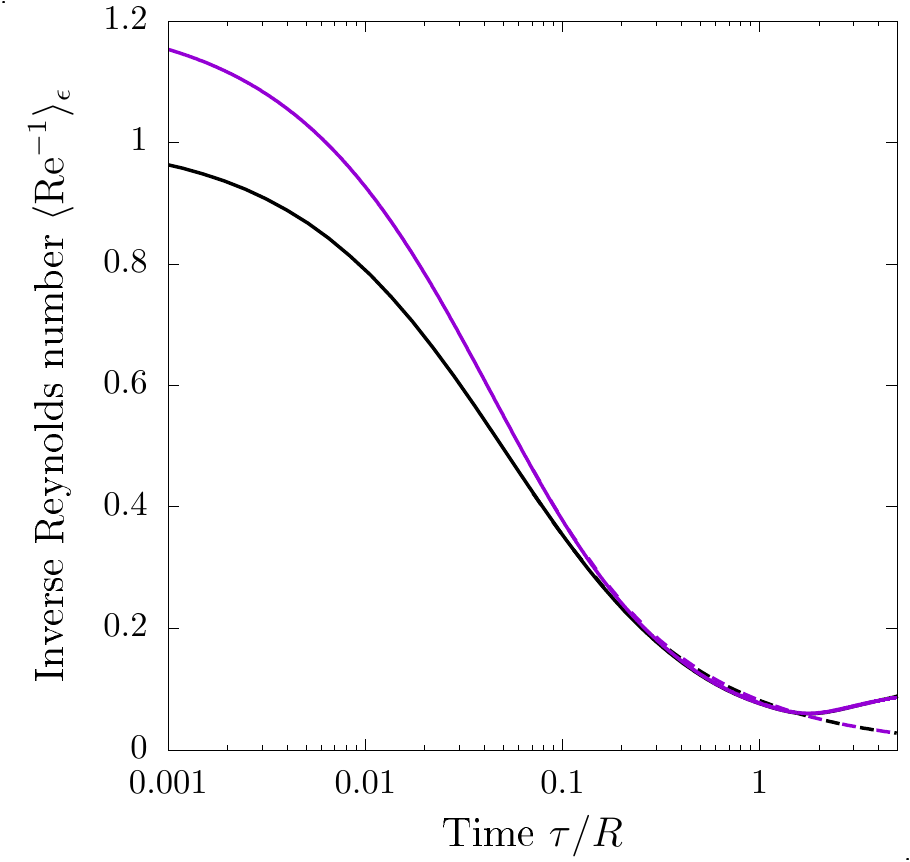}  &
    \includegraphics[width=.18\textwidth]{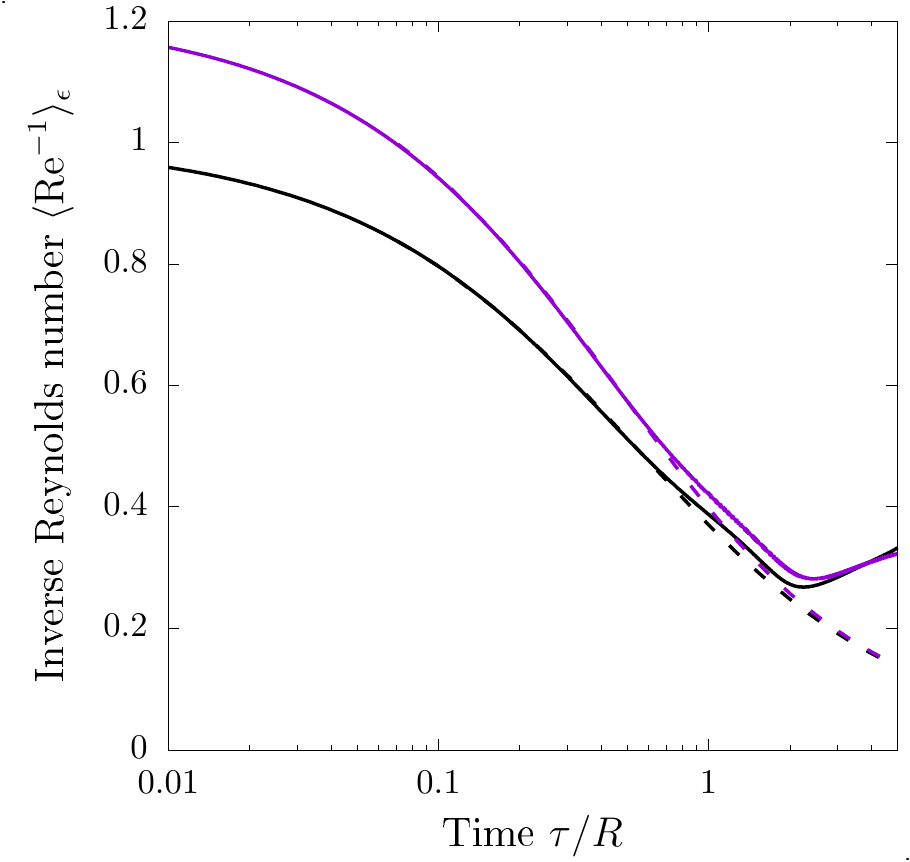}& 
    \includegraphics[width=.18\textwidth]{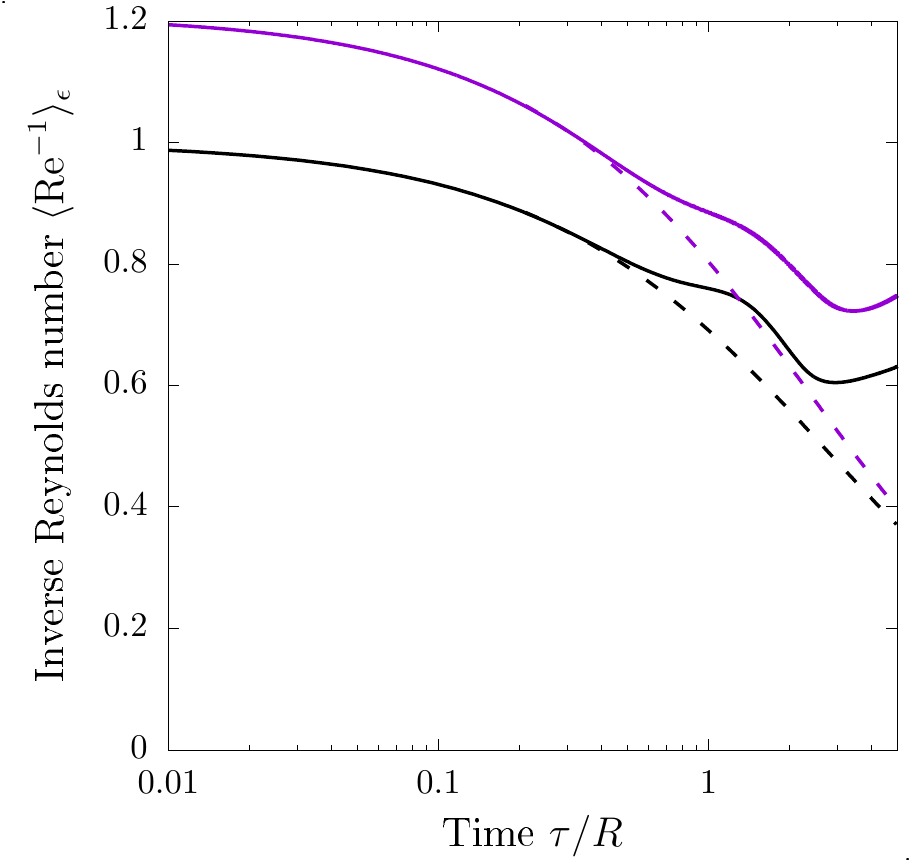}
    \end{tabular}
    \caption{Time evolution of (from top to bottom) transverse energy $\d E_{\rm tr}/\d\eta$ [cf. Eq.~\eqref{eq:obs_dEtrdeta}], elliptic flow $\varepsilon_p$ [cf. Eq.~\eqref{eq:obs_epsp}], transverse flow velocity $\eavg{u_\perp}$ [cf. Eq.~\eqref{eq:obs_uT}] and inverse Reynolds number $\eavg{\mathrm{Re}^{-1}}$ [cf. Eq.~\eqref{eq:Reinv_def}] in kinetic theory (black) and scaled viscous hydrodynamics (purple). The time axis is scaled logarithmically in all plots. The plots showing elliptic flow $\varepsilon_p$ feature an inset plot of the same quantity plotted in log-log scale. The plots of flow velocity also show the pre-flow result from Table~\ref{tbl:pre-flow} for the early-time limit for $\eavg{u_\perp} / (\Delta\tau/R)$ ($0.614$ for kinetic theory and $0.606$ for scaled hydrodynamics). Bjorken scaling results are shown with dashed lines for $\d E_{\rm tr} / \d\eta$ (top) and $\langle {\rm Re}^{-1} \rangle_\epsilon$ (bottom).}
    \label{fig:time_evolution_kinetic_theory_scaled_hydro}
\end{figure*}

We will now examine the time evolution of certain characteristic transverse space integrated observables in both kinetic theory and the scaled hydrodynamics scheme that was proposed in Sec.~\ref{sec:initialization_scaled_hydro} as a countermeasure to the unphysical pre-equilibrium behaviour of hydrodynamics 
discussed in Sec.~\ref{sec:early:pre-equilibrium}. This will provide additional insights into the system's behaviour but also reveal how well the scaled hydro scheme works at different opacities. Figure~\ref{fig:time_evolution_kinetic_theory_scaled_hydro} shows comparisons of the time evolution of transverse energy $\d E_{\rm tr}/ \d\eta$, elliptic flow $\varepsilon_p$, average transverse flow velocity $\langle u_\perp \rangle_\epsilon$ and average inverse Reynolds number $\langle \mathrm{Re}^{-1}\rangle_\epsilon$ in both models at three different opacities. Since we are using a fixed initial profile, we plot $\varepsilon_p$ without normalization to the initial state eccentricity $\epsilon_2$. As an illustration of the motivation for the scaling scheme in hydrodynamics, for $\d E_{\rm tr}/ \d\eta$ and $\langle \mathrm{Re}^{-1}\rangle_\epsilon$ we also compare with the time evolution in the absence of transverse expansion, where we describe the system as a collection of local Bjorken flows.

The time evolution of transverse energy $\d E_{\rm tr}/ \d\eta$ closely follows results from Bjorken flow scaling at early times, as predicted in Sec.~\ref{sec:early:pre-equilibrium}. In Bjorken flow scaling, this observable starts out being constant in the free-streaming period of kinetic theory, while 
in hydrodynamics, it follows a positive power law,
cf. Eq.~\eqref{eq:eq_dEtrdeta_early}. 
From there, in both cases the time evolution smoothly transitions to a late time equilibrium power law $\d E_{\rm tr}/ \d\eta\sim \tau^{-1/3}$. The timescale of this transition depends on the opacity and is smaller at smaller $\eta/s$.  In RTA\footnote{In general, the equilibration timescale scales with $(\eta/s)^{3(1-\gamma/4)/2}$, with $\gamma$ as defined in Eq.~\eqref{eq:bjorken_gamma}. Numerically, the exponent $1.30$ for viscous hydrodynamics is close to the one for RTA.}, it scales as $\tau_{\rm eq}\sim (\eta/s)^{4/3}$~\cite{Ambrus:2021fej}. By construction, results from scaled hydrodynamics agree with kinetic theory results in the late time limit of Bjorken flow scaling. The time evolution in full simulations follows this behaviour up to times $\tau\sim R$, when effects of transverse expansion become significant. The rapid dilution due to transverse expansion decreases interaction rates and causes $\d E_{\rm tr}/ \d\eta$ to approach a constant value. For large opacities like 
$4\pi\eta / s = 0.5$,
the Bjorken flow equilibrium where both models agree sets in long before transverse expansion and even afterwards the results will stay in agreement. Intermediate opacities around 
 $4\pi\eta / s = 3$
mark the transition region where results for $\d E_{\rm tr}/ \d\eta$ from both models just barely come into agreement before 
approaching
a constant value. At small opacities like in the case of $4\pi\eta/s=10$, the onset of transverse expansion interrupts the Bjorken flow scaling period before 
the two model descriptions have come into agreement. 
The residual discrepancy then persists throughout the evolution of the system and leads to a mismatch of final state observables,
which becomes worse as $\eta / s$ is increased.

The second line of Figure~\ref{fig:time_evolution_kinetic_theory_scaled_hydro} shows the time evolution of the elliptic flow coefficient $\varepsilon_p$. 
Again, like in the case of $\d E_{\rm tr}/ \d\eta$, 
because of
the decrease of interaction rates due to the dilution 
caused by
transverse expansion, 
$\varepsilon_p$ reaches 
a late-time plateau. Thus, at all opacities, $\varepsilon_p$ builds up in a timeframe of 
$\tau \lesssim 2R$. Contributions from early times are negligible, such that effectively the buildup starts at $\tau \gtrsim 0.1R$.
As indicated in the log-log insets, the kinetic theory curves exhibit at early times an approximate power-law increase, $\varepsilon_p \propto \tau^{8/3}$. In contrast, the scaled hydro curve for $\varepsilon_p$ first dips to negative values. For $4\pi \eta / s =0.5$, when equilibration is achieved before the onset of transverse expansion, the scaled hydro curve merges into the RTA one as implied by the discussion in Sec.~\ref{sec:initialization_scaled_hydro}. At small opacity ($4\pi \eta / s = 10$), the merging process is interrupted by transverse expansion.
The scaled hydro result for $\varepsilon_p$ is in perfect agreement with kinetic theory at large opacities and stays in good agreement at intermediate opacities. Due to a smaller overall interaction rate, the $\varepsilon_p$-response decreases with decreasing opacity. For small opacities, a negative trend in the early time behaviour of hydrodynamics causes discrepancies with kinetic theory. This trend will become dominant at even smaller opacities, resulting in negative values of the late time plateaus.

As discussed in Sec.~\ref{sec:early:pre-flow}, at early times, $\langle u_\perp \rangle_\epsilon$ builds up linearly with the elapsed time $\Delta\tau=\tau-\tau_0$ in kinetic theory. For finite initialization time $\tau_0$, the detailed behaviour in hydrodynamics is slightly different, but almost indistinguishable from linearity in $\Delta\tau$. Hence, we plot the ratio $\frac{\langle u_\perp \rangle_\epsilon}{\Delta\tau/R}$ and indicate the early time limit using horizontal dashed blue lines. The plots confirm that there are slight differences in the early time behaviour of the flow velocities in hydrodynamics and kinetic theory, however they come into agreement on similar timescales as $\d E_{\rm tr}/ \d\eta$. This is partly owing to the fact that early time contributions to the total $\langle u_\perp \rangle_\epsilon$ are negligible. $\langle u_\perp \rangle_\epsilon$ enters a period of superlinear rise during transverse expansion. While this period ends earlier at larger opacities due to dilution of the system and transition to free-streaming, the total rise of $\frac{\langle u_\perp \rangle_\epsilon}{\Delta\tau/R}$ is nevertheless larger. Comparing  hydrodynamic results to kinetic theory results, the late time free-streaming does not seem to be accurately reproduced, as hydrodynamics underestimates $\langle u_\perp \rangle_\epsilon$. Problems in the late time behaviour are less relevant for the other observables we discuss, as they tend to plateau at late times. This late time discrepancy between hydrodynamics and kinetic theory is thus a phenomenon that mainly affects $\langle u_\perp \rangle_\epsilon$ among the observables that were tracked in this work, and is not related to pre-equilibrium.

Finally, we look at the time evolution of the average inverse Reynolds number, which is a measure of the size of non-equilibrium effects in the system. We normalized this in such a way that 
in RTA, its initial value is equal to one (note that 
on the hydro attractor, 
${\rm Re}^{-1} \sim 1.212$ when $\tau_0 \rightarrow 0$). 
Like for $\d E_{\rm tr}/ \d\eta$, the two model descriptions will come into agreement in the late time limit of Bjorken flow scaling, on timescales that are larger for smaller opacities. Due to equilibration, in this period $\langle \mathrm{Re}^{-1}\rangle_\epsilon$ experiences a phase of rapid decay towards $0$, as expected 
since $\mathrm{Re}^{-1}$ measures
non-equilibrium effects. The effect of transverse expansion on this observable is not straightforwardly understood, except for the fact that due to the additional dilution, $\langle \mathrm{Re}^{-1}\rangle_\epsilon$ must be larger in full simulations than in Bjorken flow scaling. For large opacities, transverse expansion seems to only slow down the approach to equilibrium. However, at intermediate opacities we see a significant rise in $\langle \mathrm{Re}^{-1}\rangle_\epsilon$. We also computed results for the limit of vanishing opacity. Here, the inverse Reynolds number remains constant at early times, but later increases due to transverse expansion, e.g. at $\tau=4R$ to a value of $\langle\mathrm{Re}^{-1}\rangle_\epsilon(\tau=4R)=1.322$. However, an increase due to transverse expansion cannot be the only late time effect, as we can see from the results at $4\pi\eta/s=10$, where the trend of this quantity changes multiple times. It first departs from the local Bjorken flow prediction at $\tau/R\sim 0.3$, but later the curve returns to decreasing at a rate comparable to that during the Bjorken flow stage. At late times, the behaviour transitions to a rise in the inverse Reynolds number. Despite this peculiar behaviour, our numerical results indicate that $\langle \mathrm{Re}^{-1}\rangle_\epsilon$ 
reaches a minimum value that is larger for smaller opacities. For very small opacities, it will not drop significantly below its initial value of $1$ before starting to rise.

For a more detailed examination of the opacity dependence of the time evolution in kinetic theory ranging from very small ($4\pi\eta/s=1000$) to very large opacities ($4\pi\eta/s=0.01$), please see Appendix~\ref{app:tevo_gdep}.

\begin{figure*}
    \centering
    \includegraphics[width=.6\textwidth]{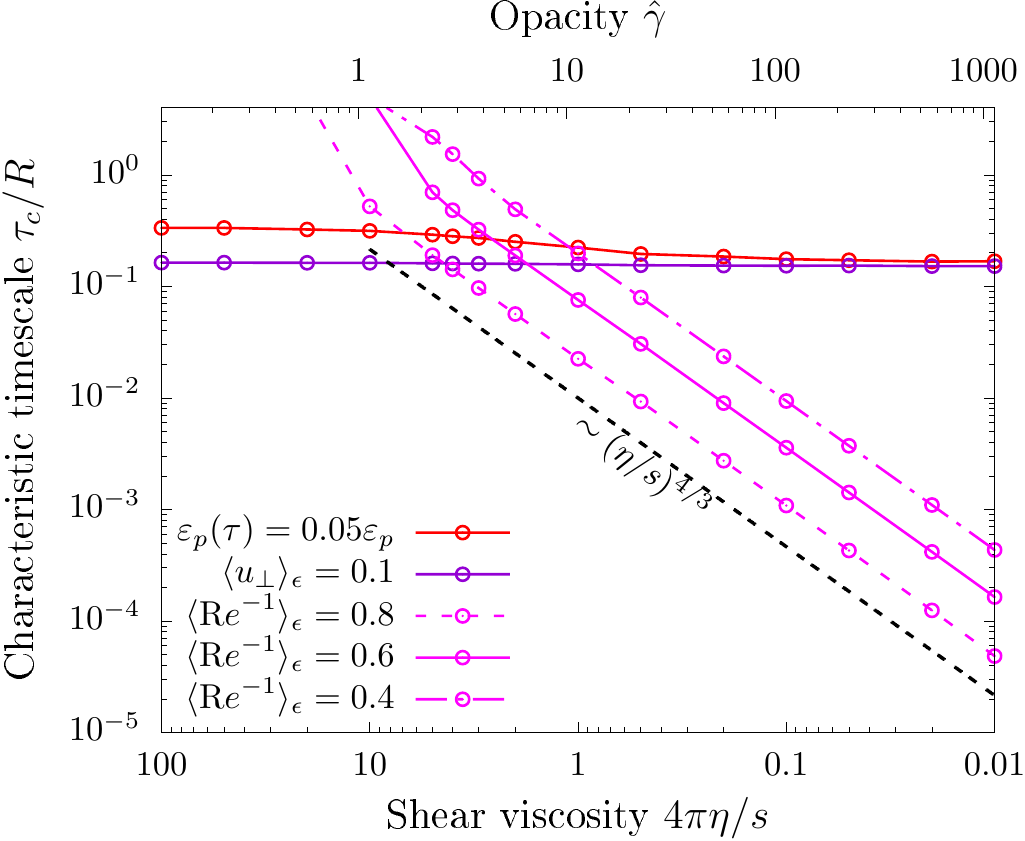}
    \caption{Opacity ($\hat{\gamma}=\frac{11.3}{4\pi\eta/s}$) dependence of the characteristic times where the elliptic flow $\varepsilon_p$ [cf. Eq.~\eqref{eq:obs_epsp}] reaches 5\% of its late time ($\tau=4R$) value (red), the transverse flow velocity [cf. Eq.~\eqref{eq:obs_uT}] builds up to a value of $\eavg{u_\perp}=0.1$ (purple), or the inverse Reynolds number [cf. Eq.~\eqref{eq:Reinv_def}] drops to a value of $\eavg{\mathrm{Re}^{-1}}=0.8$ (pink, dashed), $0.6$ (pink, solid) or $0.4$ (pink, long-short dashed). The buildup in transverse flow velocity marks the transition from the Bjorken flow scaling regime to the regime of transverse expansion, while the drop in inverse Reynolds number marks the region where hydrodynamics is applicable.}
    \label{fig:ts_per_etas}
\end{figure*}

After examining the time evolution of these observables and establishing some understanding about the implications of their buildup, we now want to invert this logic. As the change in these observables carries information on the state of the system, e.g. the progress of its equilibration or the onset of transverse expansion, we want to track the first times  when these observables reach a specific milestone of their time evolution as a function of opacity. Figure~\ref{fig:ts_per_etas} shows plots of kinetic theory results for these curves for five different milestone criteria. Specifically, we tracked when the average transverse flow velocity reaches a value of $0.1$ as a criterion for the onset of transverse expansion, the time when the elliptic flow response builds up to $5\%$ of its 
maximum value 
at the given opacity as a criterion for the beginning of the buildup of flow, and the time when the average inverse Reynolds number reaches values of $0.4$, $0.6$ and $0.8$, which tells us to what degree hydrodynamization has progressed. As it turns out, the curve for the flow velocity criterion is almost perfectly flat at a value of $\tau_c\approx 0.15R$, meaning that the early time buildup of $\langle u_\perp \rangle_\epsilon$ is mostly independent of the opacity. The elliptic flow criterion is met at similar times as the flow velocity criterion at large opacities, but at slightly later times $\tau_c\approx 0.3R$ for small opacities. Despite the general timeframe of $\varepsilon_p$-buildup being independent of opacity, it seems to start slightly earlier at larger opacities. The most interesting criterion curves are those for the average inverse Reynolds number. The 
 system's
adherence to early time Bjorken flow scaling leads to a power law behaviour
$\tau_c\propto(\eta/s)^{4/3}$ for all three of these curves at large opacities.
The curves deviate from this power law when the criterion is not reached before transverse expansion sets in at times $\tau\sim R$. For small opacities, the criteria are never met, as the average inverse Reynolds number reaches a minimum value larger than the criterion value, as already stated in the discussion of Figure~\ref{fig:time_evolution_kinetic_theory_scaled_hydro}. 
The behaviour of $\d E_{\rm tr} / \d \eta$ resembles that of $\d E_\perp / \d \eta$, which we already discussed in our previous paper~\cite{Ambrus:2021fej}. 
Similarly to $\langle \mathrm{Re}^{-1}\rangle_\epsilon$, it follows Bjorken flow scaling at early times, resulting in a similar power law behaviour.

\subsection{Time evolution in hybrid schemes} \label{sec:time_evolution_hybrid}

Another way to alleviate discrepancies due to the behaviour of hydrodynamics in the pre-equilibrium phase as discussed in Sec.~\ref{sec:early:pre-equilibrium} is to model the time evolution via a hybrid scheme, switching from a kinetic theory based description at early times to hydrodynamics at later times, i.e. initializing the hydrodynamic simulation with the energy-momentum tensor computed from the kinetic theory based time evolution. This requires to fix a criterion for when to switch descriptions.

As we argue that hydrodynamics becomes viable only after some timescale related to equilibration, we also expect the accuracy of hybrid scheme results to depend on the switching times. Due to the opacity dependence of equilibration, it might be beneficial to choose switching times as a function of opacity. Hence we tested both a hybrid scheme with fixed switching times at two different times $\tau=0.4\ $fm and $\tau=1\ $fm, which are in the range of switching times typically used in phenomenological descriptions, and with dynamically determined switching times. 

In order to tie this definition to the phenomenon of equilibration, we determine the dynamical switching times on the basis of the 
decrease of the average inverse Reynolds number $\langle \mathrm{Re}^{-1}\rangle_\epsilon$, i.e. we switch as soon as this quantity first reaches a specific value. Specifically, we chose the values $\langle \mathrm{Re}^{-1}\rangle_\epsilon=0.8$, $0.6$ and $0.4$  (sometimes we will consider switching also when $\langle {\rm Re}^{-1} \rangle_\epsilon$ drops below 0.2). 
In the case of a transversally homogeneous system, Figure~\ref{fig:attractor_curves} shows that these values for the inverse Reynolds number correspond to various degrees of hydrodynamization of the system. 
Specifically, ${\rm Re}^{-1} = 0.8$ ($\tilde{w} \simeq 0.2$) corresponds to the onset of hydrodynamization. When ${\rm Re}^{-1} = 0.6$ ($\tilde{w} \simeq 0.6$), the system significantly progressed through the hydrodynamization process,
while when ${\rm Re}^{-1} = 0.4$ ($\tilde{w} \simeq 1$), the system has hydrodynamized and the kinetic theory and hydrodynamics attractor curves are almost merged.
Due to the relation \eqref{eq:wt} between $\tilde{w}$ and the Bjorken time $\tau$, the characteristic times $\tau_c$ when ${\rm Re}^{-1}$ drops below a certain threshold increase with $4\pi\eta / s$ (see Sec.~\ref{sec:time_evolution_kinetic_theory_scaled_hydro} for a detailed discussion).

\begin{figure*}
    \centering
    \begin{tabular}{ccc}
    \small $4\pi\eta/s=0.5$&\small $4\pi\eta/s=3$&\small $4\pi\eta/s=10$\\
        \includegraphics[width=.32\textwidth]{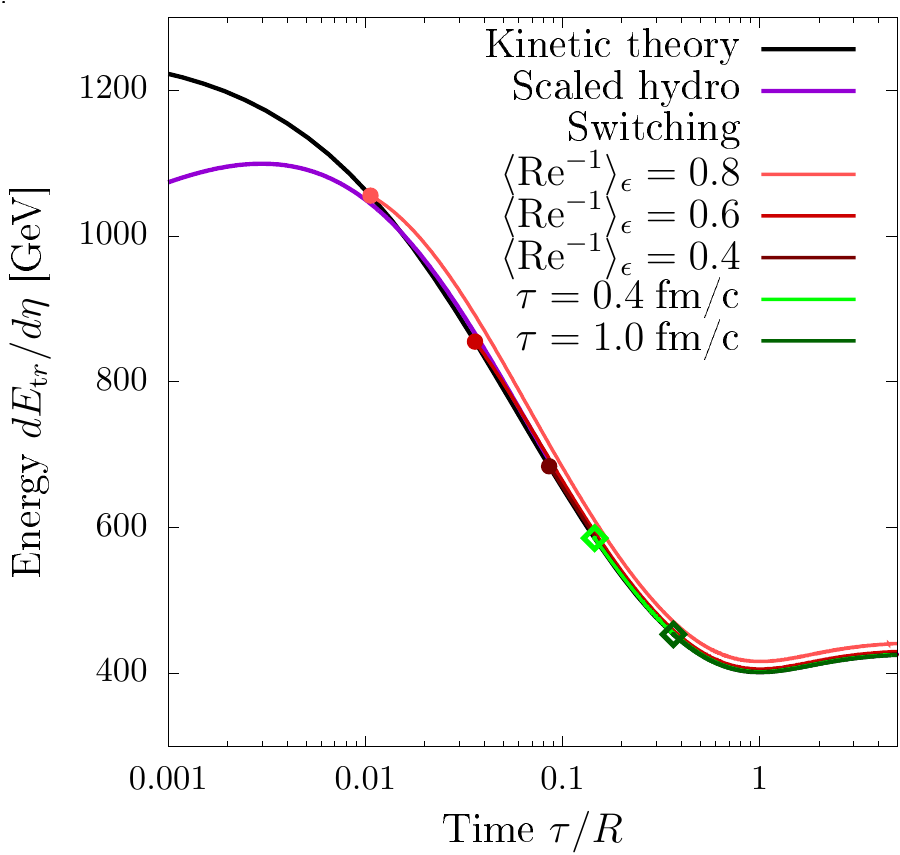}&
        \includegraphics[width=.32\textwidth]{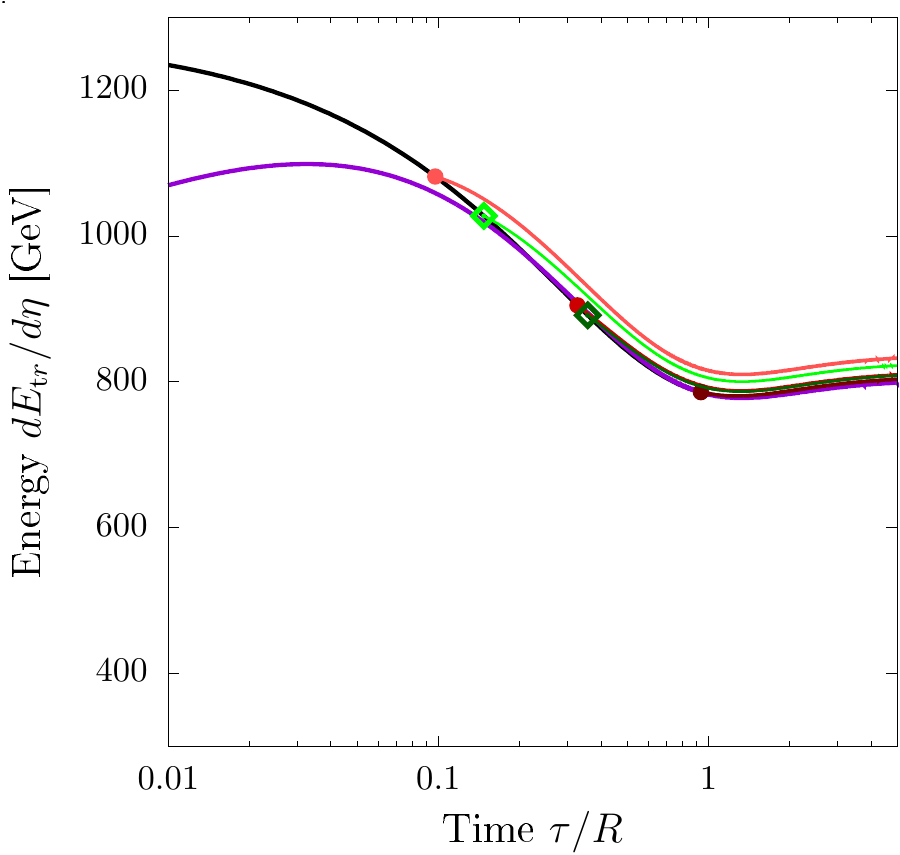}&
        \includegraphics[width=.32\textwidth]{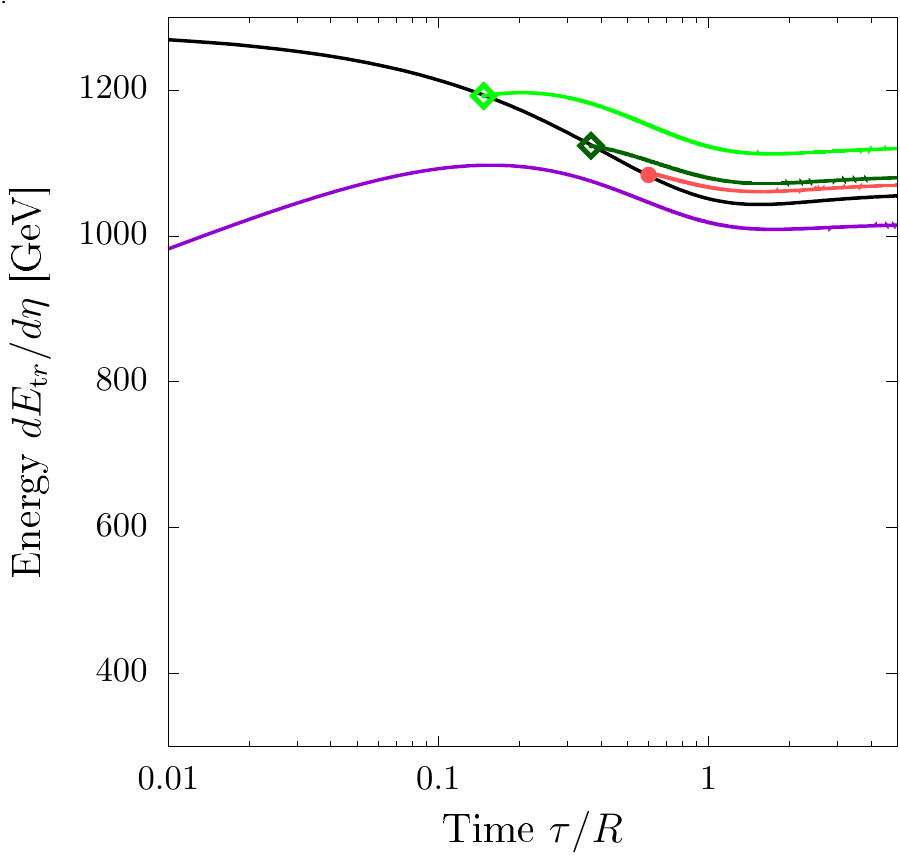}\\
        \includegraphics[width=.32\textwidth]{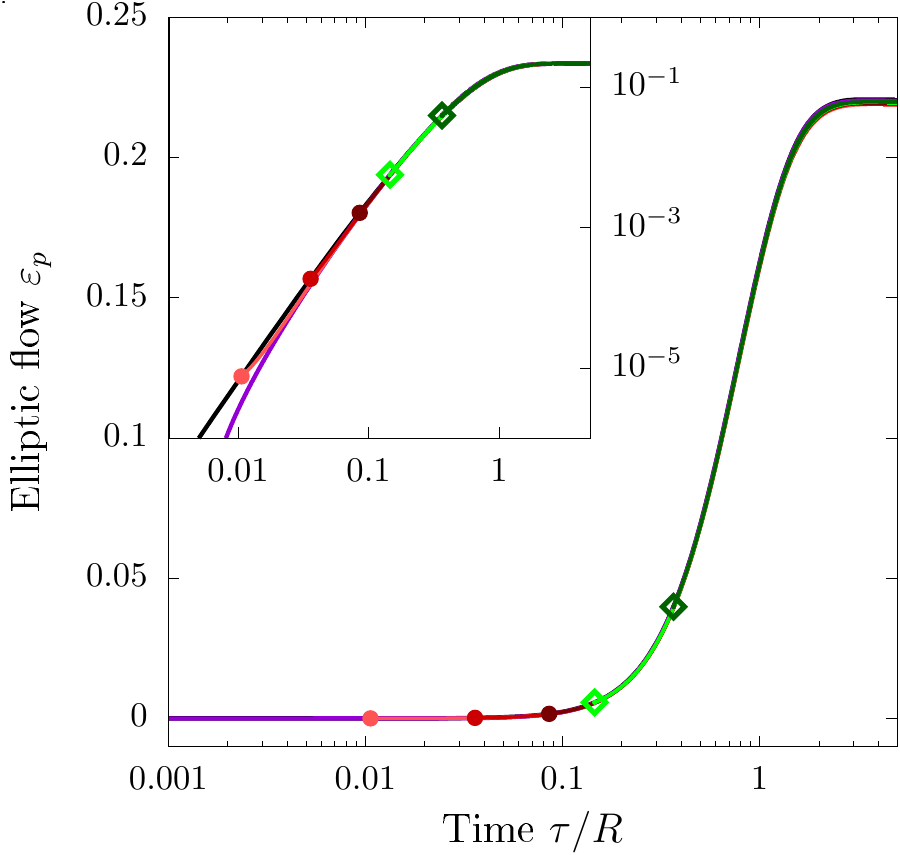}&
        \includegraphics[width=.32\textwidth]{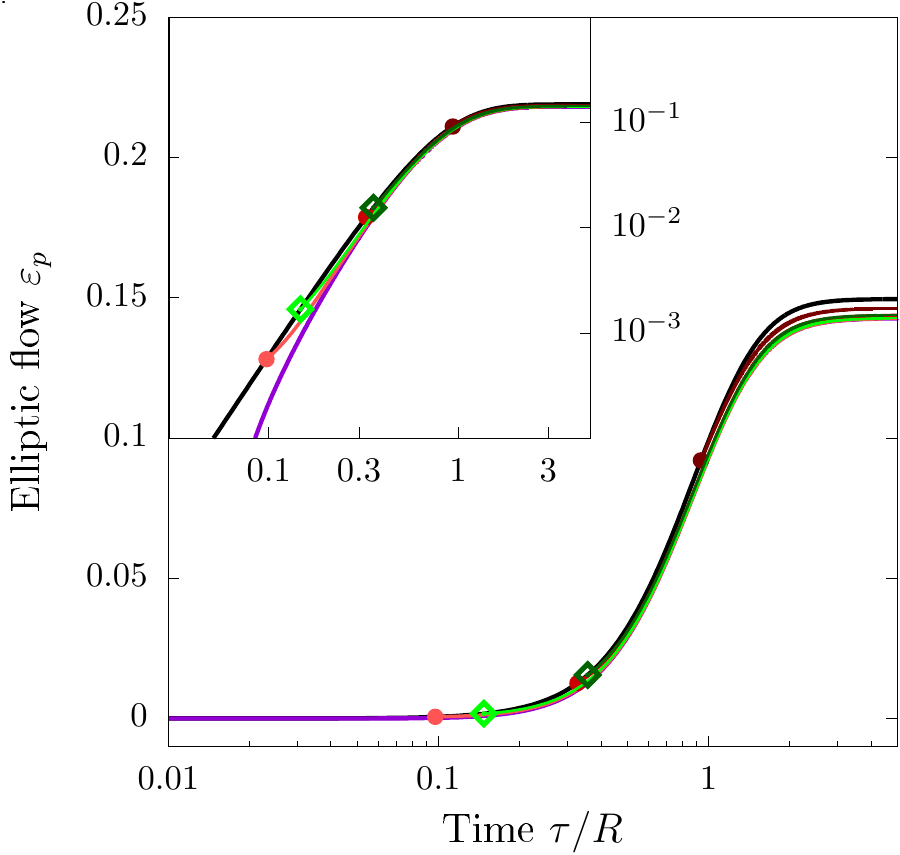}&
        \includegraphics[width=.32\textwidth]{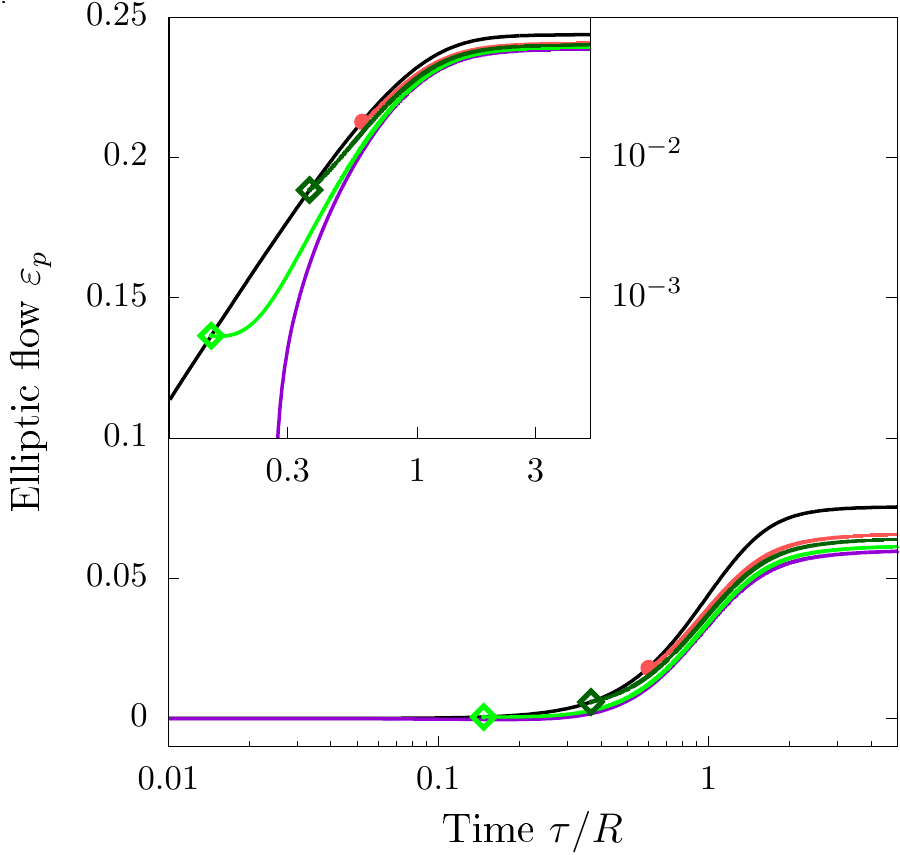}\\
        \includegraphics[width=.32\textwidth]{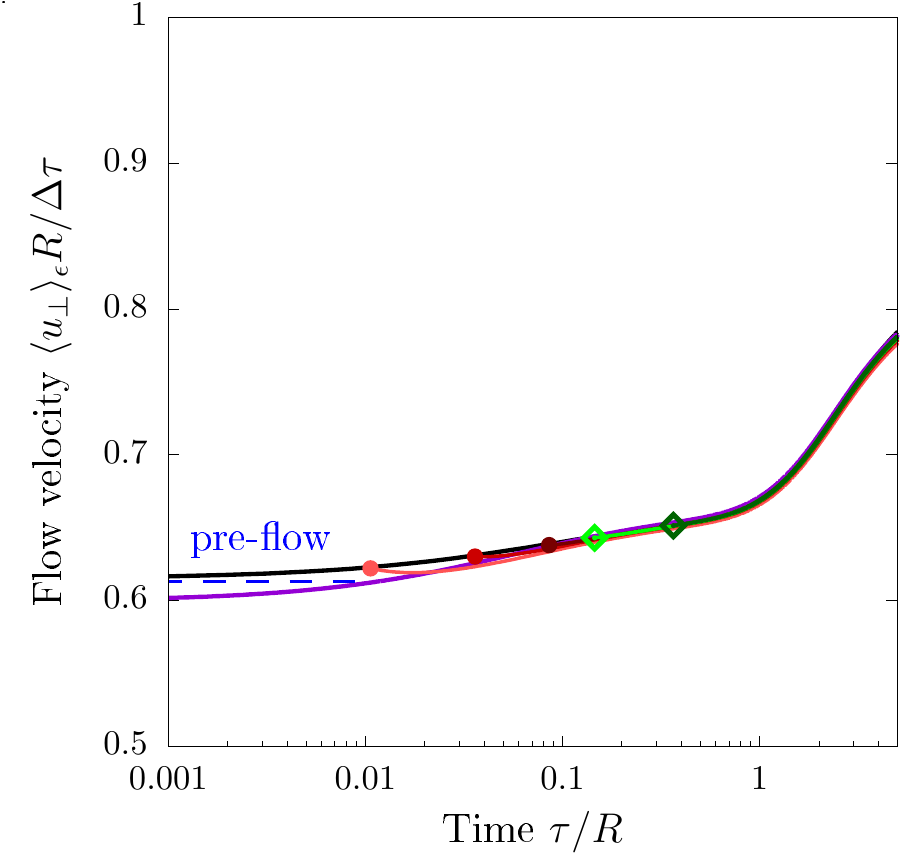}&
        \includegraphics[width=.32\textwidth]{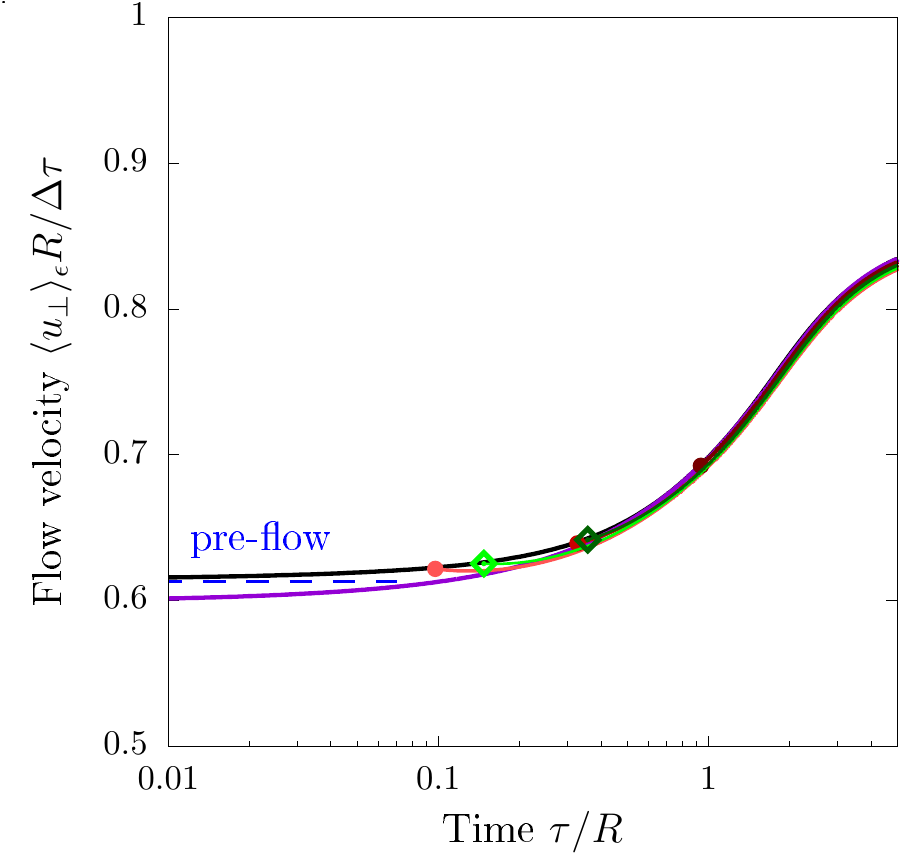}&
        \includegraphics[width=.32\textwidth]{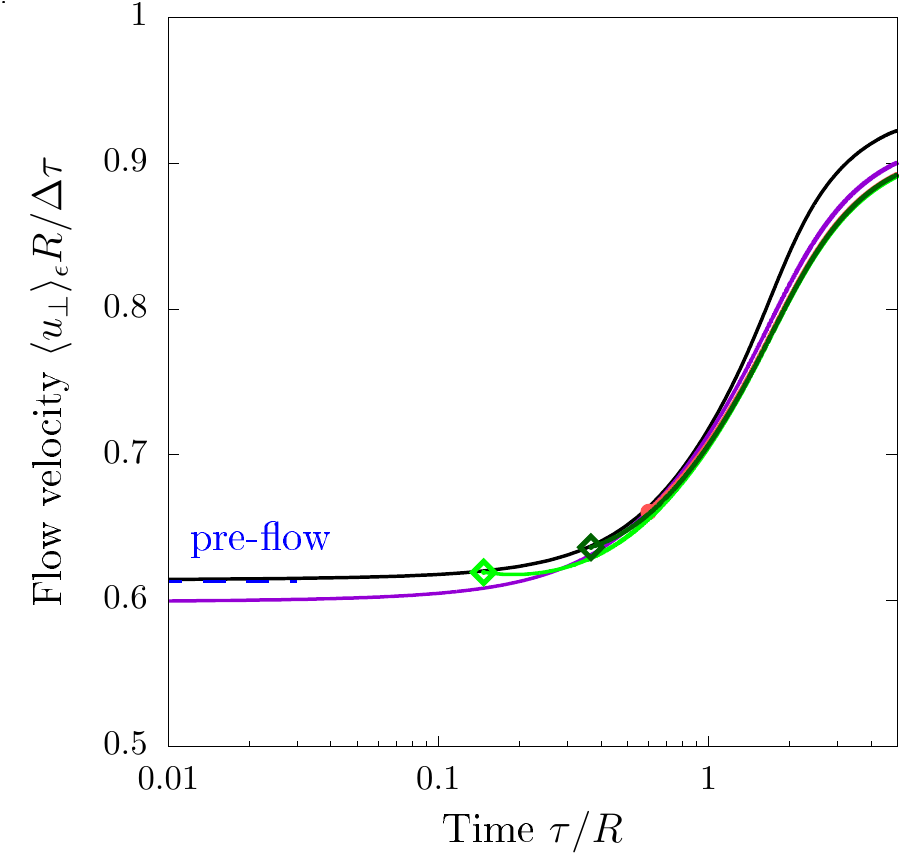}
    \end{tabular}
    
    \caption{Time evolution of transverse energy $\d E_{\rm tr}/\d\eta$ [top, cf. Eq.~\eqref{eq:obs_dEtrdeta}], elliptic flow $\varepsilon_p$ [middle, cf. Eq.~\eqref{eq:obs_epsp}] and transverse flow velocity $\eavg{u_\perp}$ [bottom, cf. Eq.~\eqref{eq:obs_uT}] in hybrid kinetic theory + viscous hydro simulations at opacities $4\pi\eta/s=0.5$ (left), $3$ (middle) and $10$ (right) when switching at different values of the inverse Reynolds number [cf. Eq.~\eqref{eq:Reinv_def}] $\eavg{\mathrm{Re}^{-1}}=0.8$ (light red), $0.6$ (red) and $0.4$ (dark red) or fixed time $\tau=0.4\,\mathrm{fm}$ (light green) and $\tau=1\,\mathrm{fm}$ (dark green). The switching points are marked with filled symbols. The time axis is scaled logarithmically. The plots showing elliptic flow $\varepsilon_p$ feature an inset plot of the same quantity plotted in log-log scale. 
    In the flow velocity plots, we also show the estimate $\langle u_\perp \rangle_{\epsilon, {\rm RTA}}=0.614\Delta\tau/R$ for the early-time buildup of pre-flow in kinetic theory (see Table~\ref{tbl:pre-flow}).
    }
    \label{fig:switching_etasdep}
\end{figure*}

The results are illustrated by the time evolution of transverse energy $\d E_{\rm tr}/ \d\eta$, elliptic flow $\varepsilon_p$ and average transverse flow velocity $\langle u_\perp \rangle_\epsilon$ compared for different choices of the switching times, as plotted in Figure~\ref{fig:switching_etasdep} at three different opacities. The early time evolution was computed with the RLB method of simulating kinetic theory. The plots also compare to results from a pure kinetic theory simulation as well as from our scaled viscous hydro scheme. Here we plot all results including the ones for elliptic flow $\varepsilon_p$ on a logarithmic scale of the time axis so that the different switching times are discernible. The $\varepsilon_p$ plots also feature an inset plot on log-log scale.
It can be seen that the curves corresponding to the hybrid setups tend to detach from the RTA curve towards lower values of $\varepsilon_p$. Since in viscous hydro, the equilibration process leads to a decrease of spatial eccentricity $\epsilon_2$ (see lower panel of Figure~\ref{fig:eccentricity_decrease}), the hybrid simulations with early switching times will lead to lower late-time values of $\varepsilon_p$ (see the discussion in the next section). 

At a small shear viscosity of $4\pi\eta/s=0.5$, all switching schemes yield accurate results for all three observables. 
Since the equilibration timescale is small for this opacity, 
 the system
has equilibrated by the time we switch descriptions such that kinetic theory and hydrodynamics are in agreement.
The $\langle\mathrm{Re}^{-1}\rangle_\epsilon$-based criteria are fulfilled early on in the system's evolution such that the dynamically chosen switching times are significantly smaller than the fixed ones. However, when comparing results from pure kinetic theory or viscous hydrodynamics, they are within the timeframe where both descriptions are in acceptable agreement. The only curve where a deviation from kinetic theory is clearly visible is the one for $\langle\mathrm{Re}^{-1}\rangle_\epsilon=0.8$, where hydrodynamization has only partly progressed before this criterion is fulfilled.

The results at $4\pi\eta/s=3$ now show that it is indeed necessary to give the choice of switching times some thought, as here we see a significant increase in accuracy of results for $\d E_{\rm tr}/ \d\eta$ and $\langle u_\perp \rangle_\epsilon$ with later switching times. For this opacity, the dynamically chosen switching times are on a similar scale as the fixed ones. We also see that the nature of any discrepancies with pure kinetic theory results is the same as in the case of hydrodynamics. As soon as we switch, the curves of these observables 
follow 
a trajectory that is qualitatively similar to the pure hydrodynamics result, meaning that $\d E_{\rm tr}/ \d\eta$ is increased, while $\langle u_\perp \rangle_\epsilon$ and $\varepsilon_p$ are decreased relative to the kinetic theory result.

The strength of the dynamically chosen switching times is well displayed for results at $4\pi\eta/s=10$. In this case, the system is still far from hydrodynamized at the two fixed switching times, leading to sizeable inaccuracies in the corresponding hybrid scheme results for all three observables, but especially for $\d E_{\rm tr}/ \d\eta$. As 
$\eavg{{\rm Re}^{-1}}$
does not drop low enough, two of the three criteria for the dynamical switching were not reached. However, the result for switching at the largest of the three values of $\eavg{\mathrm{Re}^{-1}}$ retains a similar level of accuracy 
as 
at smaller shear viscosity and is a significant improvement to fixed time switching results.

Overall, we find that while switching at fixed time is conceptionally straightforward and always possible, 
the accuracy of this scheme strongly depends on the opacity and
results at  small opacity show large deviations from full kinetic theory. On the other hand, switching based on $\eavg{\mathrm{Re}^{-1}}$ is not always possible because this quantity does not drop to the desired values at small opacities, but whenever it is possible, the accuracy of the result can be estimated beforehand and depends only little on opacity. In other words, the dynamical definition yields the earliest possible switching time for a desired accuracy, and whenever $\eavg{\mathrm{Re}^{-1}}$ does not drop enough for it to be determined, hydrodynamics is not viable in the first place.

Finally, we also tested hybrid schemes with the same switching times but with an early time evolution computed in K{\o}MP{\o}ST. We found that due to its limited range of applicability, some of the later switching times could not be viably reached with this description. But whenever we were able to obtain results, they were in good agreement with the results from the previously discussed scheme, except for some systematic errors in $\varepsilon_p$ and $\langle u_\perp \rangle_\epsilon$. These results are presented in more detail in Appendix~\ref{app:time_evolution_hybrid_kompost}, together with analogous time dependence plots to those presented in Figure~\ref{fig:switching_etasdep}.

\section{Opacity dependence of final state observables in different time evolution models}\label{sec:opacity_dependence}

The previous section's comparison of the time evolution in different models has provided insights into the nature of different sources of discrepancies and at what opacities to expect them. For a detailed opacity-resolved analysis, it is convenient to study the dependence of final-state observables on a wide range of opacity, from the free-streaming regime to ideal fluid behaviour. In Sec.~\ref{sec:opacity:scaled}, we present opacity dependencies in kinetic theory, naive viscous hydrodynamics and scaled viscous hydrodynamics. Section~\ref{sec:opacity:hybrid} discusses results for hybrid simulation schemes.

\subsection{Scaled and naive hydrodynamics compared to kinetic theory}\label{sec:opacity:scaled}

\begin{figure}
    \centering
    \includegraphics[width=.45\textwidth]{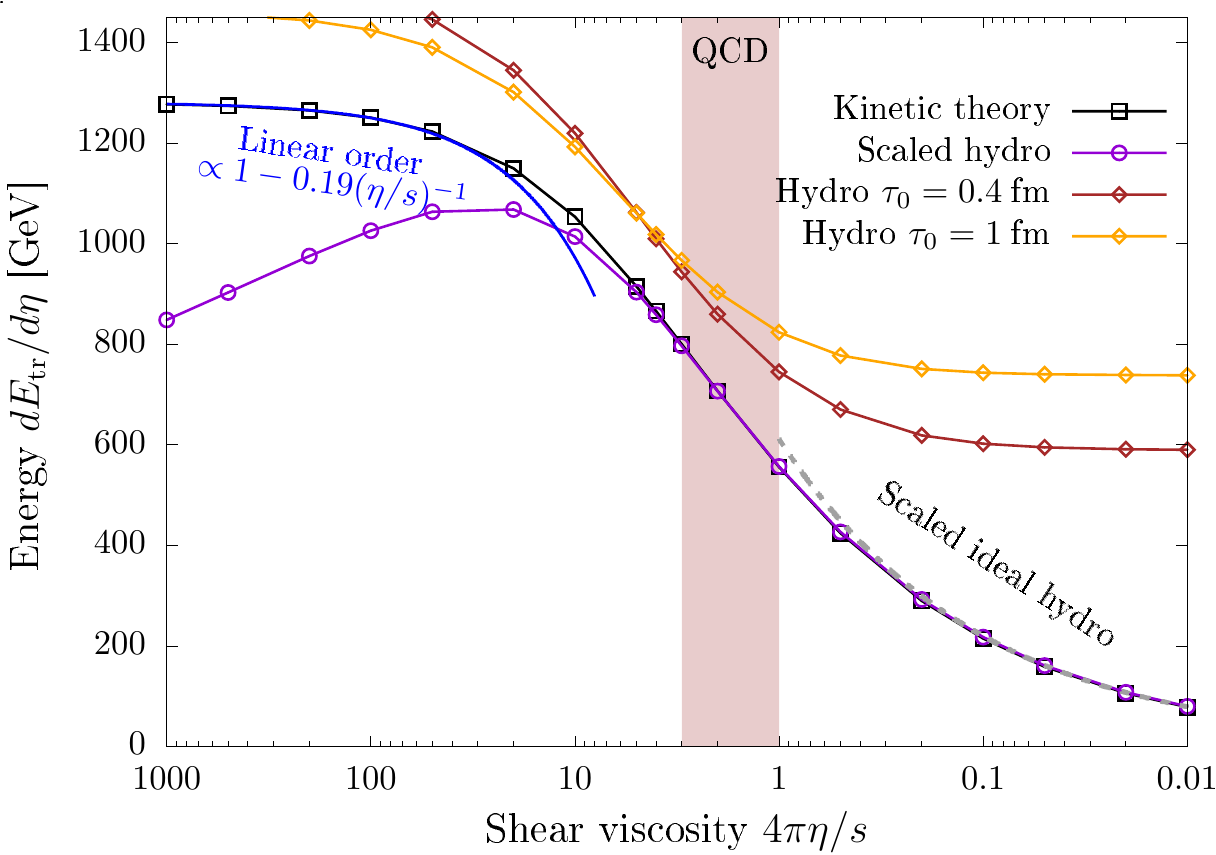}\\
    \includegraphics[width=.45\textwidth]{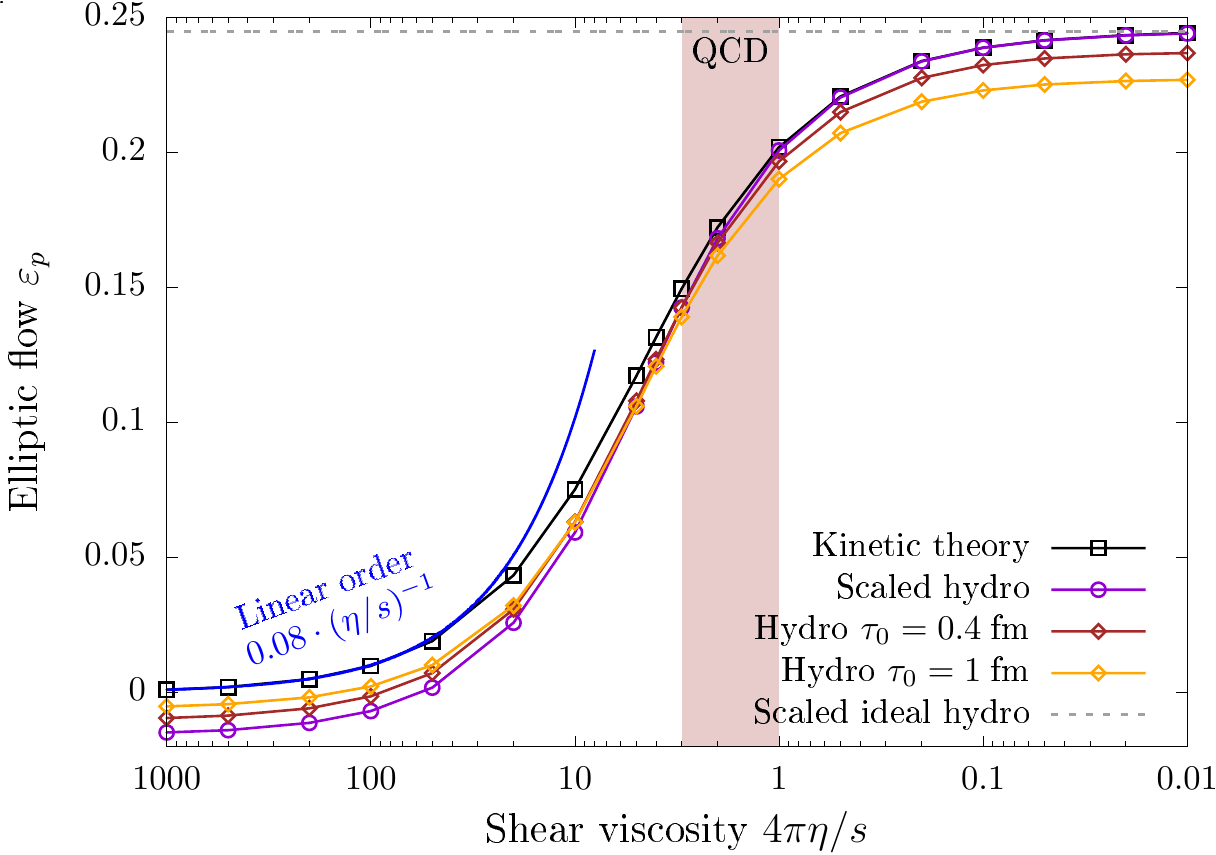}\\
    \includegraphics[width=.45\textwidth]{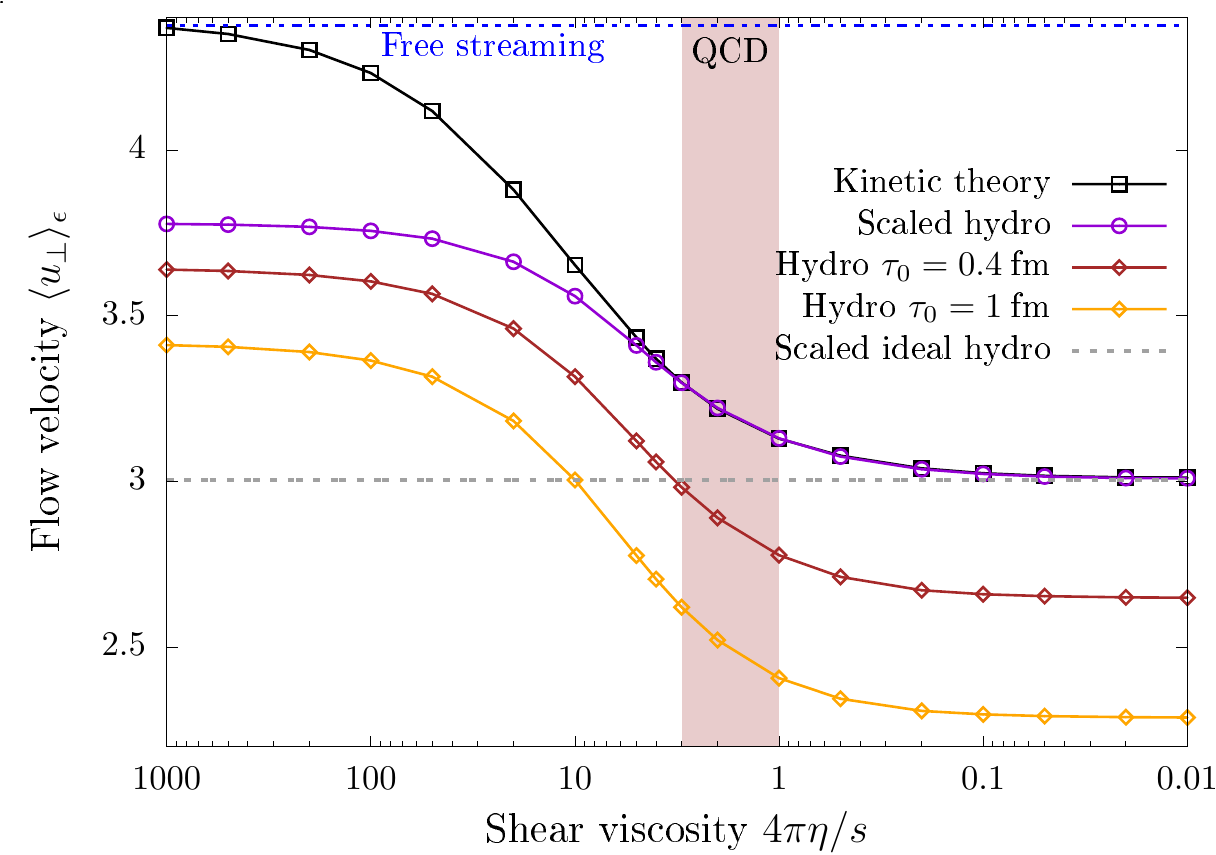}
    \caption{
    Opacity- ($\eta/s$-) dependence of the final ($\tau=4R$) values of transverse energy $\d E_{\rm tr}/\d\eta$ [top, cf. Eq.~\eqref{eq:obs_dEtrdeta}], elliptic flow $\varepsilon_p$ [middle, cf. Eq.~\eqref{eq:obs_epsp}] and transverse flow velocity $\langle u_\perp\rangle_\epsilon$ [bottom, cf. Eq.~\eqref{eq:obs_uT}] for kinetic theory (black), scaled hydro (purple) and naive hydro at two different initialization times $\tau_0=0.4\,\mathrm{fm}$ (brown) and $1\,\mathrm{fm}$ (yellow). Also plotted are the small opacity limits of an opacity-linearized result (blue) in the top two plots, the free-streaming result (blue, dashed) in the bottom plot as well as the 
    opacity-scaled ideal hydrodynamics results (grey, dashed). The latter follows a $(\eta/s)^{4/9}$ scaling law for $\d E_{\rm tr}/d\eta$ as per the initialization scheme in Eq.~\eqref{eq:scaling_id}. The ideal hydro results are $611\;\mathrm{GeV} \cdot(4\pi\eta/s)^{4/9}$ for $\d E_{\rm tr}/d\eta$, $0.244$ for $\varepsilon_p$ and $3.01$ for $\eavg{u_\perp}$.
    The red shaded region shows the realistic values for QCD according to Bayesian estimates.}
    \label{fig:master_naive}
\end{figure}

First, we assess the performance of scaled hydrodynamics as described in Sec.~\ref{sec:initialization_scaled_hydro} when compared to a common naive initialization scheme of hydrodynamics, where the simulation is started at a time $\tau_0$ where hydrodynamization is likely to have set in, with the same initial condition for $\tau_0 \epsilon(\tau_0,\xT)$ as we are using for kinetic theory simulations initialized in the early time free-streaming limit. Fig~\ref{fig:master_naive} shows the opacity dependences of transverse energy $\d E_{\rm tr}/ \d\eta$, elliptic flow $\varepsilon_p$ and average transverse flow velocity $\langle u_\perp \rangle_\epsilon$ in kinetic theory, scaled hydrodynamics and naive hydrodynamics initialized on the hydrodynamic attractor at two different times $\tau_0=0.4\,$fm and $\tau_0=1\,$fm, which are in the range of values typically used in phenomenological descriptions.

For all three observables, the kinetic theory results smoothly interpolate between limiting cases of small and large opacities. For $\d E_{\rm tr}/ \d\eta$ and $\varepsilon_p$, we compare at small opacities to results from the linear order opacity expansion that is introduced in Sec.~\ref{sec:opacity_expansion}. Results from full kinetic theory are in excellent agreement with these approximations for $4\pi\eta/s\gtrsim 20$. In the case of $\langle u_\perp \rangle_\epsilon$, we present results for the free-streaming limit $\hat{\gamma}\to 0$, to 
which
the full kinetic theory results converge at small opacities.

On the other end of the opacity spectrum, the results from both kinetic theory and scaled viscous hydrodynamics converge to those of scaled ideal hydrodynamics in the limit $\eta / s \rightarrow 0$. Even though $\eta / s = 0$ by definition in ideal hydrodynamics, we represent the scaled ideal hydro results as a function of $4\pi \eta / s$ in the equivalent RTA simulation [see discussion around Eq.~\eqref{eq:scaling_id}], leading to a power-law dependence $\d E_{\rm tr} / \d\eta \propto (4\pi \eta / s)^{4/9}$, which is confirmed by the scaled viscous hydrodynamics and kinetic theory results (this result was derived from early-time Bjorken scaling in \cite{Ambrus:2021fej}).
The curves for $\varepsilon_p$ and $\langle u_\perp \rangle_\epsilon$ converge at large opacities to the ideal hydrodynamics limit that was obtained with a scaled initial condition. This is not a priori obvious but rather an achievement of our proposed scheme. Ideal hydrodynamics is the large opacity limit of kinetic theory only after hydrodynamization. At any finite opacity, kinetic theory simulations feature a pre-equilibrium period which is absent in ideal hydro. In this period, the ellipticity $\epsilon_2$ decreases in kinetic theory, such that with the same initial condition, it would result in a smaller elliptic flow $\varepsilon_p$ than in ideal hydro. The agreement is only reached after scaling the hydro initial condition as discussed in 
Sec.~\ref{sec:initialization_scaled_hydro}.

Comparing now to hydrodynamic results, for all three obserables, the large opacity limits of scaled hydrodynamics and kinetic theory are in excellent agreement. Going to small opacities, all observables are underestimated in hydro, 
as
will be further discussed in the following. Agreement holds for 
$4\pi\eta/s\lesssim 3$.

On the other hand, for naive hydrodynamics
initialized at $\tau_0 = 0.4\ {\rm fm}$ and $1.0\ {\rm fm}$, the opacity dependence curves 
show qualitatively similar behaviour to kinetic theory
but remain in quantitative disagreement for all opacities.
This is obvious in the case of $\d E_{\rm tr}/ \d\eta$, which is significantly overestimated. We find that the large opacity power law is not captured. There are different reasons for this in the two limiting cases of large and small opacity. For small opacities $4\pi\eta/s\gtrsim 10$, despite the initialization time being large, it is still smaller than the equilibration timescale and the simulation will partly undergo a pre-equilibrium phase. As we have seen, in  the hydrodynamic description of this phase $\d E_{\rm tr}/ \d\eta$ increases before the onset of transverse expansion, while staying constant in kinetic theory, so it is overestimated in hydro. For the smaller initialization time $\tau_0=0.4\,$fm, the system is in pre-equilibrium for a longer time compared to $\tau_0=1\,$fm. This is why results for $\tau_0=0.4\,$fm yield a larger final value of $\d E_{\rm tr}/ \d\eta$ at small opacities. On the other hand, for large opacities $4\pi\eta/s \lesssim 3$, the system would have been in equilibrium for a significant amount of time if it had been initialized at an earlier time. In the equilibrated phase before transverse expansion, $\d E_{\rm tr}/ \d\eta$ drops proportionally to $\tau^{-1/3}$. The larger the initialization time, the more of this period is cut out of the simulation, resulting in a larger final value. This is why the curve for initialization at $\tau_0=1\,$fm is above the one for $\tau_0=0.4\,$fm at large opacities, resulting in a crossing of the two at intermediate opacities $4\pi\eta/s\sim 5$. The equilibration timescale becomes smaller and smaller at larger and larger opacities, meaning that for fixed initialization time more and more of the $\tau^{-1/3}$-scaling period is cut out. This is why the large opacity power law is not reproduced in naive hydrodynamics. 

These problems are cured in scaled hydrodynamics. It is initialized at very early times, so no part of the time evolution is lost. The discrepancies due to hydrodynamic pre-equilibrium behaviour are cured via scaling the initial energy density as discussed in 
Sec.~\ref{sec:initialization_scaled_hydro}
such that agreement with kinetic theory is reached only after equilibration. However, for small opacities $4\pi\eta/s \gtrsim 3$, the underlying assumption of a timescale separation of equilibration and transverse expansion no longer holds. In this case, scaled hydrodynamics underestimates $\d E_{\rm tr}/ \d\eta$, as transverse expansion interrupts its approach to kinetic theory behaviour before agreement is reached.

Of the three presented observables, $\varepsilon_p$ in naive hydrodynamics shows the weakest deviations from kinetic theory 
. 
This is in alignment with our expectations, as we know that hydro has been very successful in phenomenological descriptions of anisotropic flow. The reasons might be that $\varepsilon_p$ builds up on timescales that are fully captured by simulations at initialization times of $\sim 1\;$fm and depends very little on the overall energy scale. But certainly, this level of agreement was not to be expected a priori and should be regarded as a coincidence. The influence of the initialization time is as follows. At small opacities $4\pi\eta/s\gtrsim 10$, a part of the early time negative trend in hydrodynamics is cut out, resulting in larger results for later initialization times. For large opacities $4\pi\eta/s \lesssim 1$, $\varepsilon_p$ already has positive contributions at early times which might be cut out, resulting in smaller final values for later initialization times. But very early initialization times cannot bring hydro into agreement with kinetic theory. As discussed in Sec.~\ref{sec:early:pre-equilibrium}, hydrodynamics initialized at very early times exhibits a larger decrease of the eccentricity during pre-equilibrium, resulting in lower final values of $\varepsilon_p$ than in kinetic theory. However, the scaling procedure counteracts this phenomenon by increasing the eccentricity in the initial state of hydrodynamic simulations, such that scaled hydrodynamics is in perfect agreement with kinetic theory at large opacities $4\pi\eta/s \lesssim 3$. For small opacities $4\pi\eta/s \gtrsim 10$, on the other hand, due to the early initialization scaled hydrodynamics features a longer period of the aforementioned early time negative buildup of $\varepsilon_p$, resulting in 
final values which are lower than in the case of the naive hydro initializations discussed above.

The flow velocity results from naive hydrodynamics again show two effects. One of them is straightforward: as this observable rises monotonically with time, for larger initialization times, there is less time for $\langle u_\perp \rangle_\epsilon$ to build up, resulting in an underestimate. This effect is cured in scaled hydrodynamics due to the early initialization. At small opacities $4\pi\eta/s\gtrsim 10$, we see an additional decrease of hydrodynamic results compared to kinetic theory due to its inability to describe the late-time free-streaming behaviour. 
This is an effect that both hydro schemes (based on naive and scaled initial conditions) have in common.

\subsection{Hybrid simulations}\label{sec:opacity:hybrid}

As described in Sec.~\ref{sec:time_evolution_hybrid}, another way to bring hydrodynamic results into agreement with kinetic theory is to use hybrid schemes switching from a  kinetic theory based early time description to hydrodynamics at later times. We tried switching both at fixed times as well as at the first times equilibration has proceeded to a given extent, which we quantified by the inverse Reynolds number dropping to a specific value. We also tested two different model descriptions for early times: full kinetic theory and K{\o}MP{\o}ST. As described in the previous section, the time evolution curves of all observables instantaneously change behaviour when the models are switched, such that switching too early will 
be affected by
the inaccurate description of pre-equilibrium in hydrodynamics. We now want to quantitatively assess the accuracy of various switching schemes as a function of opacity.

\begin{figure*}
    \centering
    \includegraphics[width=.385\textwidth]{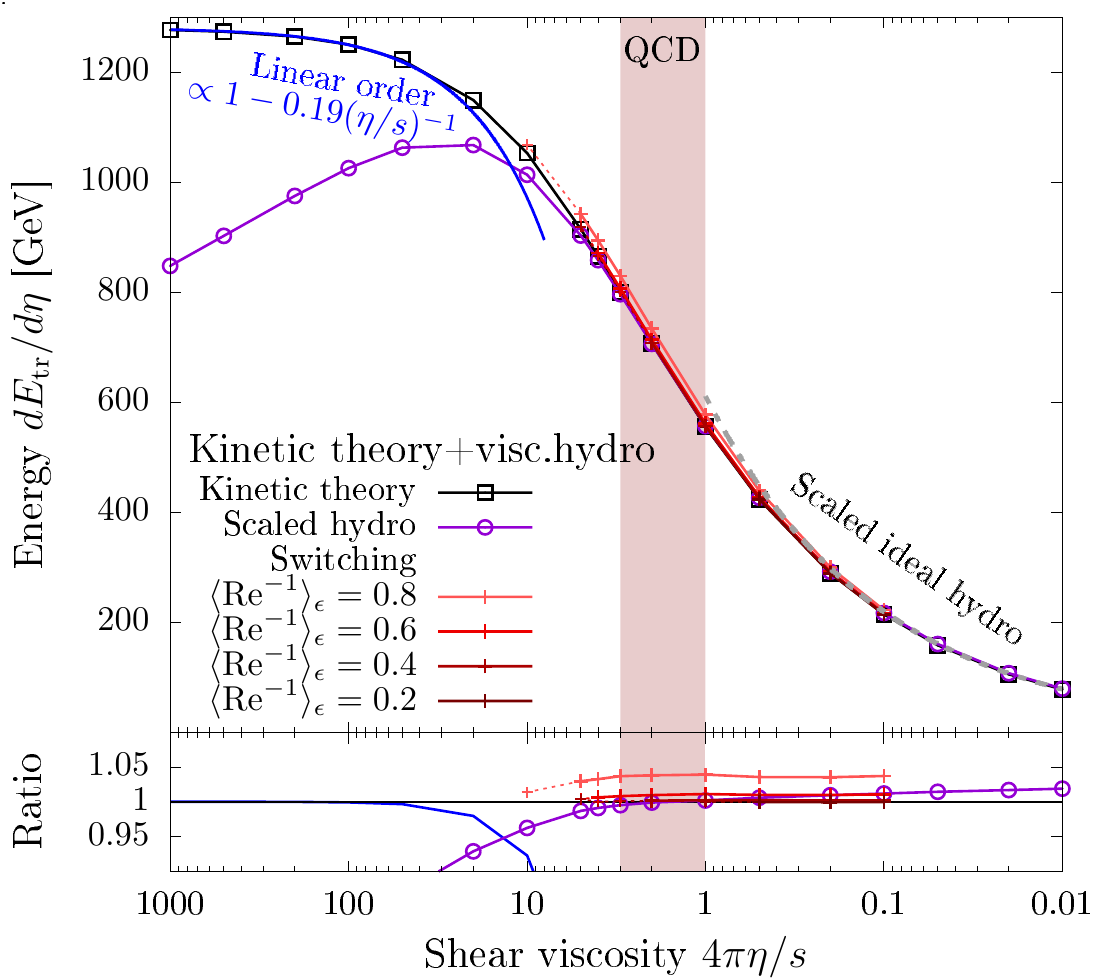}
    \includegraphics[width=.385\textwidth]{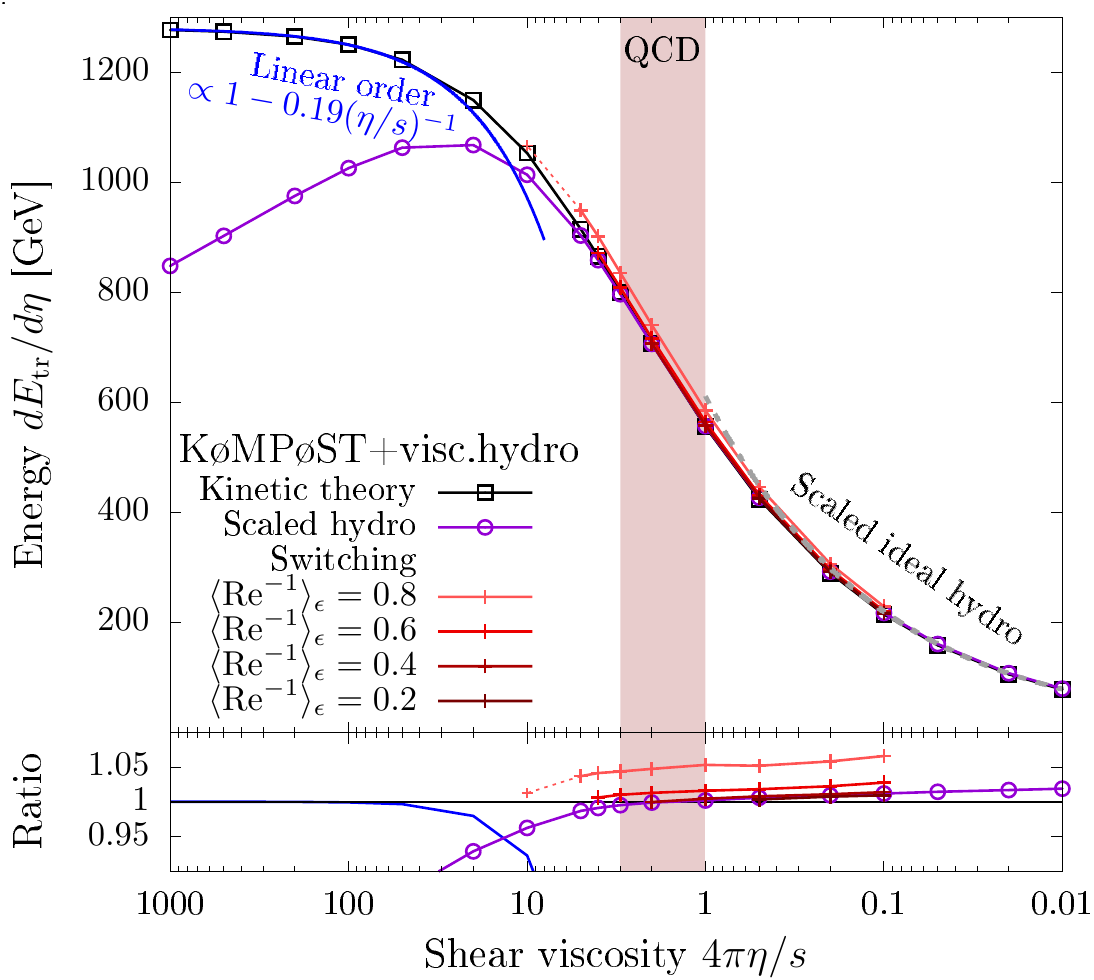}\\
    \includegraphics[width=.385\textwidth]{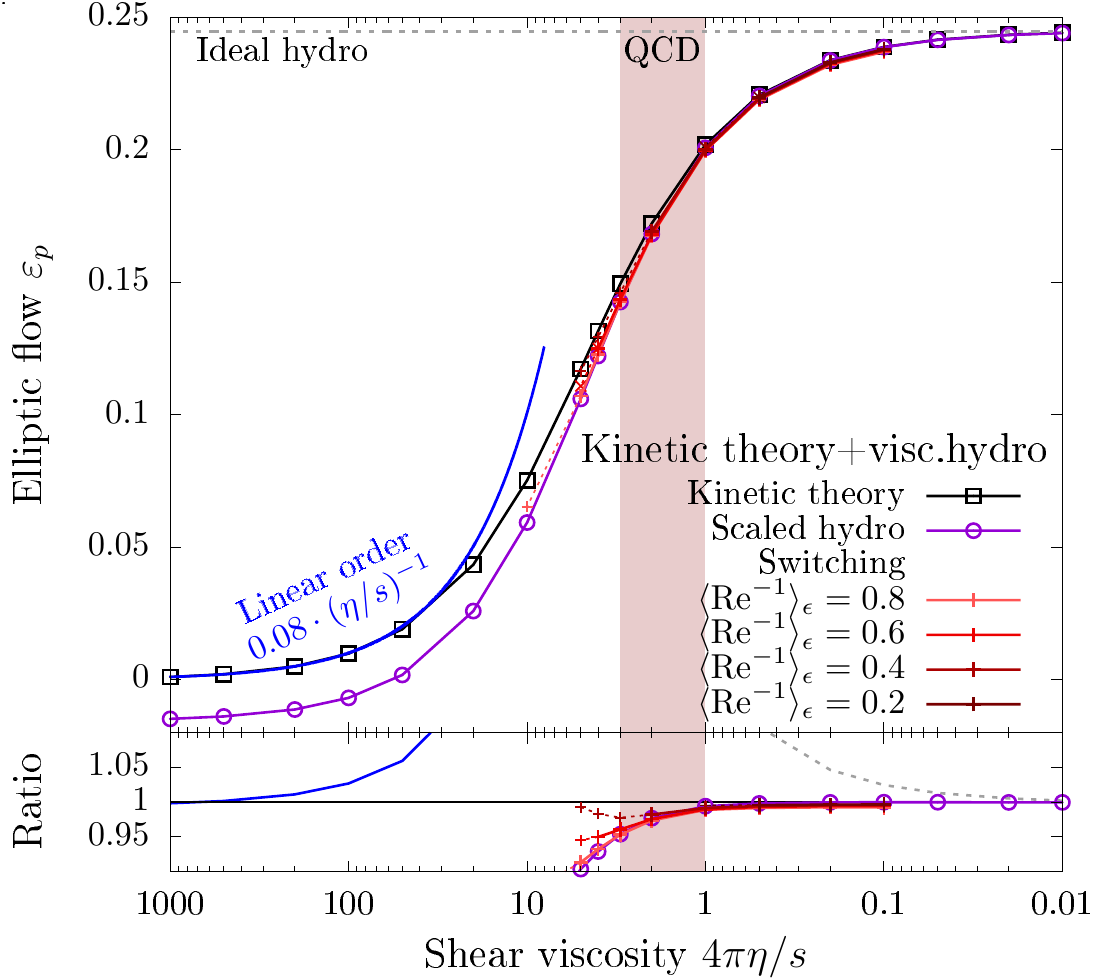}
    \includegraphics[width=.385\textwidth]{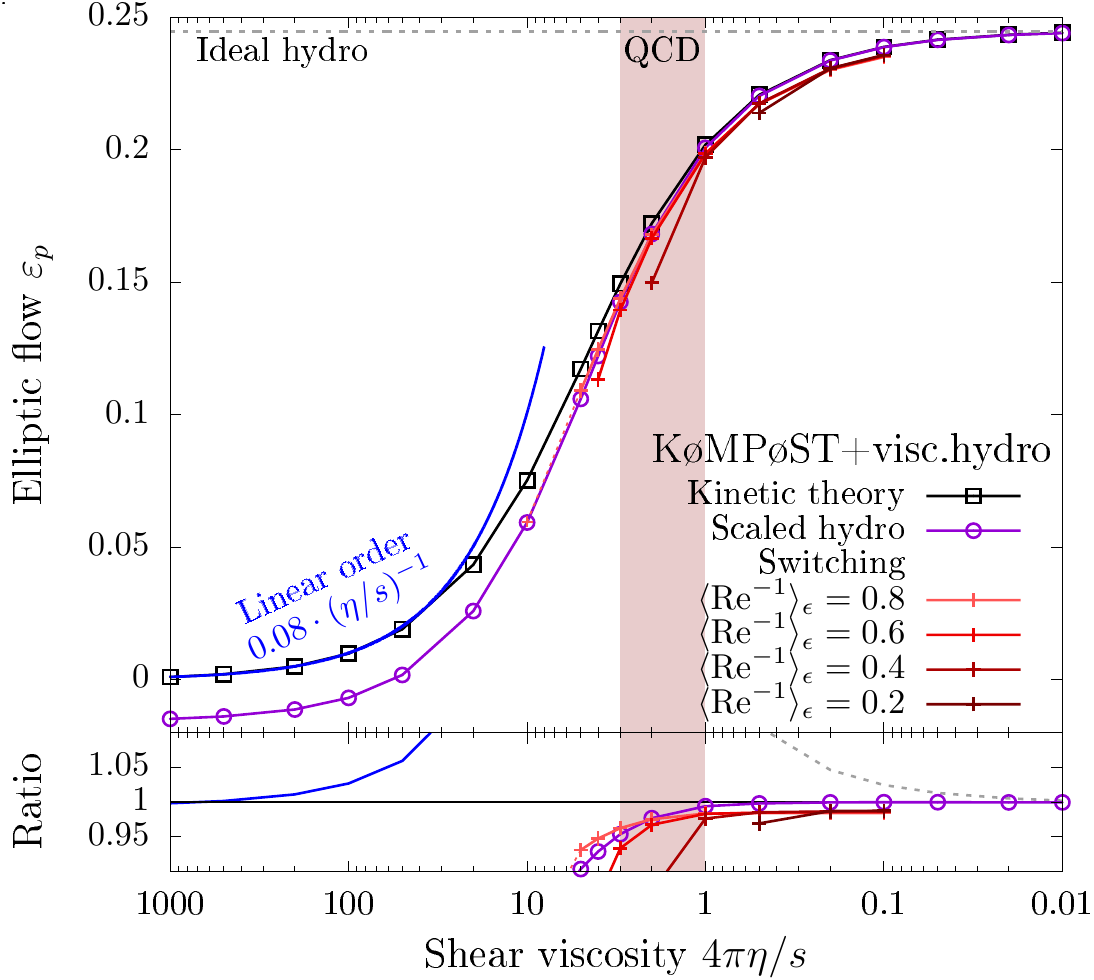}\\
    \includegraphics[width=.385\textwidth]{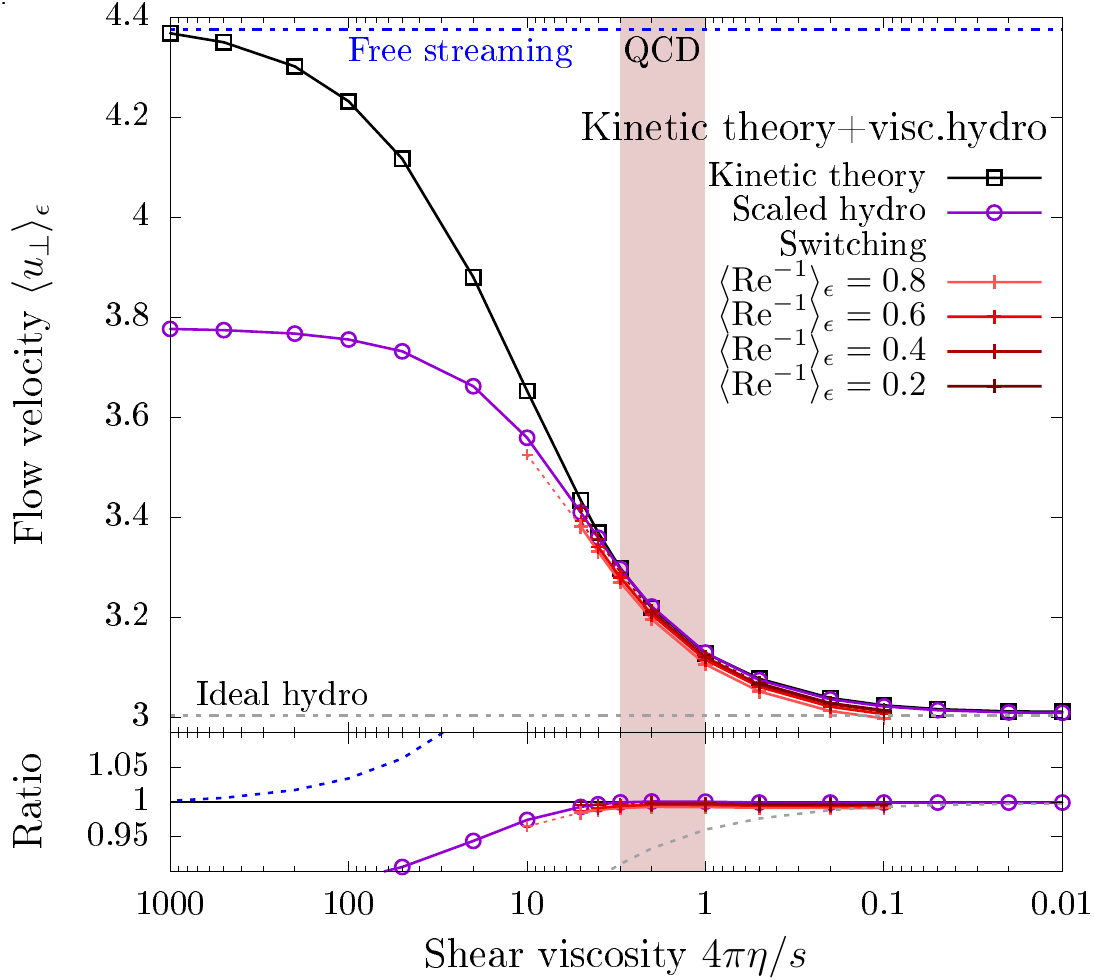}
    \includegraphics[width=.385\textwidth]{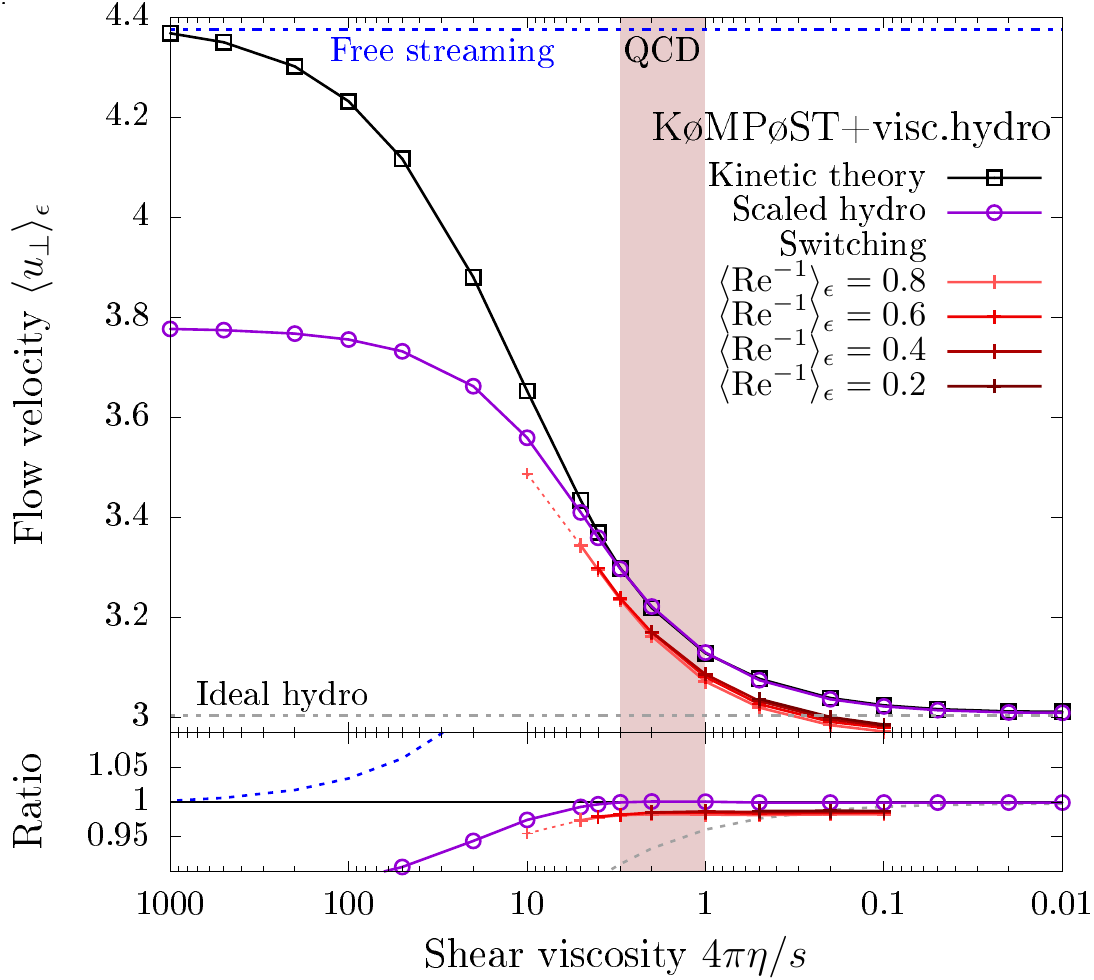}
    \caption{Opacity- ($\eta/s$-) dependence of the final ($\tau=4R$) values of transverse energy $\d E_{\rm tr}/\d\eta$ [top, cf. Eq.~\eqref{eq:obs_dEtrdeta}], elliptic flow $\varepsilon_p$ [middle, cf. Eq.~\eqref{eq:obs_epsp}] and transverse flow velocity $\langle u_\perp\rangle_\epsilon$ 
   [bottom, cf. Eq.~\eqref{eq:obs_uT}] in hybrid kinetic theory + viscous hydro (left) and K{\o}MP{\o}ST + viscous hydro simulations (right) when switching at different values of the inverse Reynolds number [cf. Eq.~\eqref{eq:Reinv_def}] $\eavg{\mathrm{Re}^{-1}}=0.8$, $0.6$, $0.4$ and $0.2$ plotted in different shades of red from light to dark. Results from simulations with switching times after $\tau=0.5R$ are plotted with smaller points ($+$) and dashed lines. 
   The results are compared to kinetic theory (black), scaled hydro (purple) and the small opacity limits of an opacity-linearized result (blue) in the top two plots, the free-streaming result (blue, dashed) in the bottom plot as well as the large opacity limit of scaled ideal hydro (grey, dashed), which scales as $(\eta/s)^{4/9}$ in the top plot. 
   The red shaded region shows the realistic values for QCD according to Bayesian estimates. The bottom part of 
    each
    plot shows the ratios of all results to those from kinetic theory.}
    \label{fig:master_hybrid_Reinv}
\end{figure*}

We first discuss results for the opacity dependence in hybrid simulations with $\langle \mathrm{Re}^{-1} \rangle_\epsilon$-based switching, which are plotted in Figure~\ref{fig:master_hybrid_Reinv}. For early switching times on the timescale of equilibration, hybrid results may reflect the inaccurate pre-equilibrium behaviour in hydrodynamics. Of course, in this case, there is no scaling of the initial condition to counteract this behaviour. However, this also means that these schemes do not suffer from discrepancies due to an incomplete approach of a scaled initial condition to kinetic theory behaviour before the onset of transverse expansion, and therefore tend to be more accurate than scaled hydrodynamics at intermediate opacities, i.e. for $4\pi\eta/s\sim 3$. However, results plotted with smaller crosses and dashed lines were obtained in simulations with switching times larger than $0.5R$, so in this case it is questionable whether these schemes could be considered hybrid results, as 
the crucial parts of the time evolution were actually described in kinetic theory.

Going into more detail, hybrid results typically overestimate $\d E_{\rm tr}/ \d\eta$ because of the hydrodynamic pre-equilibrium increase after switching. $\varepsilon_p$ is underestimated, however, the hydrodynamic negative early time trend is alleviated, such that results from kinetic theory + viscous hydro are typically larger than scaled hydro results. Hybrid results show a consistent underestimation of $\langle u_\perp \rangle_\epsilon$, but on a relative scale this error is negligible. This could be due to hydrodynamic flow velocities typically being smaller than those in kinetic theory at early times, causing a dip in $\langle u_\perp \rangle_\epsilon$ relative to kinetic theory after switching.

Comparing kinetic theory + viscous hydrodynamics in the left column of the figure to K{\o}MP{\o}ST + viscous hydrodynamics in the right column, one obvious difference is that in the latter, some of the results for smaller opacities are missing, because there the $\langle \mathrm{Re}^{-1} \rangle_\epsilon$-based switching times were too late to be reached with K{\o}MP{\o}ST.\footnote{For large evolution times, K{\o}MP{\o}ST crashes in the setup stage when computing the Green's functions. This is because they are only implemented for a finite number of points in momentum space  and have to be convolved with a Gaussian smearing kernel $\exp(-\sigma^2|\mathbf{k}|/2)$. But the Green's functions scale in $|\mathbf{k}|(\tau-\tau_0)$ such that for too large of an evolution time this smearing is no longer sufficient.}
Where it does work, it produces almost the same results for $\d E_{\rm tr}/ \d\eta$ as kinetic theory. The underestimation of $\langle u_\perp \rangle_\epsilon$ is slightly more severe in K{\o}MP{\o}ST. It does seem to have a systematic component on top of the one related to switching early. But the total deviation is still negligible. The largest difference is seen in $\varepsilon_p$, which is not built up at all in K{\o}MP{\o}ST simulations, thus there is a significantly larger underestimation at smaller opacities, where a larger part of the time evolution is described in K{\o}MP{\o}ST.

Next, we shift our attention to results from hybrid schemes at fixed switching times $\tau_s=0.4\,$fm and $\tau_s=1\,$fm, which are presented in Figure~\ref{fig:master_hybrid_fix}. As expected from the discussion of the time evolution in Sec.~\ref{sec:time_evolution_hybrid}, again kinetic theory + viscous hydrodynamics yields perfectly accurate results at large opacities $4\pi\eta/s\lesssim 1$ and improves on scaled hydrodynamics at intermediate opacities $4\pi\eta/s\sim 3$, but less so than for dynamically chosen switching times. The upshot is that hybrid schemes with fixed switching times are applicable for arbitrarily small opacities. However, here the results for the three tracked observables show similar problems to those obtained in naive hydrodynamics simulations discussed earlier in this section. Due to incomplete equilibration at early switching times, $\d E_{\rm tr}/ \d\eta$ increases after switching. $\varepsilon_p$ suffers from the early time negative trend in hydrodynamics, but slightly less than scaled hydrodynamics. $\langle u_\perp \rangle_\epsilon$ is again only slightly underestimated in hybrid schemes when compared to scaled hydrodynamics due to the different pre-equilibrium. This is an improvement over
naive hydrodynamics, as instead of starting at late times with no flow velocity, the early time buildup is described in kinetic theory.  Both schemes suffer equally from the inability of hydrodynamics to describe flow velocities in the late time free-streaming limit.

Also for fixed switching times, K{\o}MP{\o}ST + viscous hydrodynamics results for $\d E_{\rm tr}/ \d\eta$ and $\langle u_\perp \rangle_\epsilon$ are in good agreement with those obtained in kinetic theory + viscous hydrodynamics simulations. We again see the effect of the absence of $\varepsilon_p$-buildup in K{\o}MP{\o}ST. Since we do not increase the duration of time evolution in K{\o}MP{\o}ST, the effect is not larger at small opacities $4\pi\eta/s \gtrsim 10$. In fact, here we see agreement with results from kinetic theory + viscous hydrodynamics, as there is no significant buildup of $\varepsilon_p$ at early times. However, at large opacities $4\pi\eta/s\lesssim 5$, this buildup starts earlier, which is why K{\o}MP{\o}ST + viscous hydrodynamics results underestimate the final values in these cases.

\begin{figure}
    \centering
    \includegraphics[width=.45\textwidth]{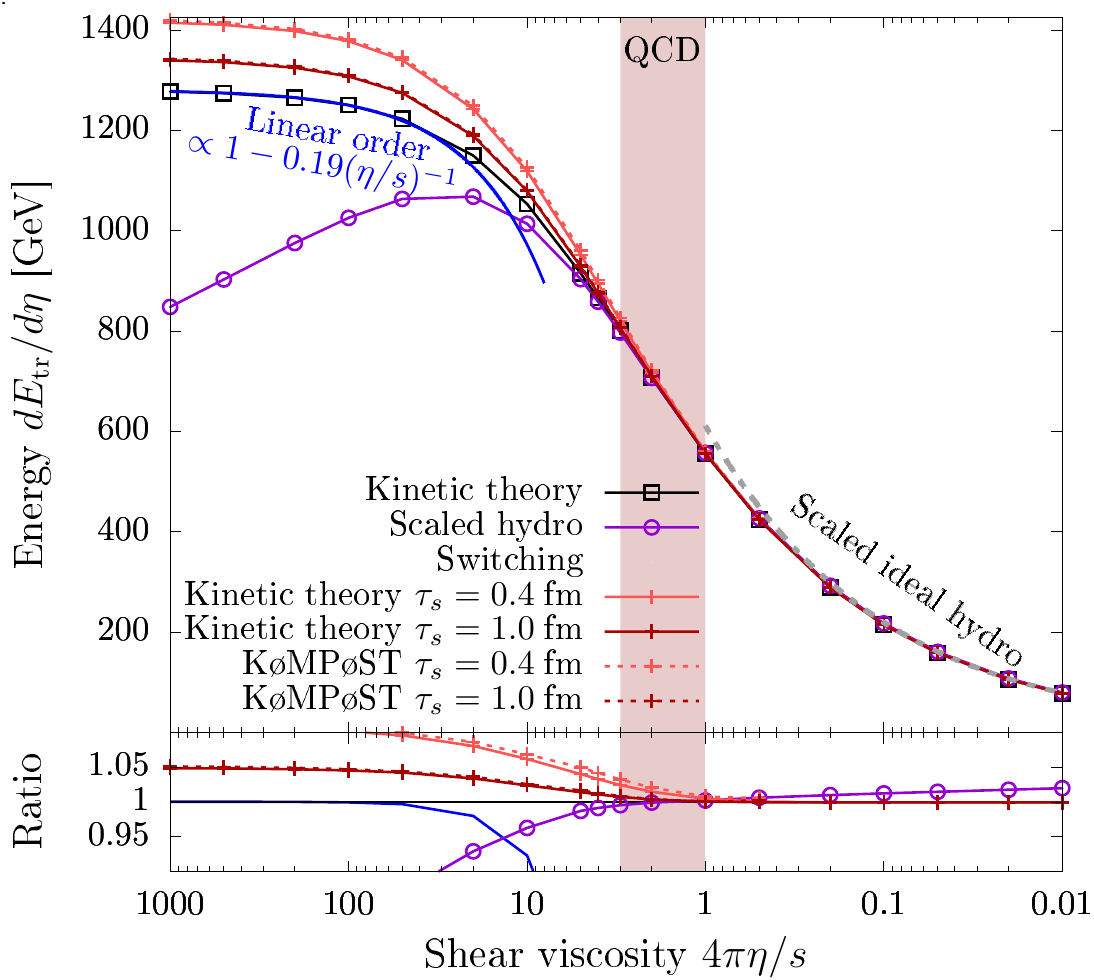}\\
    \includegraphics[width=.45\textwidth]{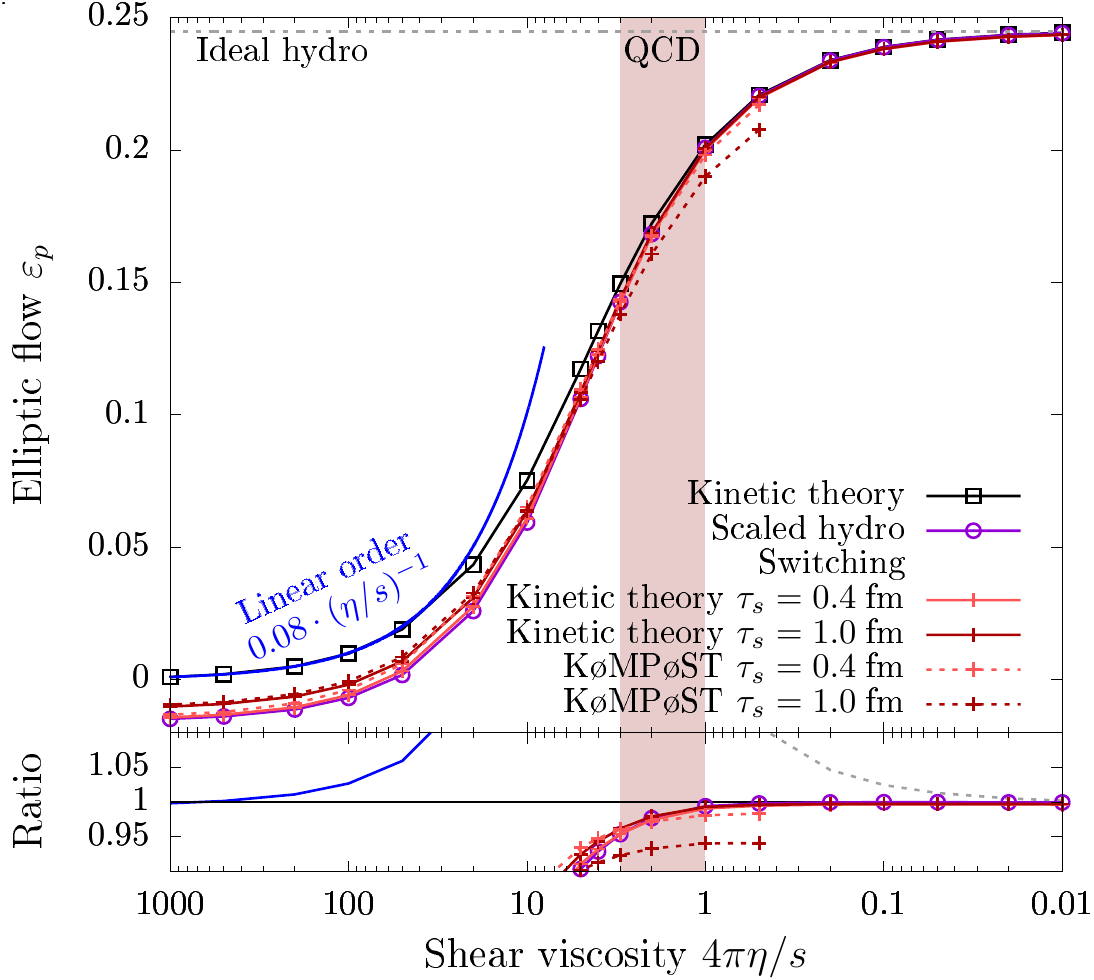}\\
    \includegraphics[width=.45\textwidth]{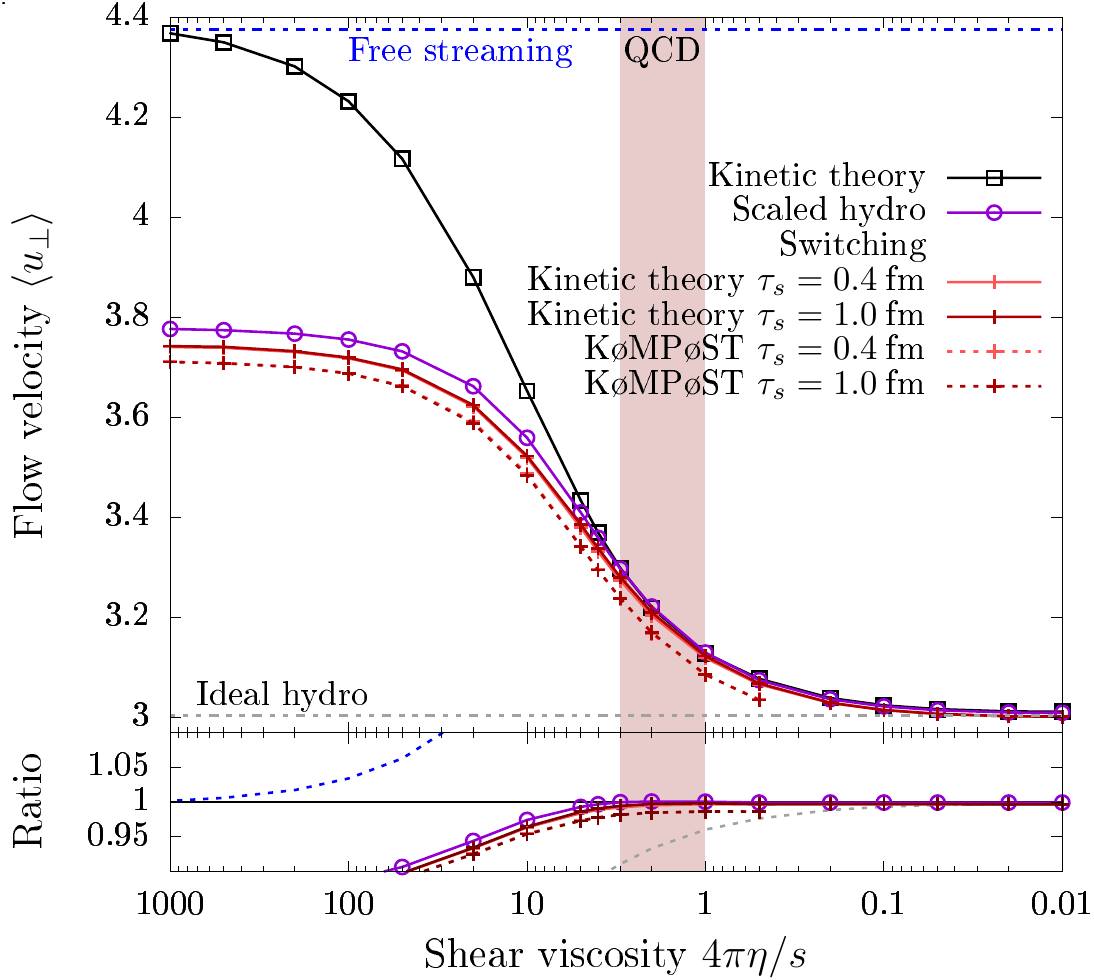}
\end{figure}
\begin{figure}
    \centering
    \caption{Opacity- ($\eta/s$-) dependence of the final ($\tau=4R$) values of transverse energy $\d E_{\rm tr}/\d\eta$ [top, cf. Eq.~\eqref{eq:obs_dEtrdeta}], elliptic flow $\varepsilon_p$ [middle, cf. Eq.~\eqref{eq:obs_epsp}] and transverse flow velocity $\langle u_\perp\rangle$ [bottom, cf. Eq.~\eqref{eq:obs_uT}] in hybrid kinetic theory + viscous hydro (solid lines) and K{\o}MP{\o}ST + viscous hydro simulations (dashed lines) when switching at fixed times $\tau=0.4\,\mathrm{fm}$ (light red) and $\tau=1\,\mathrm{fm}$ (dark red). The results are compared to kinetic theory (black), scaled hydro (purple) and the small opacity limits of an opacity-linearized result (blue) in the top two plots, the free-streaming result (blue, dashed) in the bottom plot, as well as to the large opacity limit of scaled ideal hydro (grey, dashed), which scales as $(\eta/s)^{4/9}$ in the top plot. The red shaded region shows the realistic values for QCD according to Bayesian estimates. The bottom part of the plot shows the ratios of all results to those from kinetic theory.}
    \label{fig:master_hybrid_fix}
\end{figure}

\section{Conclusions}\label{sec:conclusion}

In this work, we examined hydrodynamic and kinetic theory simulations of hadronic collisions. Within a simplified model setup based on RTA and using a fixed initial profile that was obtained as an average of events in the $30-40\%$ centrality class of Pb-Pb collisions, we scanned the dynamical behaviour on the whole range in interaction rates as parametrized by the opacity $\hat{\gamma}$ defined in Eq.~\eqref{eq:ghat}, which for our fixed profile is inversely proportional to shear viscosity, $\hat{\gamma}=11.3/(4\pi\eta/s)$. This study was based on results for the transverse energy $\d E_{\rm tr} / \d\eta$,
elliptic flow $\varepsilon_p$, radial flow $\eavg{u_\perp}$ and shear stress as measured via the inverse Reynolds number $\eavg{\mathrm{Re}^{-1}}$. At small opacities $4\pi\eta/s\gtrsim 20$, kinetic theory agrees with results from a linearization in opacity. Here, the system is too dilute for hydrodynamics to be applicable, which was confirmed quantitatively in Sec.~\ref{sec:opacity:scaled}: the time evolution of transverse energy, radial flow and shear stress is in significant disagreement in hydrodynamic simulations compared to kinetic theory. For large opacities
$4\pi\eta/s\lesssim 0.1$, 
in the limit of high interaction rates, kinetic theory is expected to converge to hydrodynamics. Our results confirm that the two descriptions are in agreement after pre-equilibrium. Going down to intermediate opacities, we found that for suitable setups of hydrodynamics, results for final state transverse energy, elliptic flow and radial flow are in good agreement with kinetic theory up to shear viscosities $4\pi\eta/s\lesssim 3$ for the examined profile, which translates to opacity values $\hat{\gamma}\gtrsim 4$.

However, hydrodynamics is not suitable for describing out-of-equilibrium behaviour in the early pre-equilibrium stage and the late time period where the microscopic description of kinetic theory approaches a free-streaming behaviour. In both of these regimes, hydrodynamic results are in quantitative disagreement with kinetic theory, which can be seen at 
the level of final state observables, as discussed in Sec.~\ref{sec:early:pre-equilibrium}. Omitting the pre-equilibrium period or naively employing hydrodynamics to describe it will yield inaccurate results. On the other hand, at late times where interactions die out, these observables no longer build up and approach constant values, such that hydrodynamic descriptions yield similar results to kinetic theory. However, the late time free-streaming stage does have an effect on radial flow, which is underestimated in hydrodynamics.

We examined two different modified setups of hydrodynamic simulations that can alleviate problems with the pre-equilibrium evolution. The first setup follows the idea of an early initialization of hydrodynamics with a scaled initial condition relative to kinetic theory to counteract the differences in 
the pre-equilibrium evolution. These differences are
predicted locally based on insights from Bjorken flow, which is accurate in systems with a timescale separation of equilibration and the onset of 
 transverse flow. 
By construction, this setup yields accurate results at large opacities $4\pi\eta/s\lesssim 3$, but fails at smaller opacities, where equilibration takes longer and is interrupted by transverse expansion. The results obtained in this setup are presented in Sec.~\ref{sec:time_evolution_kinetic_theory_scaled_hydro} and~\ref{sec:opacity:scaled}.

The second setup is a hybrid simulation switching from kinetic theory based descriptions at early times to hydrodynamics for later times. In these schemes, as described in Sec.~\ref{sec:time_evolution_hybrid}, we saw an immediate change of the time evolution behaviour at the moment that we switch the dynamical descriptions. Thus, the accuracy of hybrid simulations depends on the extent to which the 
kinetic theory and hydrodynamic descriptions of the system's time evolution have come into agreement by the time 
of the switch. 
This approach to agreement between 
the two descriptions is what we call hydrodynamization. We found that this criterion can in practice be quantified via the inverse Reynolds number. Figure~\ref{fig:attractor_curves} shows that the system is partly hydrodynamized when $\mathrm{Re}^{-1}=0.8$, significantly hydrodynamized when $\mathrm{Re}^{-1}=0.6$ and has reached almost perfect agreement of the two descriptions at $\mathrm{Re}^{-1}=0.4$. The accuracy of hybrid simulations when switching at a fixed value of $\eavg{\mathrm{Re}^{-1}}$ can be estimated beforehand and is almost independent of the opacity. As detailed in Sec.~\ref{sec:opacity:hybrid}, results from simulations with late switching times are accurate at high opacities $4\pi\eta/s \lesssim 1$ and can slightly improve on our first setup at intermediate opacities $4\pi\eta/s\sim 3$. At small opacities $4\pi\eta/s \gtrsim 20$, $\eavg{\mathrm{Re}^{-1}}$ does not drop below $0.8$, meaning the system does not equilibrate enough for hydrodynamics to become applicable at any point during the system's evolution.

For the early time kinetic theory description in hybrid models, we used both full kinetic theory and the compact K{\o}MP{\o}ST code. The latter uses a linearization scheme in perturbations around local homogeneity to propagate the energy-momentum tensor according to the 
Boltzmann equation under the relaxation time approximation (the original version \cite{Kurkela:2018wud,Kurkela:2018gitrep} is based on the QCD effective kinetic theory \cite{Arnold:2002zm}).
We first tested the performance of K{\o}MP{\o}ST as detailed in Sec.~\ref{sec:validation_of_kompost} and found that it can accurately reproduce full kinetic theory results for transverse energy, radial flow and isotropic shear stress, but due to the linearization it significantly underestimates elliptic flow.  It is by construction limited to times on the order of $0.5R$.  In hybrid simulations with switching times in this regime, transverse energy and radial flow results reach similar accuracy as when employing full kinetic theory. However, the underestimation of elliptic flow causes discrepancies when the switching time is non-negligible compared to the timescale of transverse expansion.  These shortcomings have already been reported in the original K{\o}MP{\o}ST paper~\cite{Kurkela:2018vqr} and will require further investigation in the future.

This work provides the baseline for analyses of hadronic collisions in frameworks based on the microscopic dynamics of kinetic theory. It is part of a series of recent efforts~\cite{Kurkela:2019kip,Kurkela:2020wwb,Roch:2020zdl,Kurkela:2021ctp, Ambrus:2021fej} to push the practical applicability of these dynamics in theoretical simulations. One important goal that has yet to be reached is to improve the codes that implement them in order to be able to also run event-by-event simulations. At the moment, the tool that is closest to fulfilling this goal is K{\o}MP{\o}ST, which we confirmed to function properly for its intended use, but it is confined to the pre-equilibrium phase of heavy-ion collisions. 

Broadly speaking, our results confirm that in principle hydrodynamics is the proper tool for describing mid-central collisions, if and only if pre-equilibrium is described correctly. Issues with this phase in hydrodynamic descriptions can be alleviated by changing the interpretation of the initial state in the way discussed in Sec.~\ref{sec:initialization_scaled_hydro}. As alluded to in Sec.~\ref{sec:early:attractor} as well as in previous works~\cite{Blaizot:2021cdv,Martinez:2010sc,Florkowski:2010cf}, appropriate changes to the evolution equations might achieve similar improvements. If such changes are not incorporated, we discussed in Sec.~\ref{sec:opacity_dependence} that hydrodynamic results can be in significant disagreement with kinetic theory. We also refer the interested reader to our companion paper~\cite{Ambrus:2022qya}, where we extract a more general criterion for the applicability of hydrodynamics and infer phenomenological conclusions for the description of the space-time dynamics of high-energy collisions.

\begin{acknowledgments}
We thank P.~Aasha, N.~Borghini, H.~Elfner, A.~Mazeliauskas,  H.~Roch, A.~Shark, and U.~A.~Wiedemann for valuable discussions. This work is supported by the Deutsche Forschungsgemeinschaft (DFG, German Research Foundation)
through the CRC-TR 211 ’Strong-interaction matter under extreme conditions’– project
number 315477589 – TRR 211. V.E.A.~gratefully acknowledges the support through a grant of the 
Ministry of Research, Innovation and Digitization, CNCS - UEFISCDI,
project number PN-III-P1-1.1-TE-2021-1707, within PNCDI III. C.W. was supported by the program Excellence Initiative–Research University of the University of Wrocław of the Ministry of Education and Science.
Numerical calculations presented in this work were performed at the Paderborn Center for Parallel Computing (PC2) and the Center for Scientific Computing (CSC) at the Goethe-University of Frankfurt and we gratefully acknowledge their support. 
\end{acknowledgments}

\appendix

\section{Relativistic lattice Boltzmann implementation details}\label{app:RLB}

The first step in applying the relativistic lattice Boltzmann (RLB) method is the parametrization of the momentum space, which we perform using two sets of coordinates, namely the spherical (subscript s) and free-streaming (subscript fs) coordinates: 
\begin{subequations}
\begin{align}
    (p_{\rm s},v_{z;{\rm s}}) &= \left(p^\tau, \frac{\tau p^\eta}{p^\tau}\right),\\
    (p_{\rm fs}, v_{z;{\rm fs}}) &= \left(p_{\rm s} \Delta_{\rm s}, \frac{\tau v_{z;{\rm s}}}{\tau_0 \Delta_{\rm s}}\right),
\end{align}
\end{subequations} 
where $\Delta_{\rm s} = [1 + (\frac{\tau^2}{\tau_0^2} - 1) v_{z;{\rm s}}^2]^{1/2}$ \cite{Kurkela:2018qeb}. The azimuthal coordinate $\varphi_p = \arctan(p^y/p^x)$ is employed in both parametrizations.

Due to the particularly simple nature of RTA, the dynamics of the 
observables introduced in Sec.~II are fully described by the reduced distribution $\F_*$ ($* \in \{{\rm s}, {\rm fs}\}$), obtained from the phase-space distribution $f$ via
\begin{equation}
    \F_*=\frac{\nu_{\rm eff}~\pi R^2 \tau_0}{(2\pi)^3}\left(\epn\right)^{-1}
    \int_0^\infty \d p_* p_*^3f \;.
    \label{eq:Fdef}
\end{equation}
Using the non-dimensionalization conventions introduced around Eq.~\eqref{eq:boltz_nondim}, 
the non-dimensional function $\mathcal{F}_{\rm s} \equiv \mathcal{F}_{\rm s}(\tilde{\tau}, \tilde{\mathbf{x}}_T, \varphi_p, v_{z;{\rm s}})$ satisfies 
\begin{multline}
 \left(\tilde{\partial}_\tau + \sqrt{1 - v_{z;{\rm s}}^2} \vT \cdot \tilde{\nabla}_\perp + \frac{1 + v_{z;{\rm s}}^2}{\tilde{\tau}}\right)\F_{\rm s} \\
 - \frac{1}{\tilde{\tau}} \frac{\partial [v_{z;{\rm s}}(1 - v_{z;{\rm s}}^2) \mathcal{F}_{\rm s}]}{\partial v_{z;{\rm s}}} =-\hat{\gamma} v^\mu u_\mu\, \tilde{T}\left(\mathcal{F}_{\rm s} - \F_{\rm s}^{\rm eq}\right)\;,\label{eq:boltz_F}
\end{multline}
while $\mathcal{F}_{\rm fs} \equiv \mathcal{F}_{\rm fs}(\tilde{\tau}, \tilde{\mathbf{x}}_T, \varphi_p, v_{z;{\rm fs}})$ obeys
\begin{multline}
 \left(\tilde{\partial}_\tau + \frac{1}{\Delta_{\rm fs}} \sqrt{1 - v_{z;{\rm fs}}^2} \vT \cdot \tilde{\nabla}_\perp\right)\mathcal{F}_{\rm fs} \\
 =-\hat{\gamma} v^\mu u_\mu\, \tilde{T}\left(\mathcal{F}_{\rm fs} - \mathcal{F}_{\rm fs}^{\rm eq}\right)\;,\label{eq:boltz_Ffs}
\end{multline}
with $\Delta_{\rm fs} = [1 - (1 - \frac{\tilde{\tau}_0^2}{\tilde{\tau}^2}) v_{z;{\rm fs}}^2]^{1/2}$.
The equilibrium functions $\mathcal{F}^{\rm eq}_*$ are given by
\begin{align}
 \mathcal{F}^{\rm eq}_{\rm s} &= 
 \Delta_{\rm fs}^4 \mathcal{F}^{\rm eq}_{\rm fs} = 
 \frac{\tilde{\tau}_0 \tilde{\epsilon}}{4\pi (v^\mu u_\mu)^4},
\end{align}
where 
\begin{align}
 v^\mu u_\mu &= \gamma \left(1 - \sqrt{1 - v_{z;{\rm s}}^2} \vT \cdot \bm{\beta}\right) \nonumber\\
 &= \gamma\left(1 - \frac{1}{\Delta_{\rm fs}} \sqrt{1 - v_{z;{\rm fs}}^2} \vT \cdot \bm{\beta}\right),
\end{align}
with $\gamma = u^\tau \equiv 1/\sqrt{1 - \beta^2}$ being the local Lorentz factor. In the above,
$\bm{\beta} = \beta (\cos\varphi_u, \sin\varphi_u)$ and $\vT = (\cos\varphi_p, \sin\varphi_p)$ are transverse-plane vectors.

Vanishing longitudinal pressure and azimuthal momentum isotropy imply the following initial state for the reduced distributions $\F_*$:
\begin{equation}
 \F_*(\tilde{\tau}_0, {\txT}, \varphi_p, v_{z;*}) = 
 \frac{\delta(v_{z;*})}{2\pi} \tilde{\tau}_0 \tilde{\epsilon}_0(\xT)
 \label{eq:F0}
\end{equation}
and depends only on the initial transverse energy distribution $\d E^{0}_\perp / \d\eta \d^2\xT = \tau_0 \epsilon_0$ [see Eq.~\eqref{eq:init_eps0}]. 
Note that at $\tau = \tau_0$, $\Delta_{\rm s} = \Delta_{\rm fs} =1$ and $(p_{\rm fs}, v_{z;{\rm fs}}) = (p_{\rm s}, v_{z;{\rm s}})$.

Due to the singular nature of the Dirac delta function $\delta(v_z)$, Eq.~\eqref{eq:F0} cannot be achieved exactly with our numerical approach. We instead employ the Romatschke-Strickland distribution with anisotropy parameter $\xi_0$,
\begin{equation}
 f_{\rm RS} = \left[\exp\left(\frac{p^\tau}{\Lambda_0}\sqrt{1 + \xi_0 v_z^2}\right)-1\right]^{-1}\;,
\end{equation}
where $\xi_0 = 0$ corresponds to the isotropic Bose-Einstein distribution, while $\xi_0 \rightarrow \infty$ is required in order to achieve Eq.~\eqref{eq:F0}. The parameter $\Lambda_0 \equiv \Lambda_0(\xT)$ represents an energy scale satisfying 
\begin{equation}
 \Lambda_0 = 2^{1/4} T_0 \left(\frac{\arctan{\sqrt{\xi_0}}}{\sqrt{\xi_0}} + \frac{1}{1 + \xi_0}\right)^{-1/4},
\end{equation}
reducing to the initial temperature $T_0$ when $\xi_0 = 0$.
 Thus, the system is initialized according to
\begin{equation}
 \mathcal{F}_{0; {\rm s}} = 
 \mathcal{F}_{0; {\rm fs}} = \frac{\tilde{\tau}_0 \tilde{\epsilon}_0}{2\pi} (1 + \xi_0 v_z^2)^2
 \left(\frac{\arctan{\sqrt{\xi_0}}}{\sqrt{\xi_0}} + \frac{1}{1 + \xi_0}\right)^{-1}.
\end{equation}

\begin{table}[]
\centering
\begin{tabular}{r|llllllll}
 $4\pi \eta / s$ & $S$ & $(\frac{\delta \tilde{\tau}}{\tilde{\tau}})_{\rm M}$ & $\delta \tilde{\tau}_{\rm M}$ & $Q_\varphi$ & $Q_z(*)$ & $\tilde{\tau}_0$ &
 $\xi_0$ & $P_L/P_T$ \\\hline
 $[0.01: 0.5]$ & $200$ & $0.05$ & $0.002$ & $80$ & $40 ({\rm s})$ & $10^{-6}$ &
 $20$ & $0.08$ \\
 $[1 : 5]$ & $100$ & $0.02$ & $0.005$ & $40$ & $200 ({\rm s})$ & $10^{-6}$ &
 $100$ & $0.02$ \\
 $[10:1000]$ & $100$ & $0.1$ & $0.005$ & $40$ & $1000 ({\rm fs})$ & $10^{-3}$ &
 $1000$ & $0.002$
\end{tabular}
\caption{Simulation parameters for the RLB solver, as employed for the ranges of $4\pi \eta / s$ displayed in the left column. The notation is explained in this appendix.}
\label{tbl:RLB}
\end{table}

We now summarize the characteristics and parameters of our RLB solver. The advection operator $\vT \cdot \nabla_\perp$ is implemented using the upwind-biased finite-difference fifth-order weighted essentially non-oscillatory (WENO) scheme \cite{Jiang:1996,Rezzolla:2013} (see Ref.~\cite{Ambrus:2018kug} for details). The spatial domain consists of a square box of size $16R$ centered on the system's center of mass and is discretized equidistantly using $S^2$ cells. Periodic boundary conditions are employed at the domain edges. When initializing the system, a background value $\epsilon_{\rm th} = 10^{-10} \times \frac{R}{\tau_0} \epsilon_{\rm ref}$ is added to the energy density to avoid numerical underflow.

The time stepping is performed by solving the equation $\tilde{\partial}_\tau \mathcal{F}_* = L[\tilde{\tau}, \mathcal{F}_*(\tilde{\tau})]$ using the third-order Runge-Kutta scheme \cite{Shu:1988,Gottlieb:1998,Rezzolla:2013}, as described in Ref.~\cite{Ambrus:2018kug}. The time step $\delta \tilde{\tau}$ is chosen dynamically as
\begin{equation}
 \delta \tilde{\tau}(\tilde{\tau}) = {\rm min} \left[\tilde{\tau} \left(\frac{\delta \tilde{\tau}}{\tilde{\tau}}\right)_{\rm M}, 
 \frac{{\rm max}_\perp(\tilde{\tau}_R)}{2},
 \delta \tilde{\tau}_{\rm M}\right],
\end{equation}
where 
${\rm max}_\perp(\tilde{\tau}_R)$ represents the maximum value of $\tilde{\tau}_R(\tilde{\tau},\tilde{\xT})$ taken over the entire flow domain at time $\tilde{\tau} = \tau / R$,
while the values of $(\delta \tilde{\tau} / \tilde{\tau})_{\rm M}$ and $\delta \tilde{\tau}_{\rm M}$ are shown in Table~\ref{tbl:RLB}.

The discretization of $\delta \varphi_p$ is done equidistantly using $Q_\varphi$ points (the employed values of $Q_\varphi$ are summarized in Table~\ref{tbl:RLB}). 

The $v_{z;*}$ degree of freedom is discretized using $Q_z$ values. When employing the spherical coordinate $v_{z;{\rm s}}$, these points are chosen according to the Gauss-Legendre quadrature rules as the roots of the Legendre polynomial of order $Q_z$, i.e., $P_{Q_z}(v_{z;j}) = 0$. When $v_{z;{\rm fs}}$ is employed, the discretization is performed equidistantly at the level of the parameter $\chi = {\rm artanh}(A v_{z;{\rm fs}})$, namely $\chi_j = (\frac{2j - 1}{Q_z} - 1) {\rm artanh}A$. In this paper, we take $1 - A = 10^{-6}$ (see Sec. IV.B of Ref.~\cite{Ambrus:2021fej} for more details).

As shown in Table~\ref{tbl:RLB}, the (s) and (fs) approaches are employed when $4\pi \eta / s \le 5$ and $4\pi \eta / s \ge 10$, respectively. Employing the (s) approach at larger values of $4\pi \eta / s$ requires increasing $Q_z$, otherwise the time evolution leads to energy-momentum tensor configurations which are incompatible with the Landau frame. Using $Q_z = 200$ gives reliable results for $4\pi \eta / s \lesssim 10$. Because the computation of the force term involving $\partial_{v_z} \mathcal{F}$ is quadratic with respect to $Q_z$ (see Sec. III.E of Ref.~\cite{Ambrus:2018kug} for details), the (s) strategy becomes inefficient when $Q_z \gtrsim 200$. 

Conversely, the (fs) approach requires larger $Q_z$ as $\tau_f / \tau_0$ is increased (we ran all simulations up to $\tau_f = 5R$). Since in the (fs) approach, the computational time scales linearly with $Q_z$, we employed $Q_z = 1000$. With our choice of parameters, this limits the lower value of $\tau_0$ to $10^{-3} R$, which is insufficient to correctly capture the early-time dynamics of the system when $4\pi \eta / s \lesssim 1$.

Finally, the choice of $\xi_0$ in preparing the initial state depends on the $v_{z;*}$ resolution offered by the chosen discretization. As $\xi_0 \rightarrow \infty$, the initial state becomes peaked around $v_z = 0$, hence the $v_{z;*}$ discretization must include sufficient points around this value. We found that the influence of the initial value of $\xi_0$ on the observables is less significant at smaller values of $4\pi \eta / s$. Thus, we employed progressively larger values of $\xi_0$ 
as we increased $\eta / s$, 
which were compatible with the discretization of $v_{z;*}$, as shown in Table~\ref{tbl:RLB}.

\section{Numerical code computing linear order results}\label{app:numerical_LO}

In this appendix, we discuss the numerical code needed for obtaining the linear order results discussed in Sec.~\ref{sec:opacity_expansion}. Sec.~\ref{app:setup_numerical_LO} discusses the conceptual setup of the code and Sec.~\ref{app:details_numerical_LO} deals with the details of how the integration is performed.

\subsection{Setup of the linear order  code}\label{app:setup_numerical_LO}

The code is set up to compute the zeroth and first order contributions to the energy-momentum tensor, which is given in terms of the phase space density as
\begin{align}
     T^{\mu\nu}=\frac{\nu_{\rm eff}}{(2\pi)^3}\int \d^2 p_\perp \int \d y~ p^\mu p^\nu f \;.
\end{align}
For simplicity, observables that are nonlinear in $T^{\mu\nu}$ with contributions from both zeroth and first order in the opacity expansion were computed only to zeroth order.

The code is set up as follows. For an arbitrary initial energy density distribution $\epsilon_0(\tau_0,\xT)$, the free-streaming energy momentum tensor is given as 
\begin{align}
     T^{(0)\mu\nu}=\frac{\tau_0}{\tau}\int \frac{\d \phi_v}{2\pi}\,v_\perp^\mu v_\perp^\nu\,  \epsilon(\tau_0,\xT-\Delta\tau\vT) \ \ ,\label{eq:Tmunu_zeroth_order}
\end{align}
where $\Delta \tau = \tau - \tau_0$.
The integral over $\phi_v$ is performed numerically, using the same stencils for all entries to prevent errors later on. Now, to go to first order in the opacity expansion, we first have to compute the zeroth order results for the restframe energy density $\epsilon(\tau,\xT)$ and the flow velocity $u^\mu(\tau,\xT)$, as they are required for evaluating the RTA collision kernel. This is achieved by numerical diagonalization of $T^{(0)\mu\nu}$.

As computed before \cite{Ambrus:2021fej}, the first order correction to the phase space distribution is given as an integral of the zeroth order collision kernel:
\begin{multline}
     f^{(1)}(\tau,\xT,\pT,y-\eta) \\
     =\int_{\tau_0}^\tau \mathrm{d}\tau ' \ \frac{C[f^{(0)}]}{p^\tau}\left(\tau ',\mathbf{x}_\perp',\pT,y'-\eta\right),
     \label{eq:LO_f1}
\end{multline}
 where $f^{(0)}$ is the free-streaming solution given in Eq.~\eqref{eq:LO_f0}. The primes on the variables indicate the use of free-streaming coordinates,
\begin{align}
 \mathbf{x}_\perp' &= \xT - \vT t(\tau,\tau',y-\eta), \nonumber \\
 y' &= \eta + {\rm arcsinh}\left(\frac{\tau}{\tau'} \sinh(y - \eta)\right),
 \label{eq:LO_fs}
\end{align}
with $t(\tau,\tau', y-\eta)$ being given in Eq.~\eqref{eq:LO_fs_t}.

From this, the first order correction to the energy-momentum tensor is obtained as
\begin{multline}
 T^{(1)\mu\nu} = \frac{\nu_{\rm eff}}{(2\pi)^3}
\int_{\pT} \int dy~ p^\mu p^\nu \int_{\tau_0}^{\tau} \d\tau'~ \\
\times \frac{C[f^{(0)}](\tau',\xT',\pT,y'-\eta)}
{p^{\tau}(\pT,y'-\eta)}\;,\label{eq:LO_Tmunu}
\end{multline}
where $\int_\pT \equiv \int d^2\pT$.
As it turns out, the observables that are to be computed to first order in opacity depend only on transverse integrals of the components of $T^{ij}$.
Thus, we need to perform a 6D integral, which can be done in part analytically, reducing the complexity of the numerical integration. For further details of the analytical preparatory groundwork for the numerical implementation, see Appendix~\ref{app:details_numerical_LO}.

The observables are now computed from these results in the following way. 
In the case of the transverse-plane energy, we have
\begin{align}
    \frac{\d E_{\rm tr}}{\d \eta}&=\tau \int_{\xT} (T^{11}+T^{22}) \nonumber\\
    &= \tau \int_{\xT} (T^{(0)11}+T^{(0)22}+T^{(1)11}+T^{(1)22}) \; .
\end{align}
As elliptic flow is given as a quotient of two transverse integrals of components of $T^{\mu\nu}$ where the numerator 
vanishes at zeroth order, the first order result is given as
\begin{align}
      e^{2i \Psi_p}
     \varepsilon_p&=\frac{\int_{\xT} (T^{11}-T^{22}+2iT^{12})}{\int_{\xT} (T^{11}+T^{22})} \nonumber\\
     &=\frac{\int_{\xT} (T^{(1)11}-T^{(1)22}+2iT^{(1)12})}{\int_{\xT} (T^{(0)11}+T^{(0)22})} \;.
\end{align}
Both of these observables depend on the transverse integral of $T^{(0)11}+T^{(0)22}$, which using~\ref{eq:Tmunu_zeroth_order} can be straightforwardly evaluated to
\begin{align}
    \int_{\xT} T^{(0)11}+T^{(0)22}=\frac{1}{\tau}\epn \;.
\end{align}
In particular, the quantity $\d E_{\rm tr}/\d\eta$, which we introduced as the analogue of $\d E_\perp/\d\eta = \int_\xT \langle p^\tau p_\perp \rangle$, is in fact identical to $\d E_\perp/\d\eta$ to zeroth order. Furthermore it is constant in time, so only the first order correction has to be computed. We furthermore compute zeroth order results for the average transverse flow velocity and the average inverse Reynolds number as
\begin{align}
    \langle u_\perp \rangle&=\frac{\int_{\xT}\epsilon^{(0)} \sqrt{\left(u_1^{(0)}\right)^2+\left(u_2^{(0)}\right)^2} }{\int_{\xT} \epsilon^{(0)}}\;,\\
    \langle \mathrm{Re}^{-1} \rangle &= \frac{\int_{\xT} \sqrt{6\pi^{(0)\mu\nu}\pi^{(0)}_{\mu\nu}}}{\int_{\xT} \epsilon^{(0)}} \nonumber\\
    &=\frac{\int_{\xT} \sqrt{6T^{(0)\mu\nu}T^{(0)}_{\mu\nu}-\frac{24}{3}\left(\epsilon^{(0)}\right)^2}}{\int_{\xT} \epsilon^{(0)}}\;.
\end{align}

\subsection{Analytical and numerical integration in the computation of linear order results}\label{app:details_numerical_LO}

As discussed in the previous appendix, obtaining numerical results for the linear order term in the energy momentum tensor requires the computation of a 6D integral. In this appendix, we explain what part of this integral is performed analytically and give the specific form of the remaining integral which the code computes numerically.

We start from the expression in Eq.~\eqref{eq:LO_Tmunu} for the purely spatial components of the energy momentum tensor, 
\begin{multline}
 \int_{\xT} T^{(1)ij} = 
\frac{\nu_{\rm eff}}{(2\pi)^3}\int_{\xT}\int_{\pT} \int \d y~ \mathbf{p}_\perp^i \mathbf{p}_\perp^j \\\times 
\int_{\tau_0}^{\tau} \d\tau' \frac{C[f^{(0)}]}{p^{\tau}}
\left(\tau',\mathbf{x}_\perp',\pT,y'-\eta\right)\;,
\label{eq:app_T1_aux}
\end{multline}
where the free-streaming coordinates $\mathbf{x}_\perp'$ and $y'$ were introduced in Eq.~\eqref{eq:LO_fs}.

The integration variables can be changed from $(\xT, y)$ to 
 $(\mathbf{x}_\perp', y')$, where
\begin{align}
 \d y' d^2\mathbf{x}_\perp' &=\frac{\tau}{\tau'}\frac{\cosh(y-\eta)}{\cosh (y'-\eta)}\d y d^2\xT \\
 &=\frac{\tau}{\tau'}\frac{\sqrt{1+\left(\frac{\tau'}{\tau}\right)^2\sinh^2 (y'-\eta)}}{\cosh (y'-\eta)}\d y d^2\xT\;.
 \label{eq:app_coord_change}
\end{align}
Right away and from this point on, we will drop the primes on all coordinates except $\tau'$ for convenience. 
The specific form of the RTA kernel is
\begin{align}
    C[f]&=-\frac{p^\mu u_\mu}{\tau_R} (f-f_{\rm eq})\;,
    \label{eq:app_Cf}
\end{align}
where $\tau_R=5 (\eta / s) T^{-1}$ is the relaxation time and
\begin{equation}
 p^\mu u_\mu = \gamma p_\perp[\cosh(y - \eta) - \vT \cdot \bm{\beta}], \label{eq:app_pu}
\end{equation}
where $\vT = \pT / p_\perp = (\cos\varphi_p, \sin\varphi_p)$ is a unit vector in the transverse plane. Similarly, $\bm{\beta} \equiv \bm{u}_\perp / u^\tau = \beta(\cos\varphi_u, \sin\varphi_u)$ is the transverse-plane fluid velocity and $\gamma = 1 / \sqrt{1 - \beta^2}$ is the local Lorentz factor.
Plugging 
 Eqs.~\eqref{eq:app_coord_change}--\eqref{eq:app_pu} into Eq.~\eqref{eq:app_T1_aux}, 
we arrive at
\begin{multline}
\int_{\xT} T^{(1)ij} = 
- \frac{\nu_{\rm eff}}{(2\pi)^3} \int_{\pT} \int_{\xT} \int \d y \int_{\tau_0}^{\tau} \frac{\tau' \d\tau'}{\tau} \\ 
\times \frac{\mathbf{v}_\perp^i
\mathbf{v}_\perp^j}{ \tau_R} \frac{\gamma[\cosh(y - \eta) - \vT \cdot \bm{\beta}]}{\sqrt{1+\left(\frac{\tau'}{\tau}\right)^2\sinh^2(y-\eta)}} \\\times
p_\perp^2 (f^{(0)}-f_{\rm eq})(\tau',\xT,\pT,y-\eta)\;.\label{eq:intTijAnsatz_redefined} 
\end{multline}
In the above, all macroscopic quantities $\tau_R$, $\gamma$ and $\bm{\beta}$ are computed from the zeroth order solution, $f^{(0)}$.

It is convenient to consider separately the contributions 
involving $f^{(0)}$ and $f_{\rm eq}$.
In the case of the former, we plug in
\begin{multline}
 f^{(0)}(\tau,\xT,\pT,y-\eta) =\frac{(2\pi)^3}{\nu_{\rm eff}}\frac{\delta(y-\eta)}{\tau{p}_\perp} \\\times 
 \frac{\d N_{0}}{ \d^2\xT\d^2 \pT\d y}\left(\xT-\vT\Delta\tau,\pT\right)\;.
\end{multline}
Using the relation
\begin{multline}
  \epsilon(\tau_0, \xT-\vT\Delta\tau)\\
  =\frac{2\pi}{\tau_0} \int \d p_\perp~p_\perp^2~\frac{\d N_{0}}{ \d^2\xT \d^2 \pT\d y}\left(\xT-\vT\Delta\tau,\pT\right) \;,\label{eq:epsilon-f-relation}
  \end{multline}
it is not difficult to obtain:
\begin{multline}
    \int_\xT T^{(1)ij}_{(0)} = -\frac{\tau_0}{\tau} \int_{\tau_0}^{\tau} \d{\tau'} \int_\xT \frac{\gamma}{\tau_R} \int \frac{\d\varphi_p}{2\pi}\\ 
    \times {\mathbf{v}_\perp^i \mathbf{v}_\perp^j} (1 - \vT \cdot \bm{\beta})
    \epsilon\left(\tau_0,\xT-\vT\Delta{\tau'} \right) \;,
    \label{eq:LO_T1ij0}
\end{multline}
where $\gamma$, $\tau_R$ and $\bm{\beta}$ are evaluated at $(\tau', \xT)$. 

For the equilibrium buildup contribution, we can use the property
\begin{multline}
 \frac{\nu_{\rm eff}}{(2\pi)^3}\int_0^\infty \d p_\perp \, p_\perp^3 f_{\rm eq}\left(\frac{p^\mu u_\mu}{T}\right)
 = 
  \frac{1}{4\pi} 
 \frac{\epsilon}{(p^\mu u_\mu/p_\perp)^4},
\end{multline}
leading to
\begin{equation}
 \int_\xT T^{(1)ij}_{\rm eq} = 
 \int_\xT \int_{\tau_0}^{\tau} \frac{\tau' \d\tau'}{\tau} \frac{\epsilon^{(0)}}{\tau_R } F^{ij}_{\rm eq},
 \label{eq:LO_T1ijeq}
\end{equation}
where 
\begin{align}
 F^{ij}_{\rm eq} &= \int \frac{\d y}{2}
\int \frac{\d\varphi_p}{2\pi}
\frac{(p^\mu u_\mu/p_\perp)^{-3}}{\sqrt{1+\left(\frac{\tau'}{\tau}\right)^2\sinh^2(y-\eta)}} \mathbf{v}_\perp^i \mathbf{v}_\perp^j \; \nonumber\\
 &= \delta^{ij} I_{3/2} + 3 \beta^i \beta^j I_{5/2}.
\end{align}
In the above, we introduced 
\begin{equation}
 I_\alpha = \frac{1}{4} \int_{-\infty}^\infty dy
 \frac{[\cosh^2(y-\eta) - \beta^2]^{-\alpha}}{\gamma^3\sqrt{1 + (\frac{\tau'}{\tau})^2 \sinh^2(y-\eta)}}.
\end{equation}
It is understood that in Eq.~\eqref{eq:LO_T1ijeq}, the quantities $\epsilon^{(0)}$, $\tau_R$, $\gamma$ and $F^{ij}_{\rm eq}$ are evaluated at $(\tau', \xT)$. 

 Altogether, $T^{(1)ij}$ can be computed using the following formula:
\begin{multline}
\int_{\xT} T^{(1)ij} = \int_{\tau_0}^{\tau} d{\tau'} \int_{\xT} \frac{1}{\tau_R} \int \frac{d\varphi_p}{2\pi} \Bigg[
\frac{\tau'}{\tau}\epsilon^{(0)}
F^{ij}_{\rm eq}\\ -
\frac{\tau_0}{\tau} \epsilon(\tau_0, \xT - \vT \Delta \tau') 
\gamma\mathbf{v}_\perp^i \mathbf{v}_\perp^j \left(1 - \vT \cdot \bm{\beta}\right) \Bigg]\;,
\label{eq:LO_T1ij}
\end{multline}
where $\gamma$, $\tau_R$, $\bm{\beta}$, $\epsilon^{(0)}$ and $F^{ij}_{\rm eq}$ are evaluated at $(\tau', \xT)$.
Note that the zeroth order results for the flow velocity $u^{(0)}_\mu$ and the rest frame energy density $\epsilon^{(0)}$ 
entering
$\tau_R$ via the temperature have been
computed in the first step of diagonalizing $T^{(0)\mu\nu}$. Thus, all quantities appearing in the above
integrand are known and the 
 remaining
4D integral can be performed numerically.

\section{Overview of time evolution at different opacities}\label{app:tevo_gdep}

\begin{figure*}
    \centering
    \begin{tabular}{cc}
       
    \includegraphics[width=.49\textwidth]{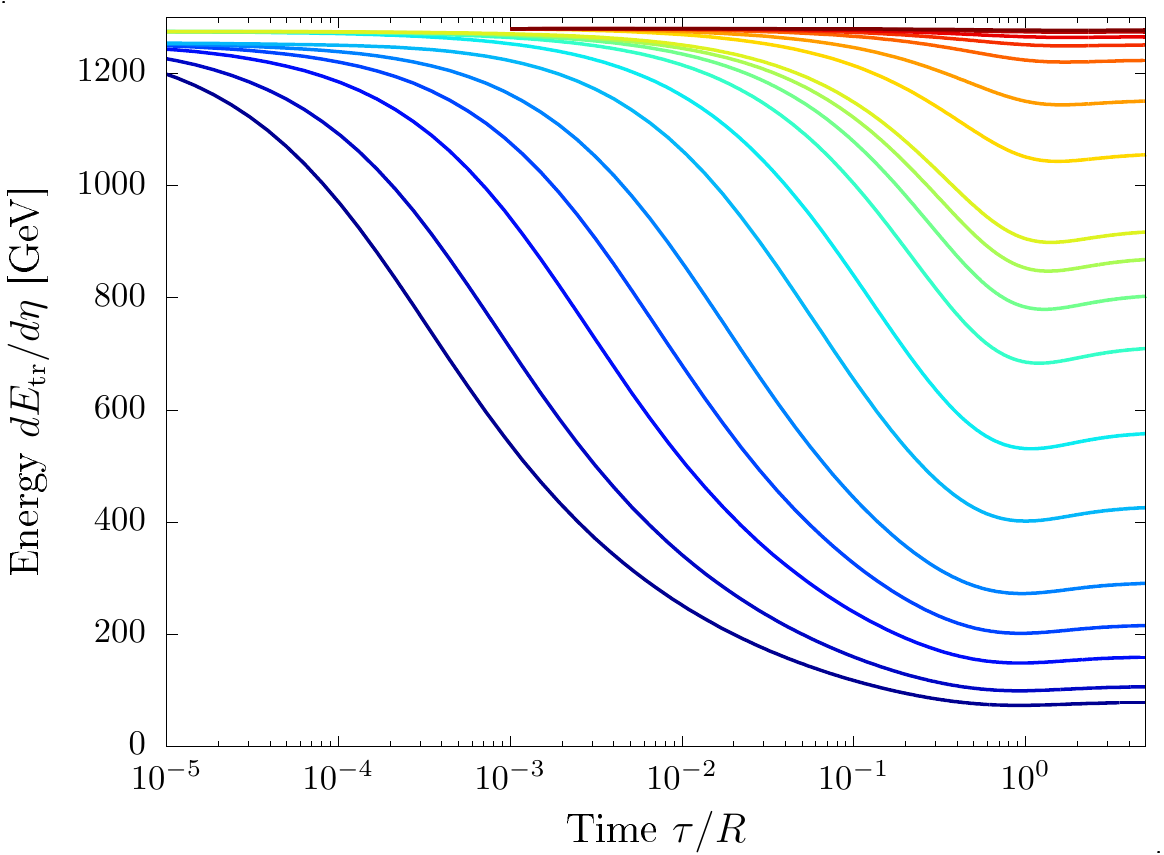}  &  \includegraphics[width=.49\textwidth]{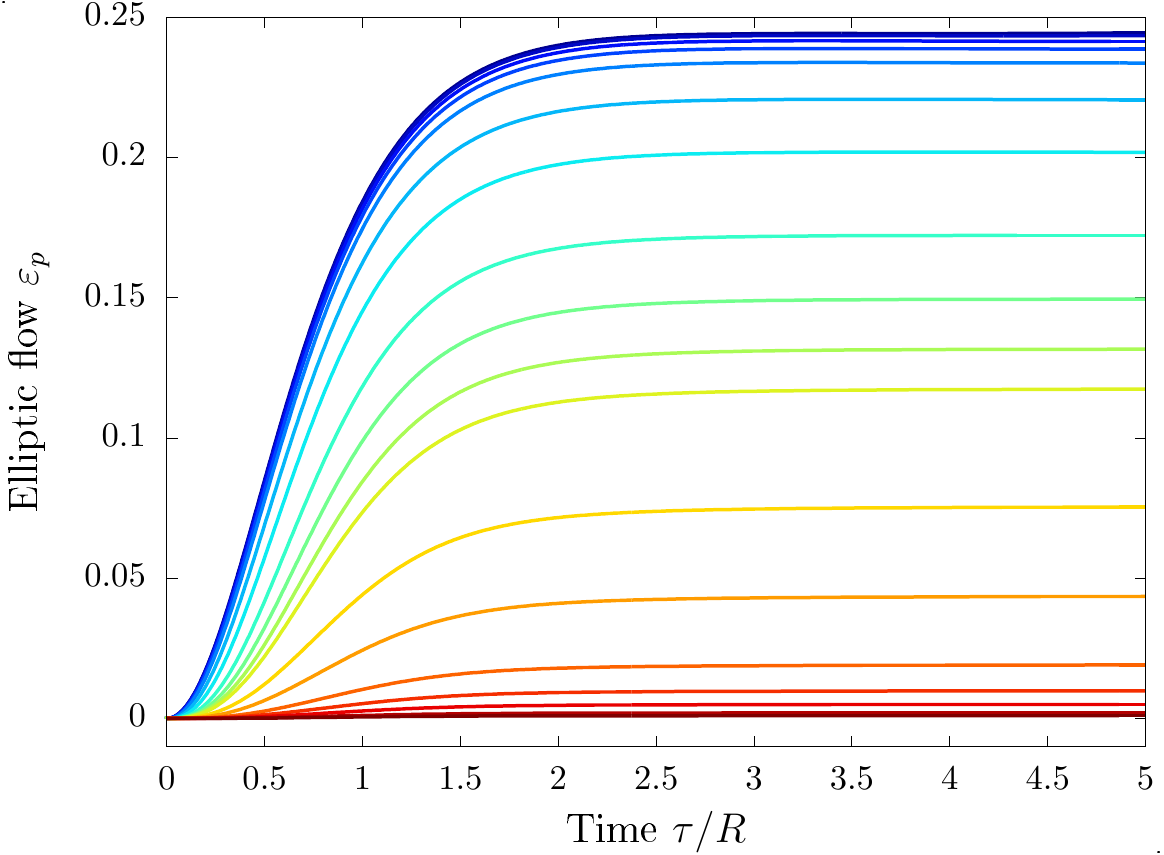} \\
     \includegraphics[width=.49\textwidth]{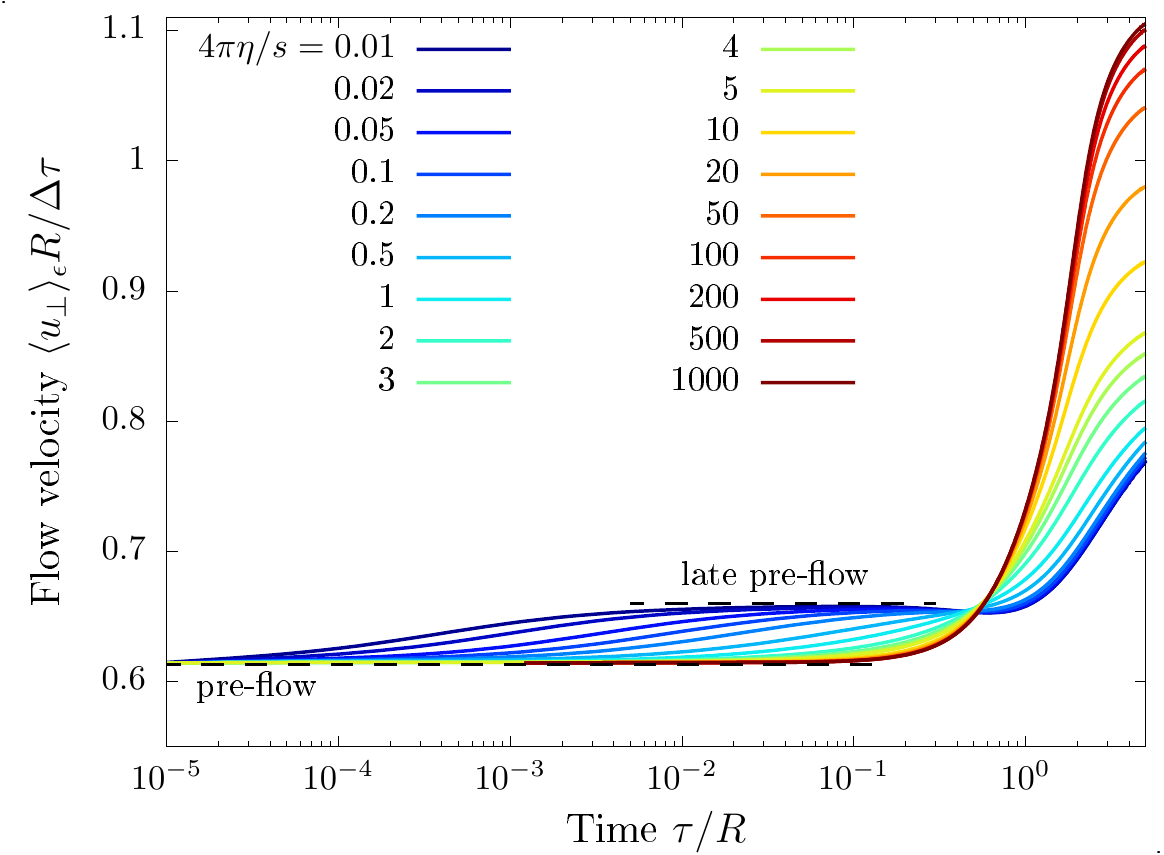}     & \includegraphics[width=.49\textwidth]{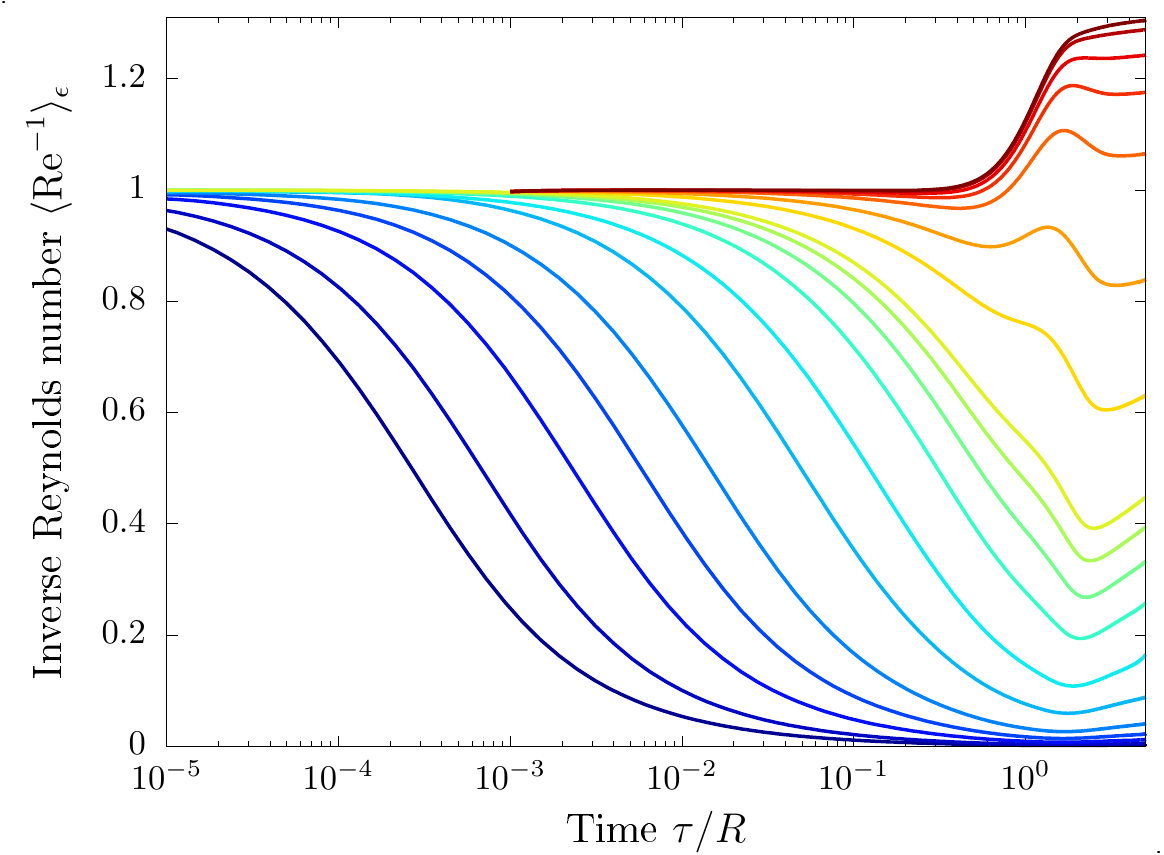} 
    \end{tabular}
    \caption{Time evolution of transverse energy $\d E_{\rm tr}/ \d\eta$ [top left, cf. Eq.~\eqref{eq:obs_dEtrdeta}], elliptic flow $\varepsilon_p$ [top right, cf. Eq.~\eqref{eq:obs_epsp}], transverse flow velocity $\langle u_\perp \rangle_\epsilon$ [bottom left, cf. Eq.~\eqref{eq:obs_uT}] and inverse Reynolds number $\langle \mathrm{Re}^{-1} \rangle_\epsilon$ [bottom right, cf. Eq.~\eqref{eq:Reinv_def}] in kinetic theory for a wide range of opacities ($\eta/s$) plotted in different colors. The plot of transverse flow velocity $\eavg{u_\perp}$ also shows the pre-flow result $\langle{u_\perp}\rangle_{\epsilon,{\rm early}}=0.614\Delta\tau/R$ according to Eq.~\eqref{eq:flow_duT_early} and the late pre-flow result $\langle u_\perp\rangle_{\epsilon,{\rm late}}=0.658\Delta\tau/R$ according to Eq.~\eqref{eq:flow_duT_late} 
    (see also Table~\ref{tbl:pre-flow})}.
    \label{fig:tevo_gdep}
\end{figure*}

In Sec.~\ref{sec:time_evolution_kinetic_theory_scaled_hydro} we compared the time evolution of the tracked observables in kinetic theory and scaled viscous hydro and pointed out some qualitative differences for results at three different opacities. To get a better overview of the opacity dependence in the time evolution, we can also compare results coming exclusively from kinetic theory on a wide range in opacity. This comparison for the time evolution of transverse energy $\d E_{\rm tr}/\d\eta$, elliptic flow $\varepsilon_p$, transverse flow velocity $\eavg{u_\perp}$ and inverse Reynolds number $\eavg{\mathrm{Re}^{-1}}$ is presented in Figure~\ref{fig:tevo_gdep} for opacities ranging from $4\pi\eta/s=0.01$ to $1000$.

For very small opacities $4\pi\eta/s\sim 1000$, the system is close to free-streaming and transverse energy $\d E_{\rm tr}/\d\eta$ is almost constant. At larger opacities, due to more work being performed 
against
the longitudinal expansion, $\d E_{\rm tr}/\d\eta$ decreases
by a larger total amount. The opacity also sets the timescale for this cooling, as it sets in earlier for larger opacities.

Elliptic flow $\varepsilon_p$ stays close to zero at small opacities $4\pi\eta/s\sim 1000$ and rises monotonically with opacity at each point in time. Qualitatively, the curves look the same at all opacities, with a buildup period at times $0.1R\lesssim \tau \lesssim 2R$ and almost constant behaviour afterwards. The onset of this buildup is slightly earlier at larger opacities, but this difference is negligible.

As expected, the transverse flow velocity $\eavg{u_\perp}$ starts with the same early-time linear behaviour for all opacities. The proportionality constant with elapsed time $\Delta\tau=\tau-\tau_0$ can be computed according to Eq.~\eqref{eq:flow_duT_early} and evaluates to $\eavg{u_\perp}=0.614\Delta\tau/R$. The larger the opacity, the earlier the system starts to deviate from this behaviour. For the largest opacities $4\pi\eta/s\lesssim 0.1$, the system is in its local Bjorken flow equilibrium state long enough for early time contributions to become negligible, such that it transitions to the late time pre-flow proportionality law. According to Eq.~\eqref{eq:flow_duT_late}, in this regime, the flow velocity is given by $\langle u_\perp \rangle_\epsilon=0.658\Delta\tau/R$. All curves exhibit their strongest rise on the timescale of transverse expansion, $\tau\sim R$. The rise is stronger at smaller opacities and in all cases contributes the most to the buildup, such that the final ($\tau=4R$) values of transverse flow velocity are also larger at smaller opacities.

The inverse Reynolds number $\eavg{\mathrm{Re}^{-1}}$ stays almost constant at early times for small opacities $4\pi\eta/s\sim 1000$, but then slightly increases due to transverse expansion. At large enough opacities $4\pi\eta/s\lesssim 10$, interactions equilibrate the system and decrease its value. This process sets in earlier at larger opacities and brings the value of the inverse Reynolds number down to almost zero for the largest opacities $4\pi\eta/s \lesssim 0.05$. In these cases, the value stays close to zero even during transverse expansion. At slightly smaller opacities $0.05 \lesssim 4\pi\eta/s\lesssim 1$, there is a small rise in inverse Reynolds number due to transverse expansion. However, this sets in later than in the case of the smallest opacities. The curves for intermediate to small opacities $1\lesssim 4\pi\eta/s\lesssim 100$ exhibit a bumpy behaviour during transverse expansion.

\section{Time evolution in K{\o}MP{\o}ST + viscous hydro simulations}\label{app:time_evolution_hybrid_kompost}

\begin{figure*}
    \centering
    \begin{tabular}{ccc}
   \small $4\pi\eta/s=0.5$&\small$4\pi\eta/s=3$&\small$4\pi\eta/s=10$\\
        \includegraphics[width=.3\textwidth]{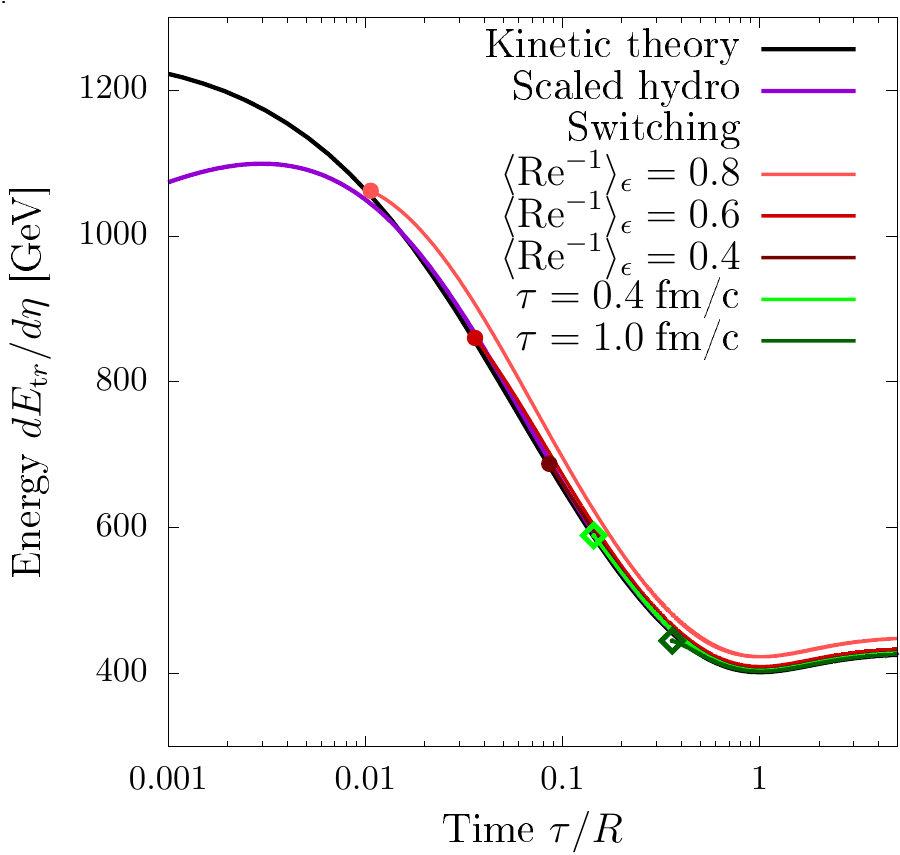}&
        \includegraphics[width=.3\textwidth]{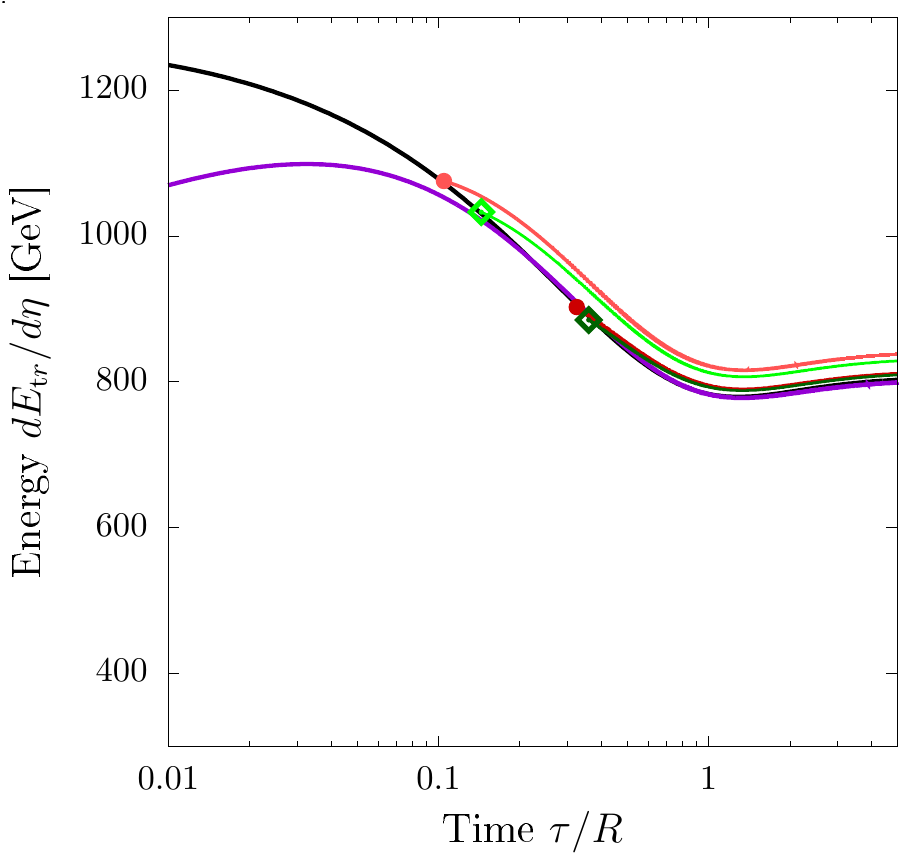}&
        \includegraphics[width=.3\textwidth]{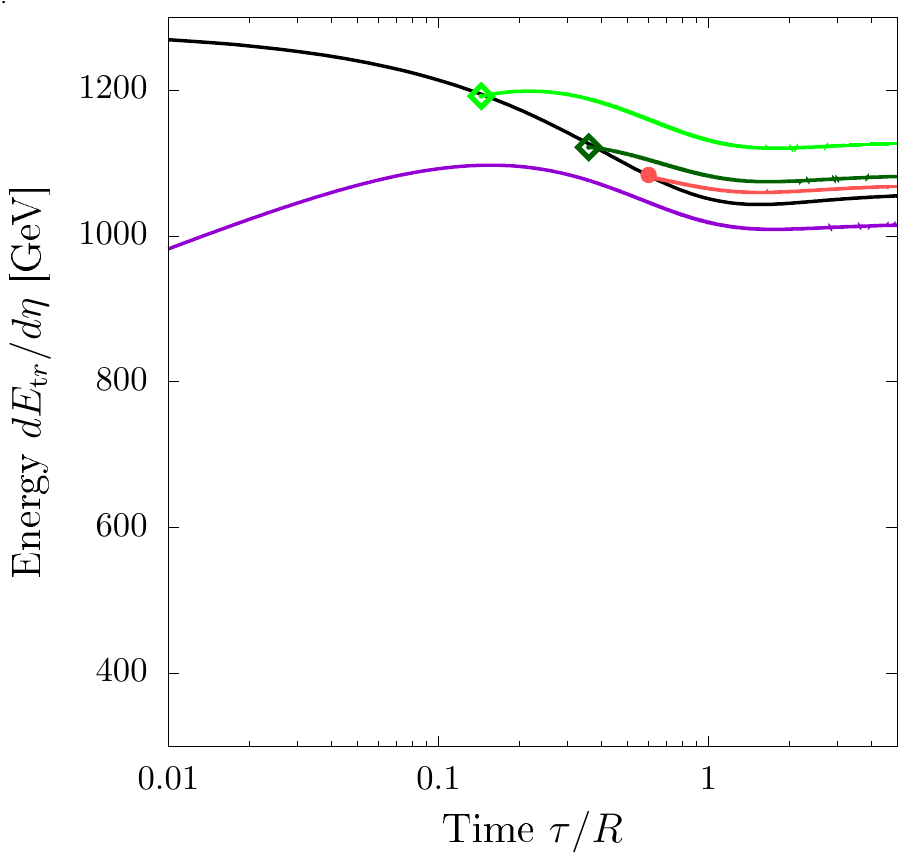}\\
        \includegraphics[width=.3\textwidth]{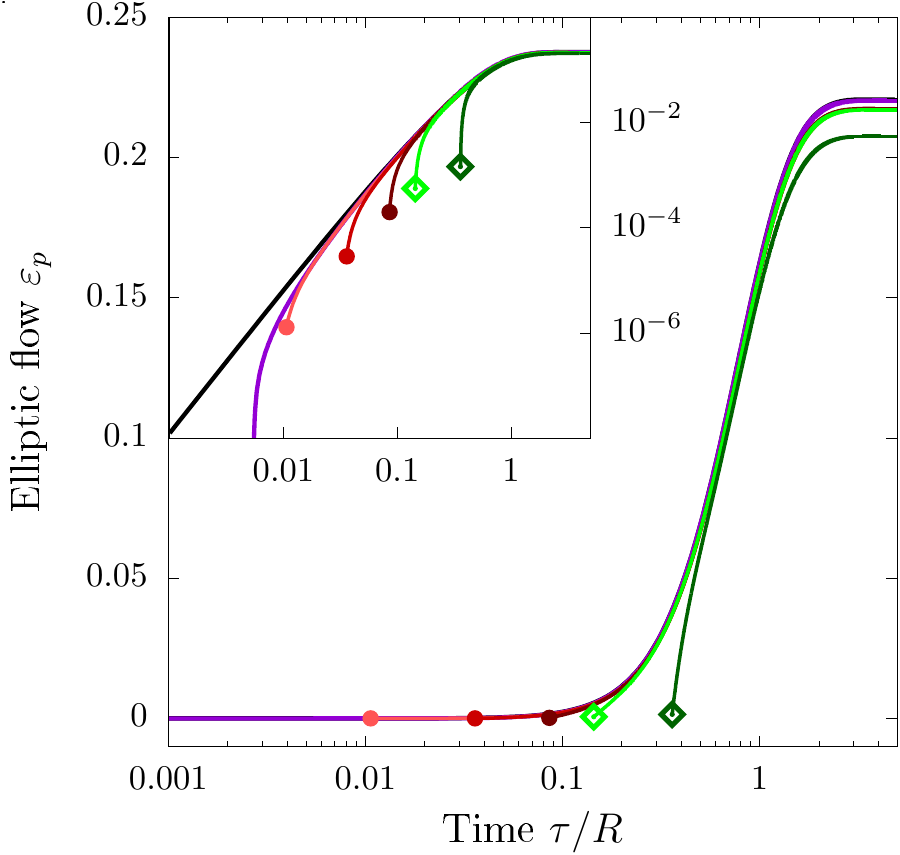}&
        \includegraphics[width=.3\textwidth]{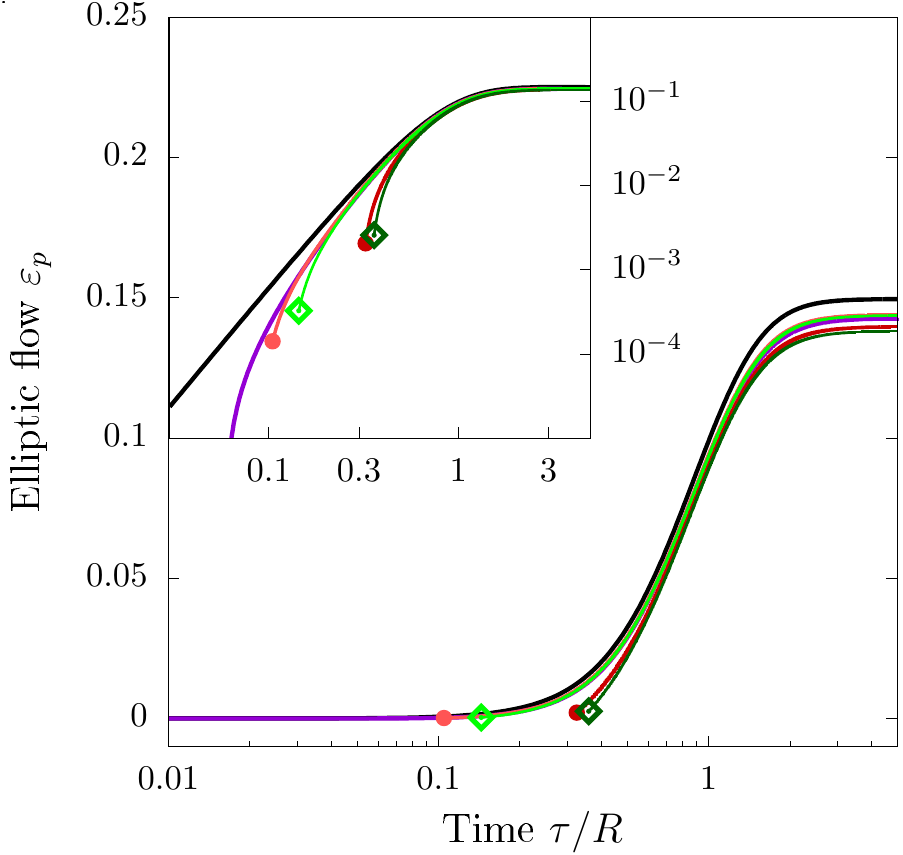}&
        \includegraphics[width=.3\textwidth]{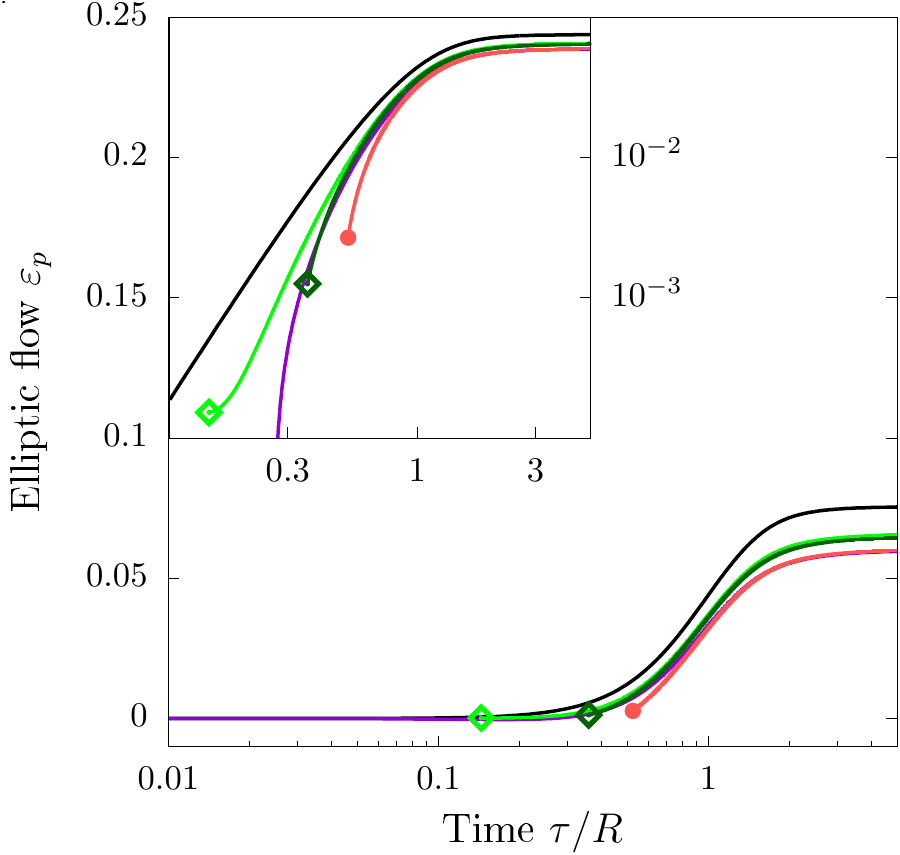}\\
        \includegraphics[width=.3\textwidth]{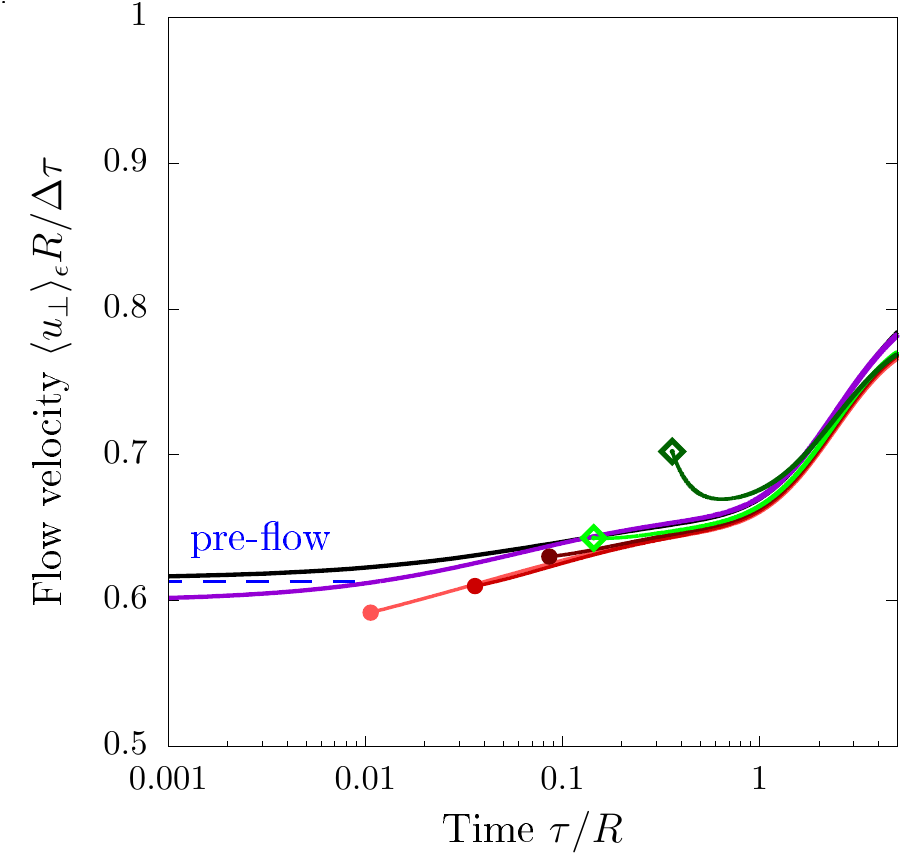}&
        \includegraphics[width=.3\textwidth]{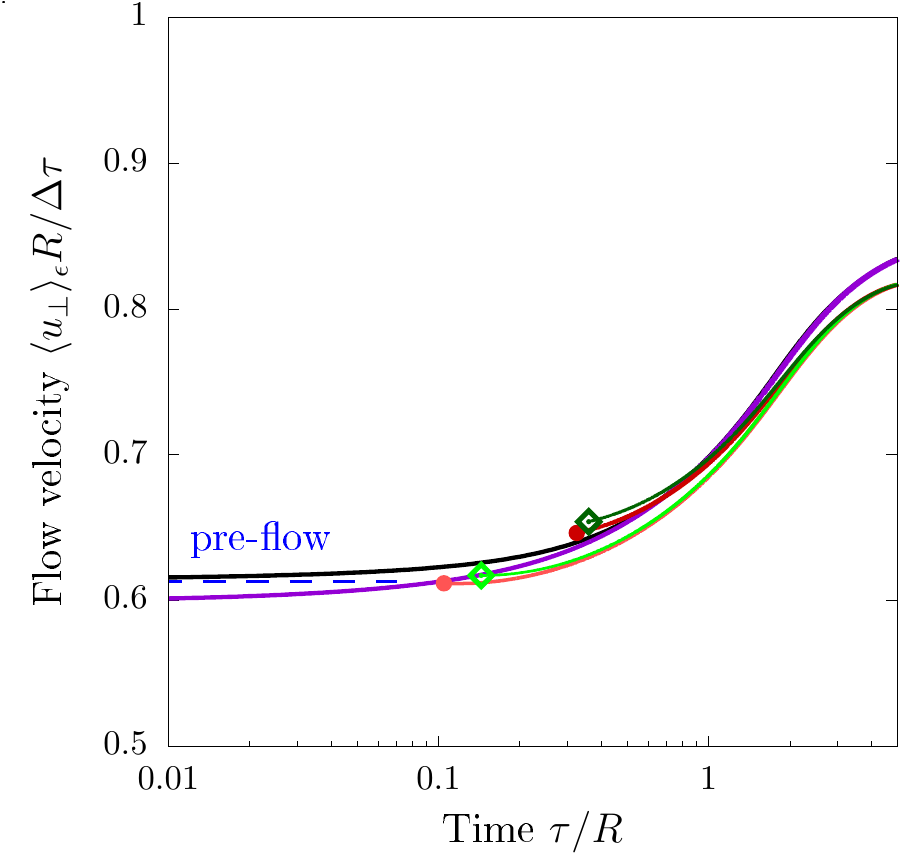}&
        \includegraphics[width=.3\textwidth]{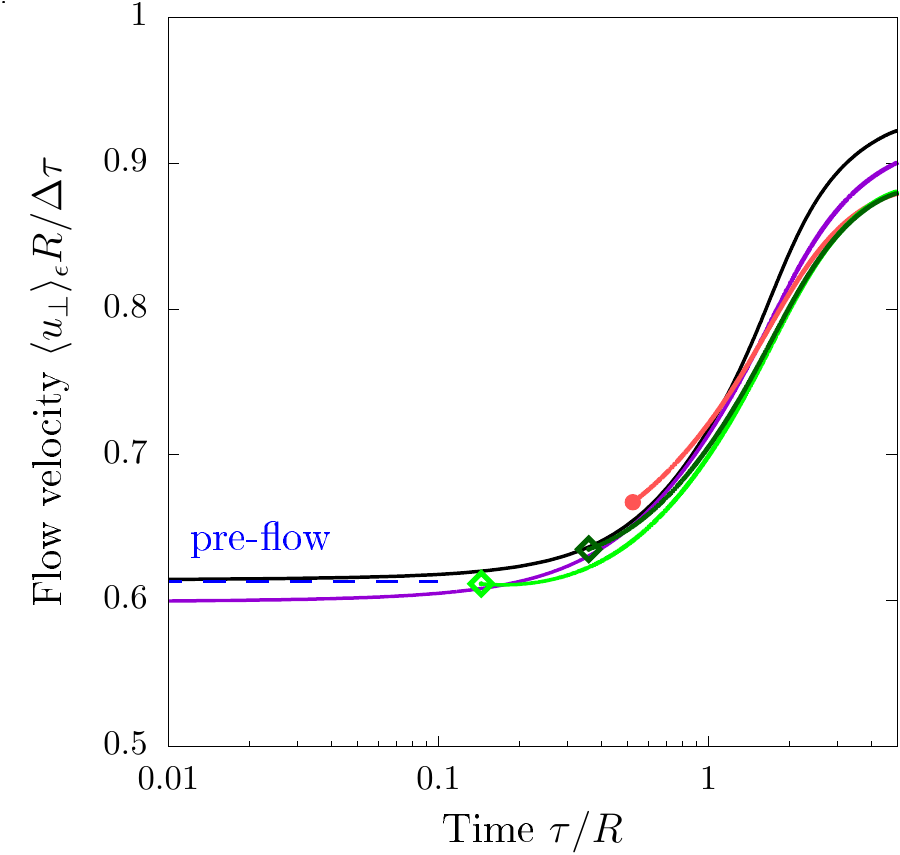}
    \end{tabular}
    
    \caption{Time evolution of transverse energy $\d E_{\rm tr}/\d\eta$ [top, cf. Eq.~\eqref{eq:obs_dEtrdeta}], elliptic flow $\varepsilon_p$ [middle, cf. Eq.~\eqref{eq:obs_epsp}] and transverse flow velocity $\eavg{u_\perp}$ [bottom, cf. Eq.~\eqref{eq:obs_uT}] in hybrid K{\o}MP{\o}ST + viscous hydro simulations at opacities $4\pi\eta/s=0.5$ (left), $3$ (middle) and $10$ (right) when switching at different values of the inverse Reynolds number [cf. Eq.~\eqref{eq:Reinv_def}] $\eavg{\mathrm{Re}^{-1}}=0.8$ (light red), $0.6$ (red) and $0.4$ (dark red) or fixed time $\tau=0.4\,\mathrm{fm}$ (light green) and $\tau=1\,\mathrm{fm}$ (dark green). The switching points are marked with filled symbols. The time axis is scaled logarithmically. The plots showing elliptic flow $\varepsilon_p$ feature an inset plot of the same quantity plotted in log-log scale. Again, the flow velocity plots also show the pre-flow result $\eavg{u_\perp}=0.614\Delta\tau/R$ according to Eq.~\eqref{eq:flow_duT_early}.}
    \label{fig:time_evolution_hybrid_kompost}
\end{figure*}

In Sec.~\ref{sec:time_evolution_hybrid} we considered
hybrid simulation frameworks as a solution for alleviating problems with pre-equilibrium in hydrodynamic simulations and discussed the time evolution mainly in hybrid kinetic theory + viscous hydro simulations. The alternative hybrid framework using K{\o}MP{\o}ST instead of full kinetic theory for the pre-equilibrium evolution has some limitations but, when applicable, yields results of similar accuracy. The time evolution of transverse energy $\d E_{\rm tr}/\d\eta$, elliptic flow $\varepsilon_p$ and transverse flow velocity $\eavg{u_\perp}$ in K{\o}MP{\o}ST + viscous hydro simulations switching at fixed time $\tau_s$ or fixed value of the inverse Reynolds number $\eavg{\mathrm{Re}^{-1}}$ is shown in
Figure~\ref{fig:time_evolution_hybrid_kompost} for three different opacities $4\pi\eta/s=0.5$, $3$ and $10$.

The values of $\d E_{\rm tr}/\d\eta$ at the time of switching are reproduced by K{\o}MP{\o}ST almost perfectly. As one would expect, the time evolution afterwards follows a very similar behaviour to kinetic theory + viscous hydrodynamics, including the inaccuracies of hydrodynamic pre-equilibrium when switching too early.

As K{\o}MP{\o}ST produces almost no elliptic flow, its value at switching time is close to zero. But the buildup during the hydro part of the simulation proceeds similarly to other simulation schemes, such that the discrepancy to kinetic theory in the final state ($\tau=4R$) is of similar size to the one at switching time. It is therefore larger at larger switching times.

The values of transverse flow velocity $\eavg{u_\perp}$ are in K{\o}MP{\o}ST slightly underestimated for small switching times and slightly overestimated for large switching times. After switching, the curves seem to bend towards the hydrodynamic curve. This bending is mainly due to the division by $\Delta\tau$. $\eavg{u_\perp}$ in the hydro phase of hybrid simulations builds up similarly 
as in pure hydrodynamic simulations. The contributions from later times are much larger than those at early times, such that the discrepancy from early times becomes negligible. At late times, results from all switching times underestimate $\eavg{u_\perp}$ by almost the same amount, similarly to hybrid kinetic theory + viscous hydro simulations.

\bibliography{references}

\end{document}